\documentclass[preprintnumbers,prd,floatfix,nofootinbib,superscriptaddress]{revtex4}

\usepackage[a4paper,
			left=20mm,
			right=20mm,
			top=20mm,
			bottom=20mm]
			{geometry} 

\usepackage[tbtags]{amsmath}  
\usepackage{mathrsfs}
\usepackage{tabularx}
\usepackage{amssymb}          
\usepackage{bm}               
\usepackage{graphicx}         
\usepackage[export]{adjustbox}
\usepackage{hhline,multirow}  
\usepackage{makecell}  
\usepackage{dcolumn}          
\usepackage{slashed}
\usepackage{datetime}
\usepackage{color}
\usepackage[dvipsnames]{xcolor}
\usepackage{slashed}
\usepackage{subfloat}

\usepackage{amscd}            
\usepackage{epsfig}
\usepackage{dcolumn}          
\usepackage[dvipsnames]{xcolor}
\usepackage[normalem]{ulem}
\usepackage{mathtools}
\usepackage{esint}
\usepackage{comment}
\usepackage[utf8]{inputenc}

\RequirePackage{color}
\RequirePackage[colorlinks=true
,urlcolor=black
,anchorcolor=black
,citecolor=black
,filecolor=black
,linkcolor=black
,menucolor=black
,pagecolor=black
,linktocpage=black
,pdfproducer=medialab
,pdfa=true
]{hyperref}

\newcommand{\nocontentsline}[3]{}
\newcommand{\tocless}[2]{\bgroup\let\addcontentsline=\nocontentsline#1{#2}\egroup}

\newcommand{\hermes}{\rm HERMES}
\newcommand{\compass}{\rm COMPASS}

\newcommand{\dynnlo}{{\tt DYNNLO}}

\newcommand{\bb}{b}

\newcommand{\F}{\hat{f}_1}

\newcommand{\kperp}{\boldsymbol{k}_\perp}
\newcommand{\modkperp}{|\boldsymbol{k}_\perp|}
\newcommand{\bT}{\boldsymbol{b}_T}
\newcommand{\modbT}{|\boldsymbol{b}_T|}

\newcommand{\Pperp}{\boldsymbol{P}_\perp}
\newcommand{\modPperp}{|\boldsymbol{P}_\perp|}
\newcommand{\qT}{\bm{q}_{T}}
\newcommand{\T}{\perp}
\newcommand{\PhT}{\bm{P}_{hT}}
\newcommand{\modqT}{|\bm{q}_{T}|}

\allowdisplaybreaks[2]

\begin{document}

\title{Unpolarized Transverse Momentum Distributions from a global fit \\ of
  Drell-Yan and Semi-Inclusive Deep-Inelastic Scattering data
  \\ \vspace{0.2cm}
\normalsize{\textmd{The \textbf{MAP} Collaboration}\footnote{The MAP acronym stands for ``Multi-dimensional Analyses of Partonic distributions''. It refers to a collaboration aimed at studying the three-dimensional structure of hadrons. The public codes released by the collaboration are available at \href{https://github.com/MapCollaboration}{https://github.com/MapCollaboration}.}}
}

\author{Alessandro Bacchetta}
\thanks{Electronic address: alessandro.bacchetta@unipv.it -- \href{https://orcid.org/0000-0002-8824-8355}{ORCID: 0000-0002-8824-8355}}
\affiliation{Dipartimento di Fisica, Universit\`a di Pavia, via Bassi 6, I-27100 Pavia, Italy}
\affiliation{INFN - Sezione di Pavia, via Bassi 6, I-27100 Pavia, Italy}

\author{Valerio Bertone}
\thanks{Electronic address: valerio.bertone@cea.fr -- \href{https://orcid.org/0000-0003-0148-0272}{ORCID: 0000-0003-0148-0272}}
\affiliation{IRFU, CEA, Universit\'e Paris-Saclay, F-91191 Gif-sur-Yvette, France}

\author{Chiara Bissolotti}
\thanks{Electronic address: cbissolotti@anl.gov --
  \href{https://orcid.org/0000-0003-3061-0144}{ORCID: 0000-0003-3061-0144}}
\affiliation{Dipartimento di Fisica, Universit\`a di Pavia, via Bassi 6, I-27100 Pavia, Italy}
\affiliation{HEP Division, Argonne National Laboratory, 9700 S. Cass Avenue, Lemont, IL, 60439 USA}

\author{Giuseppe Bozzi}
\thanks{Electronic address: giuseppe.bozzi@unica.it -- \href{https://orcid.org/0000-0002-2908-6077}{ORCID: 0000-0002-2908-6077}}
\affiliation{Dipartimento di Fisica, Universit\`a di Cagliari,
  Cittadella Universitaria, I-09042
  Monserrato (CA), Italy}
\affiliation{INFN - Sezione di Cagliari, Cittadella Universitaria, I-09042
  Monserrato (CA), Italy}

\author{Matteo Cerutti}
\thanks{Electronic address: matteo.cerutti@pv.infn.it -- \href{https://orcid.org/0000-0001-7238-5657}{ORCID: 0000-0001-7238-5657}}
\affiliation{Dipartimento di Fisica, Universit\`a di Pavia, via Bassi 6, I-27100 Pavia, Italy}
\affiliation{INFN - Sezione di Pavia, via Bassi 6, I-27100 Pavia, Italy}

\author{Fulvio Piacenza}
\thanks{Electronic address: fu.piacenza@gmail.com} 
\affiliation{Dipartimento di Fisica, Universit\`a di Pavia, via Bassi 6, I-27100 Pavia, Italy}

\author{Marco Radici}
\thanks{Electronic address: marco.radici@pv.infn.it -- \href{https://orcid.org/0000-0002-4542-9797}{ORCID: 0000-0002-4542-9797}}
\affiliation{INFN - Sezione di Pavia, via Bassi 6, I-27100 Pavia, Italy}

\author{Andrea Signori}
\thanks{Electronic address: andrea.signori@unipv.it -- \href{https://orcid.org/0000-0001-6640-9659}{ORCID: 0000-0001-6640-9659} -- Currently at Universit\`a di Torino, I-10125 Torino, Italy}
\affiliation{Dipartimento di Fisica, Universit\`a di Pavia, via Bassi 6, I-27100 Pavia, Italy}
\affiliation{INFN - Sezione di Pavia, via Bassi 6, I-27100 Pavia, Italy}

\begin{abstract}
We present an extraction of unpolarized transverse-momentum-dependent parton
distribution and fragmentation functions based on more than two thousand data points
from several experiments for two different processes: semi-inclusive
deep-inelastic scattering and Drell--Yan production. The baseline analysis is
performed using the Monte Carlo replica method and resumming large logarithms
at N$^3$LL accuracy. The resulting description of the data is very
good ($\chi^2/N_{\rm dat} = 1.06$). For semi-inclusive deep-inelastic scattering,
predictions for multiplicities are normalized by factors that cure the discrepancy with data
introduced by higher-order perturbative corrections.
\end{abstract}

\maketitle
\newpage
\tableofcontents
\newpage
\section{Introduction}
\label{s:intro}

In this paper, we present an extraction of Transverse Momentum Distributions
(TMDs)~\cite{Rogers:2015sqa,Diehl:2015uka,Angeles-Martinez:2015sea,Scimemi:2019mlf}
using more than two thousand data points from several experiments for two different
kinds of processes: Semi-Inclusive Deep Inelastic Scattering (SIDIS) and production of
Drell--Yan (DY) lepton pairs, significantly improving our previous analysis~\cite{Bacchetta:2017gcc}.

Building maps of the internal partonic structure of nucleons is a crucial step
towards understanding the interactions between quarks and gluons and the phenomenon of confinement.
A steady progress in the last decades has led to more and more refined versions of such
maps. The TMDs extracted in this article encode information about
the three-dimensional distributions of quarks in momentum space.
The level of sophistication of a TMD extraction essentially depends on
two ingredients: the amount of analyzed data from different processes, namely
how ``global'' the experimental information from which TMDs are extracted,
and the perturbative accuracy reached in the theoretical formalism.

The extraction of TMDs is based on TMD factorization
theorems, which provide a precise definition of the objects to be
extracted, establishing their universality and evolution equations.
In this case, the accuracy of the calculation is defined by the amount of
large logarithms being resummed, that in turn defines the powers of
the strong coupling $\alpha_s$ to be included in the perturbative
quantities~\cite{Collins:2011zzd, Bacchetta:2019sam, Bozzi:2003jy, Muselli:2017bad}.

Of particular relevance is also the combination of analyzed data from
different processes. Similarly to the extraction of collinear Parton
Distribution Functions (PDFs), we can talk about a global fit,
\textit{i.e.}, a fit that leverages the universality of the parton distributions and
uses different processes to constrain them. However, in SIDIS two types of
TMDs enter the cross section: the TMD PDFs, describing how partons are
arranged in the nucleon, and TMD Fragmentation Functions (FFs), describing how
a parton produces a final-state hadron. The knowledge of TMDs, in particular
of TMD FFs, would be greatly improved by using data from electron-positron
annihilations into two almost back-to-back hadrons~\cite{Bacchetta:2015ora}.
Unfortunately, this data is presently not available. There are measurements for the
inclusive production of single hadrons~\cite{Belle:2019ywy} but, in this case, transverse momenta
need to be defined with respect to the thrust axis: a careful description of
the latter is non trivial, and a rigorous factorization theorem for this process has been discussed
only recently (see, \textit{e.g.}, Refs.~\cite{Kang:2020yqw,Makris:2020ltr,Boglione:2021wov}
and references therein). Therefore, for TMD extractions we currently talk about a global fit
when data from SIDIS and DY processes are included.

In the last ten years, several extractions of TMDs have been
presented~\cite{Signori:2013mda,Anselmino:2013lza,Echevarria:2014xaa,
Sun:2014dqm,DAlesio:2014mrz,Bacchetta:2017gcc,Scimemi:2017etj,Bertone:2019nxa,
Scimemi:2019cmh,Bacchetta:2019sam,Hautmann:2020cyp, Bury:2022czx}.
Most of them suffered some shortcomings: they were either obtained in a
parton-model framework without QCD corrections, or they took into account only
a limited set of data, or did not perform a full fit. TMDs were also studied
in a different framework, the so-called parton-branching
approach~\cite{BermudezMartinez:2018fsv,BermudezMartinez:2019anj,BermudezMartinez:2020tys}.

At present, only two works have reached the stage of combining SIDIS and DY
data in a full-fledged global TMD fit: the above mentioned extraction of Ref.~\cite{Bacchetta:2017gcc},
henceforth named PV17, and the extraction of Ref.~\cite{Scimemi:2019cmh}, henceforth named SV19.

The PV17 extraction reached the NLL accuracy, was based on the calculation
of observables at mean kinematics in each bin, and did not manage to describe
the normalization of all datasets.  In this work, we push the accuracy of the
analysis to what we will refer to as N$^3$LL$^{-}$ (only NNLO collinear FFs are currently missing in order to reach full N$^3$LL accuracy).\footnote{While this work was being
completed, further efforts, including a work by the MAP collaboration, were made to include
part of the NNLO corrections~\cite{Borsa:2022vvp, Khalek:2022vgy} in the extraction
of FFs.}
Apart from the increase in perturbative accuracy, in this work we also
include many measurements published after 2017.

The SV19 extraction reached the same perturbative accuracy and included
essentially the same datasets. Our work has crucial differences in the
selection of specific data points (we include a much larger number of
points), the implementation of TMD evolution, the choice of nonperturbative
components, and the handling of normalization for SIDIS data.

As we will describe in detail, we found it particularly difficult to describe
in a satisfactory way the normalization of SIDIS data obtained in
fixed-target experiments at moderate to low scales. When the analysis is
performed at NLL accuracy, we can describe well shape and
normalization of SIDIS data, but the description of high-energy DY
data is very
poor. Going to N$^2$LL and N$^3$LL$^-$, the description of DY data
significantly improves, but we fail to reproduce the normalization of SIDIS
data, mainly due to the ${\cal O} (\alpha_s)$ corrections to the hard
factor. As a possible solution, we choose to adjust the normalization of the TMD
predictions by comparing their integral upon transverse
momentum and the corresponding
collinear formula. With this procedure, we fix the normalization a priori, in
a way that is independent of the results of the fit.

Our baseline fit is performed at N$^3$LL$^-$, using 2031 data
points and obtaining $\chi^2/N_{\rm dat} = 1.06$.  We also discuss variations of this baseline fit by changing the theoretical accuracy
and the selected data.

The paper is organized as follows. In Sec.~\ref{s:formalism} we describe the
theoretical framework used in the analysis. In Sec.~\ref{s:data} we explain how experimental data have been selected. Sec.~\ref{s:results} presents the
results of our extraction. Finally, in Sec.~\ref{s:conclusions} we draw our
conclusions.

\section{Formalism}
\label{s:formalism}

\subsection{Drell--Yan}
\label{ss:DY_Z}

\begin{figure}
\centering
\includegraphics[width=0.6\textwidth]{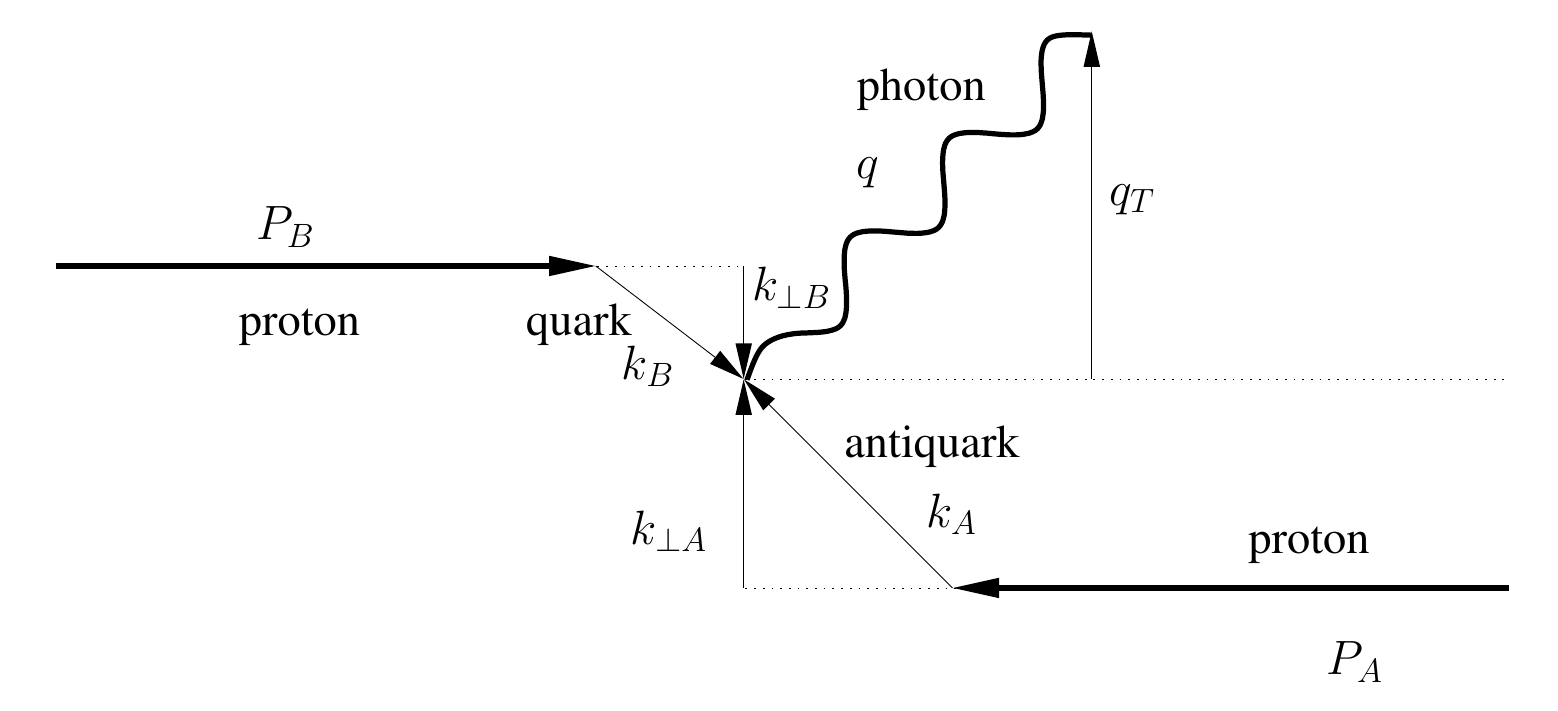}
\caption{Diagram describing the relevant momenta involved in a DY
  event. Two partons from two hadrons collide. They have transverse momenta
  ${\kperp}_A$ and  ${\kperp}_B$ (not measured). They produce a virtual
  photon with (measured) transverse momentum $\qT = {\kperp}_A + {\kperp}_B$ with respect to the hadron
  collision axis.}
\label{f:trans_momenta_DY}
\end{figure}

In the Drell--Yan (DY) process
\begin{equation}
\label{e:DY_Z}
h_A(P_A) + h_B(P_B)\ \longrightarrow\ \gamma^*/Z(q) +
X \longrightarrow \ell^+(l)+ \ell^-(l^\prime) + X \; ,
\end{equation}
two hadrons $h_A$ and $h_B$ with four-momenta $P_A$ and $P_B$, respectively,
collide with center-of-mass energy squared $s = (P_A+P_B)^2$, producing a neutral vector boson $\gamma^*/Z$ with four-momentum $q$ and large invariant mass $Q=\sqrt{q^2}$. The vector boson eventually decays into a lepton and an antilepton with four-momenta constrained by momentum conservation, $q = l + l'$.
The involved momenta and the respective transverse components are summarized in Fig.~\ref{f:trans_momenta_DY}.

The hadronic four momenta $P_A$ and $P_B$ can be chosen to identify the
longitudinal direction $z$ and define the transverse momentum
$\qT$ of the $\gamma^* / Z$.
The rapidity of the neutral boson (or, equivalently, of the lepton pair) is
defined as
\begin{equation}
\label{e:gZ_rapidity}
y = \frac12 \ln \left( \frac{q_0+q_z}{q_0-q_z} \right) \, .
\end{equation}

For our purposes, we need the cross section for this process initiated by unpolarized hadrons and integrated over the azimuthal angle of the exchanged boson.
That cross section can be written in terms of two structure functions, $F_{UU}^1$ and $F_{UU}^2$~\cite{Boer:2006eq,Arnold:2008kf}.
In the limit $M^2 \ll Q^2$ (with $M$ the mass of the incoming hadrons) and $\qT^2 \ll Q^2$, the $F_{UU}^2$ is suppressed.
Accordingly, the cross section reads
\begin{equation}
\label{e:DYZ_xsec}
\frac{d\sigma^{\text{DY/Z}}}{d|\qT|\, dy\, dQ} = \frac{16 \pi^2 \alpha^2}{9 Q^3}\, |\qT|\, {\cal P}\, F_{UU}^1 \big( x_A,x_B,|\qT|,Q \big) \, ,
\end{equation}
where $\alpha$ is the electromagnetic coupling, ${\cal P}$ is the phase space
factor to account for potential cuts on the lepton kinematics, which turns out
to have a relevant impact when high-precision data are taken into account
(see, \textit{e.g.}, a recent analysis in Ref.~\cite{Chen:2022cgv}).\footnote{In the presence of cuts on single-lepton variables, an additional parity-violating term contributes to the cross section~\cite{Boer:1999mm}. However, in Ref.~\cite{Bacchetta:2019sam} it was shown that this contribution is negligible in the experimental conditions considered in this analysis.}
At low transverse momentum $\qT^2 \ll Q^2$, the structure function can be expressed as a convolution over the partonic transverse momenta of two TMD PDFs:
\begin{equation}
\begin{split}
\label{e:FUU1_def}
F_{UU}^1 &\big( x_A,x_B,|\qT|,Q \big)
\\
& =
x_A\, x_B\, {\cal H}^{\text{DY}}(Q,\mu)\sum_a c_a(Q^2) \int d^2 {\kperp}_A\, d^2 {\kperp}_B\,
f_1^a(x_A,{\kperp^2}_A;\mu,\zeta_A)\, f_1^{\bar{a}}(x_B,{\kperp^2}_B;\mu,\zeta_B)\,
\delta^{(2)}({\kperp}_A + {\kperp}_B - \qT) \\
& = \frac{x_A x_B}{2\pi}\, {\cal H}^{\text{DY}}(Q,\mu)\sum_a c_a(Q^2)
\int_0^{+\infty} d|\bT| |\bT| J_0\big( |\bT| |\qT| \big)
\hat{f}_1^a(x_A,\bT^2;\mu,\zeta_A)\,
\hat{f}_1^{\bar{a}}(x_B,\bT^2;\mu,\zeta_B).
\end{split}
\end{equation}
In the above equation, ${\cal H}^{\text{DY}}$ is the hard factor, which can be
computed order by order in the strong coupling $\alpha_s$ and is equal to 1 at
leading order.\footnote{In the
present work,
we follow the definition of Ref.~\cite{Collins:2017oxh}.}
This function encodes the virtual part of the hard scattering and depends on the hard scale $Q$ and on the renormalisation scale $\mu$.
The unpolarized TMDs are denoted by
$f_1$. They depend on the renormalization scale $\mu$ and the rapidity scale
$\zeta$. The rapidity scales must obey the relation
$\zeta_A \zeta_B = Q^4$. Throughout the paper, we will set $\mu^2=\zeta_A= \zeta_B=Q^2$.

The following definition of the Fourier transform of the TMD PDFs has been used:\footnote{Notice that in Ref.~\cite{Bacchetta:2017gcc} the Fourier transform was defined with an extra $1/(2 \pi)$ factor.}
\begin{equation}
\label{eq:FTdef}
  \begin{split}
\F^a \big( x, \modbT; \mu, \zeta \big) &= \int d^2 \bm{k}_\T \, e^{i
      \bm{\bb}_T \cdot \bm{k}_\T  } \, f_1^a \big( x, \bm{k}_\T^2; \mu,
    \zeta \big) \,
    \\
      &= 2 \pi \int_0^{\infty} d \modkperp \,\modkperp  J_0(\modbT \modkperp) \, f_1^a \big( x, \kperp^2; \mu,
    \zeta \big). \,
 \end{split}
\end{equation}

The structure of the TMD PDFs will be addressed in details in Sec.~\ref{ss:TMDs}.
The transverse momentum of the active quark and antiquark are denoted as ${\kperp}_{A,B}$.
At low transverse momenta, the two variables $x_{A,B}$ take the values:
\begin{align}
\label{e:x12_values}
x_A & = \frac{Q}{\sqrt{s}} e^{y} \, ,
&
x_B & = \frac{Q}{\sqrt{s}} e^{-y} \, .
\end{align}

The summation over $a$ in Eq.~\eqref{e:FUU1_def} runs over the active quarks and antiquarks at the scale $Q$, and $c_a(Q^2)$ are the respective electroweak charges given by
\begin{equation}
\label{e:EW_charges}
c_a(Q^2) = e_a^2 - 2 e_a V_a V_\ell \, \chi_1(Q^2) + (V_\ell^2 + A_\ell^2)\, (V_a^2 + A_a^2)\, \chi_2(Q^2)\; ,
\end{equation}
with
\begin{align}
\label{e:EW_chi_functions}
\chi_1(Q^2) &= \frac{1}{4 \sin^2\theta_W \cos^2\theta_W } \frac{Q^2 ( Q^2 -  M_Z^2 )}{ (Q^2 - M_Z^2)^2 + M_Z^2 \Gamma_Z^2} \; ,\\
\chi_2(Q^2) &= \frac{1}{16 \sin^4\theta_W\cos^4\theta_W} \frac{Q^4}{ (Q^2 - M_Z^2)^2 + M_Z^2 \Gamma_Z^2} \; ,
\end{align}
where $e_a$, $V_a$, and $A_a$ are the electric, vector, and axial charges of the flavor $a$, respectively;
$V_\ell$ and $A_\ell$ are the vector and axial charges of the lepton $\ell$;
$\sin\theta_W$ is the weak mixing angle; $M_Z$ and $\Gamma_Z$ are mass and width of the $Z$ boson.

As discussed in Sec.~\ref{s:data} and summarized in Tab.~\ref{t:dataDY}, for DY production the observable provided by the experimental collaborations is the (normalized) cross section differential with respect to $|\qT|$.
For each bin delimited by the initial ($i$) and final ($f$) values of kinematical variables, the experimental values are compared with the following theoretical quantity:
\begin{align}
\label{e:DY_coll_observable}
{\cal O}^{th}_\text{DY,\,1}(|\qT|_{i,f}, y_{i,f}, Q_{i,f})
& =\ \fint_{|\qT|_i}^{|\qT|_f} d|\qT|\, \int_{y_i}^{y_f} dy\, \int_{Q_i}^{Q_f} dQ\,
\frac{d\sigma^{\text{DY/Z}}}{d|\qT|\, dy\, dQ}\, ,
\end{align}
where the $\fint$ symbol represents the integral divided by the width of the integration range. Hence, Eq.~\eqref{e:DY_coll_observable} corresponds to the cross section in Eq.~\eqref{e:DYZ_xsec} averaged over the transverse momentum and integrated over rapidity and invariant mass of the exchanged boson.
The normalized cross section is obtained by dividing both sides of Eq.~\eqref{e:DY_coll_observable} by the appropriate fiducial cross section, which is computed by employing the $\dynnlo$ code~\cite{Catani:2007vq,Catani:2009sm}.\footnote{See \href{https://www.physik.uzh.ch/en/groups/grazzini/research/Tools.html}{https://www.physik.uzh.ch/en/groups/grazzini/research/Tools.html}}\\

The low-energy fixed-target experiments included in this analysis (E288, E605, E772, see Tab.~\ref{t:dataDY}) measure the following cross section
\begin{equation}
\label{e:DY_FT_xsec_qTyQ}
E\, \frac{d\sigma^{\text{DY}}}{d^3 \bm{q}} = \frac{1}{2\pi\, |\qT|}\, \frac{d\sigma^{\text{DY}}}{d|\qT|\, dy},
\end{equation}
where $E$ and $\bm{q}$ are the energy and the three-momentum of the photon, respectively.

Given Eq.~\eqref{e:DY_FT_xsec_qTyQ}, in principle the experimental value in a given bin needs to be compared against the following theoretical quantity:
\begin{align}
\label{e:DY_ft_observable}
{\cal O}^{th}_\text{DY,\,ft}(|\qT|_{i,f}, y_{i,f}, Q_{i,f})
& =\ \fint_{|\qT|_i}^{|\qT|_f} d|\qT|\, \fint_{y_i}^{y_f} dy\, \int_{Q_i}^{Q_f} dQ\,
\frac{1}{2\pi\, |\qT|}\, \frac{d\sigma^{\text{DY}}}{d|\qT|\, dy\, dQ} \, .
\end{align}
However, since all the considered fixed-target experiments do not provide bins of $|\qT|$ but just the average transverse momentum values $|\overline{\qT}|$, the integration over $|\qT|$ is not considered.
Moreover, the E288 provides only the average value $\overline{y}$ for the rapidity. Accordingly, the theoretical quantity considered for that experiment reads
\begin{equation}
\label{e:DY_E288_observable}
{\cal O}^{th}_\text{DY,\,E288}(|\overline{\qT}|, \overline{y}, Q_{i,f})
 = \frac{1}{2\pi\, |\overline{\qT}|}\, \int_{Q_i}^{Q_f} dQ\,
 {\frac{d\sigma^{\text{DY}}}{d|\qT|\, dy\, dQ}}\bigg|_{y=\overline{y}, \, |\qT|=|\overline{\qT}|} \, .
\end{equation}

The E605 and E772 low-energy fixed-target experiments (see Tab.~\ref{t:dataDY}) use, in place of the rapidity $y$, the variable $x_F$, which is connected to the other kinematic variables as follows:
\begin{align}
\label{e:eta_xf}
y(x_F,Q) & = \sinh^{-1}\bigg(\frac{\sqrt{s}}{Q}\frac{x_F}{2}\bigg),
&
x_A & = \sqrt{\frac{Q^2}{s} + \frac{x_F^2}{4}} + \frac{x_F}{2},
&
x_B & = x_A - x_F\, .
\end{align}
Using Eq.~\eqref{e:eta_xf}, one obtains
\begin{equation}
\label{e:DY_FT_xsec_qTxFQ}
E\, \frac{d\sigma^{\text{DY}}}{d^3 \bm{q}} = \frac{2 E}{\pi\, \sqrt{s}}\, \frac{d\sigma^{\text{DY}}}{d\qT^2\, dx_F} \, .
\end{equation}
The E772 collaboration provides bins in $x_F$ and average transverse momentum values $|\overline{\qT}|$. Accordingly, in that case the experimental values are compared against the following theoretical quantity:
\begin{align}
\label{e:DY_E772_observable}
{\cal O}^{th}_\text{DY,\,E772}(|\overline{\qT}|, {x_F}_{i,f}, Q_{i,f})
& =\ \int_{Q_i}^{Q_f} dQ\, \fint_{{x_F}_i}^{{x_F}_f} dx_F\,
\frac{2 E}{\pi\, \sqrt{s}}\,  \frac{d\sigma^{\text{DY}}}{d\qT^2\, dx_F\, dQ}\bigg|_{|\qT|=|\overline{\qT}|} \\
\nonumber
& \approx \frac{\overline{Q}\, \cosh(\overline{y})}{\pi\, |\overline{\qT}|\, \sqrt{s}\, ({x_F}_f - {x_F}_i)}\,
\int_{y({x_F}_i, \overline{Q})}^{y({x_F}_f, \overline{Q})} dy\, \int_{Q_i}^{Q_f}\, dQ\, \frac{d\sigma^{\text{DY}}}{d|\qT|\, dy\, dQ}\bigg|_{|\qT|=|\overline{\qT}|} \, ,
\end{align}
where
\begin{align}
\label{e:average_Q_and_y}
\overline{Q} & = (Q_i + Q_f)/2 \, ,
&
\overline{y} & = [ y({x_F}_i,\overline{Q}) + y({x_F}_f,\overline{Q}) ] /2 \, .
\end{align}
For the sake of simplicity, we replaced $y$ and $Q$ with $\overline{y}$ and $\overline{Q}$
in the prefactor in front of the cross section and pull it out of the integral.

The E605 experiment, instead, provides average values for both transverse momentum and $x_F$ and its data are compared against the following theoretical quantity:
\begin{align}
\label{e:DY_E605_observable}
{\cal O}^{th}_\text{DY,\,E605}(|\overline{\qT}|, \overline{x_F}, Q_{i,f})
& \approx \frac{\overline{Q}\, \cosh(\overline{y})}{\pi\, |\overline{\qT}|\, \sqrt{s} }\,
\int_{Q_i}^{Q_f}\, dQ\, \frac{d\sigma^{\text{DY}}}{d|\qT|\, dy\, dQ}\bigg|_{|\qT|=|\overline{\qT}|, \, y=\overline{y}} \, ,
\end{align}
where, in this case, $\overline{y} = y(\overline{x_F}, \overline{Q})$
with $\overline{x_F}=(x_{Fi}+x_{Ff})/2$.

\subsection{Semi-Inclusive Deep-Inelastic Scattering (SIDIS)}
\label{ss:SIDIS}

In SIDIS, a lepton with momentum $l$ scatters off a hadron target $N$ with mass $M$ and four momentum $P$.
In the final state, the scattered lepton momentum $l^\prime$ is measured together with one hadron $h$ with mass $M_h$ and four momentum $P_h$. The other products of the scattering are undetected.
Thus the reaction reads
\begin{equation}
\label{e:SIDIS}
\ell(l) +  N(P) \rightarrow\ \ell(l^\prime) + h(P_h) + X\, .
\end{equation}
The (space-like) four-momentum transfer is $q = l-l^\prime$, with $Q^2 \equiv
-q^2 > 0$.
We use the
standard SIDIS variables~\cite{Bacchetta:2006tn,Boglione:2019nwk}:
\begin{align}
\label{e:kin_invariants}
x &= \frac{Q^2}{2\,P\cdot q} \, , &
y &= \frac{P \cdot q}{P \cdot l} \, , &
z &= \frac{P \cdot P_h}{P\cdot q} \, , &
\gamma &= \frac{2Mx}{Q} \, .
\end{align}

\begin{figure}
\centering
\includegraphics[width=0.6\textwidth]{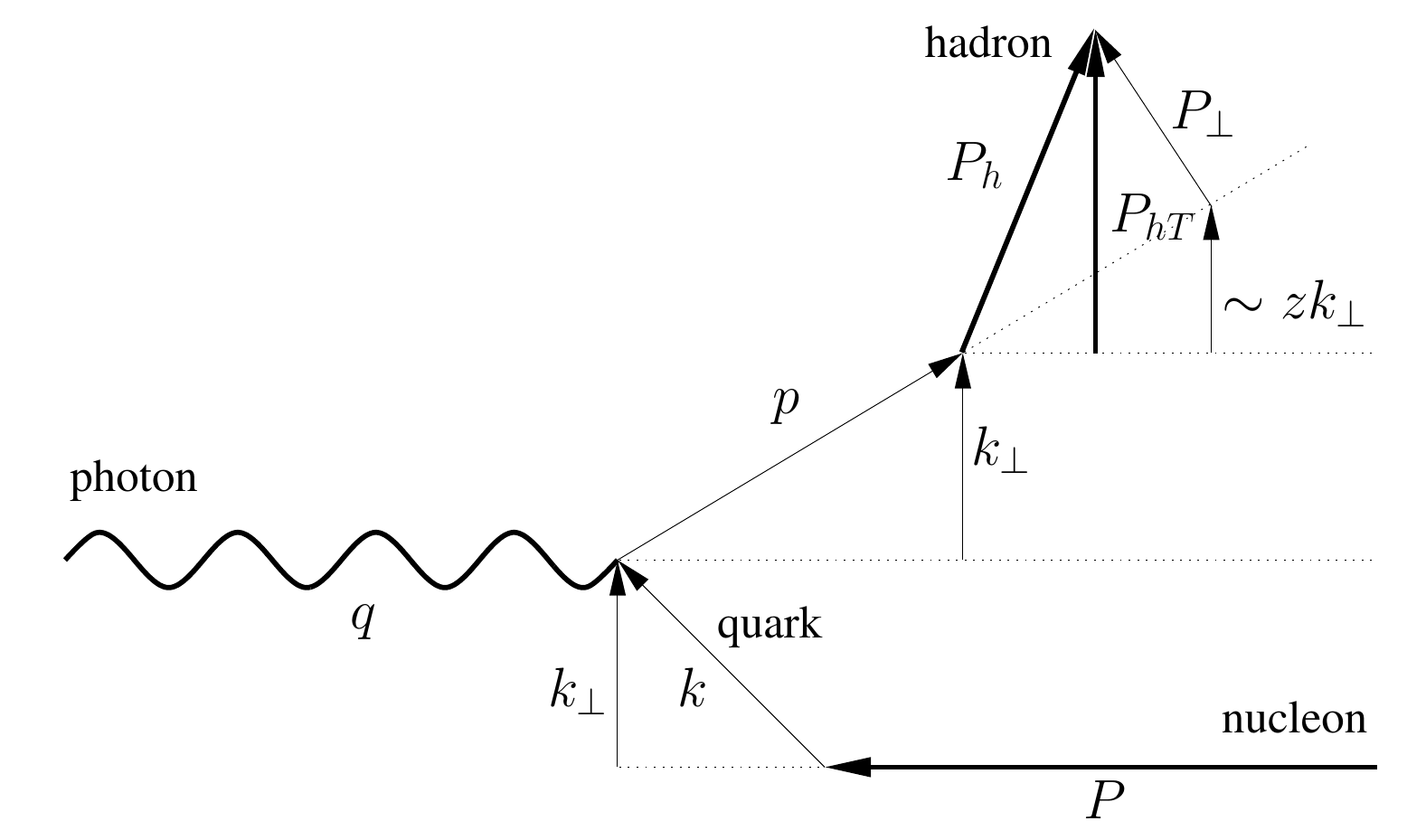}
\caption{Diagram describing the relevant momenta involved in a SIDIS event in the Breit (nucleon-photon) frame. A virtual photon with momentum $q$ (defining the reference axis) strikes a parton with momentum $k$ inside a
  nucleon with momentum $P$. The parton has a transverse momentum $\kperp$ (not measured). The
  struck parton with momentum $p= k+q$ fragments into a hadron with momentum $P_h$, which acquires a further transverse momentum $\Pperp$ (not measured) with respect to the fragmenting quark axis.
 The total measured transverse-momentum of the
  final hadron is $\PhT$. When $Q^2$ is very large, the longitudinal
  components are all much larger than the transverse components. In this regime,
  $\PhT \approx z \kperp + \Pperp$.}
\label{f:trans_momenta_SIDIS}
\end{figure}

For transverse momenta, we will follow the definitions and  notations discussed
in Ref.~\cite{Bacchetta:2004jz,Boer:2011fh} (see also Fig.~\ref{f:trans_momenta_SIDIS}).
In particular, we define $\PhT$ as the hadron transverse momentum in the Breit frame, where $P$ and $q$ form a light-cone basis; as a consequence, $P_h$ has only transverse components and $|\PhT|^2 = -P_{hT}^2$, which is frame independent. Similarly, we define $\qT$ as the photon transverse momentum in the hadron frame, where $P$ and $P_h$ form a light-cone basis; as a consequence, $q_T$ has only transverse components and $|\qT|^2 = -q_T^2$, which is also frame independent. The two momenta are related by~\cite{Bacchetta:2008xw,Bacchetta:2019qkv}
\begin{equation}
\label{e:qT_vs_PhT_general}
q_T^\mu = -\frac{P^\mu_{hT}}{z} - 2 x \frac{\modqT^2}{Q^2} P^\mu \, .
\end{equation}

In the following, we will always work assuming that the invariant mass of the photon is large compared to the target and hadron masses ($M^2, M_h^2 \ll Q^2$) and to the transverse momenta $\qT$ and $\PhT$ ($\qT^2,\, \PhT^2 \ll Q^2$).
We neglect any power corrections that vanish in this limit, both
kinematic and dynamical (higher twist), apart from some
modifications to the normalization of the SIDIS observables (that could be
seen as the effect of power corrections, see Sec.~\ref{ss:norm_SIDIS} for more details).
In this limit, Eq.~\eqref{e:qT_vs_PhT_general} reduces to
\begin{equation}
\label{e:qT_vs_PhT}
\qT \approx  -\frac{\PhT}{z} \, .
\end{equation}

Several studies have been made concerning higher-twist corrections of various
origin (see, \textit{e.g.}, Refs.~\cite{Mulders:2019mqo,Boglione:2019nwk,Accardi:2020iqn} for recent
works).
A careful study of the impact of power corrections to the case of
unpolarized TMDs has been discussed in Ref.~\cite{Scimemi:2019cmh}. Lately,
important advances in the study of higher-twist TMD factorization
have been published~\cite{Bacchetta:2019qkv,Vladimirov:2021hdn,Ebert:2021jhy,Rodini:2022wki}, but they do not directly affect the observables we consider here.

The differential cross section for SIDIS can be witten in terms of two structure functions, $F_{UU,T}$ and $F_{UU,L}$~\cite{Bacchetta:2006tn}.
The subscripts refer to the lepton, the target, and the photon polarization, respectively.
The second structure function is formally a twist four contribution and is suppressed in the limit considered here, thus we neglect it.
The differential cross section at small transverse
momentum~\cite{Bacchetta:2006tn,Bacchetta:2017gcc} reads
\begin{equation}
\label{e:SIDIS_xsec}
\frac{d\sigma^{\text{SIDIS}}}{dx\, dz\, d|\qT|\, dQ} = \frac{8\pi^2\, \alpha^2\, z^2\, |\qT|}{x\, Q^3}\, \frac{y^2}{2(1-\epsilon)}\, \bigg(1 + \frac{\gamma^2}{2x}\bigg)\, F_{UU,T}\big( x,z,|\qT|,Q \big) \, ,
\end{equation}
where $\alpha$ is the QED coupling constant and $\epsilon$ is the photon flux factor~\cite{Bacchetta:2006tn}.
Neglecting target mass corrections ${\cal O}(\gamma)$ and ${\cal O}(\gamma^2)$,
the prefactors can be approximated as
\begin{align}
\label{e:Yplus_def}
y &\approx \frac{Q^2}{xs},
&
\frac{y^2}{(1-\epsilon)}\, \bigg(1 + \frac{\gamma^2}{2x}\bigg) &\approx Y_+ =
1 + \bigg( 1 - \frac{Q^2}{xs}  \bigg)^2 \, .
\end{align}

Since we are interested only in the small-transverse-momentum limit, in
Eq.~\eqref{e:SIDIS_xsec} we have neglected the contributions from fixed-order
calculations at high $|\qT|$~\cite{Gonzalez-Hernandez:2018ipj,Wang:2019bvb}
and the matching of these on TMD
factorization~\cite{Arnold:1990yk,Collins:2016hqq,Echevarria:2018qyi}.

The unpolarized SIDIS structure function $F_{UU,T}$ is defined as~\cite{Bacchetta:2006tn}
\begin{align}
\begin{split}
\label{e:FUUT_def}
F_{UU,T}&\big( x,z,|\qT|,Q \big)
\\
& =
x\, {\cal H}^{\text SIDIS}(Q,\mu) \sum_a e_a^2  \int d^2 \kperp\, \int \frac{d^2 \Pperp}{z^2}\,
f_1^a(x,\kperp^2;\mu,\zeta_A)\, D_1^{a \to h}(z,\Pperp^2;\mu,\zeta_B)\,
\delta^{(2)}(\kperp + \Pperp/z + \qT)
\\
& = \frac{x}{2\pi}\, {\cal H}^{\text SIDIS}(Q,\mu) \sum_a e_a^2
\int_0^{+\infty} d|\bT| |\bT| J_0\big( |\bT| |\qT| \big)
\hat{f}_1^a(x,\bT^2;\mu,\zeta_A)\,
\hat{D}_1^{a \to h}(z,\bT^2;\mu,\zeta_B) \, ,
\end{split}
\end{align}
where the sum runs over quarks and antiquarks $a$. The hard factor ${\cal
H}^{\text{SIDIS}}$ can be
computed order by order in the strong coupling $\alpha_s$ and is equal to 1 at
leading order.\footnote{In the
present work,
we follow the definition of Ref.~\cite{Collins:2017oxh}.}
The variable $\kperp$ is the transverse momentum of the struck quark with respect to the nucleon axis,
whereas $\Pperp$ is the transverse momentum of the produced hadron $h$ with respect to the fragmenting quark axis (see Fig.~\ref{f:trans_momenta_SIDIS}).

The variable $\bT$ is conjugated via Fourier transform to the transverse momentum $\qT$.
$f_1^a(x,\kperp^2;\mu,\zeta_A)$ and $D_1^{a \to h}(z,\Pperp^2;\mu,\zeta_B)$ are the
unpolarized TMD PDF for a quark $a$ in a nucleon and the unpolarized TMD FF for a quark with
flavor $a$ fragmenting into a hadron with flavor $h$, respectively;
$\hat{f}_1^a(x,\bT^2;\mu,\zeta_A)$ and $\hat{D}_1^{a \to h}(z,\bT^2;\mu,\zeta_B)$ are
their Fourier transforms. The former is defined in Eq.\eqref{eq:FTdef},
the latter is
defined as
\begin{align}
 \begin{split}
\label{eq:FTdefFF}
\hat{D}_1^{a \to h}\big( z, \bT^2; \mu, \zeta \big) &=
  \int \frac{d^2 \Pperp}{z^2} \, e^{-i
      \bT  \cdot  \Pperp/z } \, D_1^a \big( z, \Pperp^2; \mu,
    \zeta \big) \,
    \\
      &= 2 \pi \int_0^{\infty} \frac{d \modPperp}{z^2} \,\modPperp  J_0(\modbT \modPperp/z) \, D_1^a \big( z, \Pperp^2; \mu,
    \zeta \big) \, .
 \end{split}
\end{align}
Their structure will be discussed in details in Sec.~\ref{ss:TMDs}.

The observable provided by the $\hermes$ and $\compass$ collaborations
is the multiplicity, namely the ratio of the one-hadron inclusive
cross section as a function of the transverse momentum of the hadron
$|\PhT|$  over the fully inclusive one:
\begin{equation}
\label{e:mult_def}
M(x,z,|\PhT|,Q)\ =\
\frac{d\sigma^{\text{SIDIS}}}{dx\, dz\, d|\PhT|\, dQ} \bigg/ \frac{d\sigma^{\text{DIS}}}{dx\, dQ} =\
\frac{1}{z}\, \frac{d\sigma^{\text{SIDIS}}}{dx\, dz\, d|\qT|\, dQ} \bigg/ \frac{d\sigma^{\text{DIS}}}{dx\, dQ} \, .
\end{equation}

The cross section for unpolarized DIS in the denominator of the multiplicities
reads
\begin{equation}
\label{e:DIS_dsigma}
\frac{d\sigma^{\text{DIS}}}{dx\, dQ} = \frac{8\pi\, \alpha_s^2}{x\, Q^3}
\frac{y^2}{2(1-\epsilon)}\,
\bigg[F_T(x,Q^2) + \epsilon F_L(x,Q^2) \bigg]
\approx
\frac{4\pi\, \alpha_s^2}{x\, Q^3}
\bigg[Y_+ F_2(x,Q^2) - y^2 F_L(x,Q^2) \bigg]\ ,
\end{equation}
where the approximation is justified by neglecting the target mass corrections.
At the perturbative order considered in this analysis, the longitudinal DIS structure function $F_L$ cannot be neglected, at variance with, \textit{e.g.}, Refs.~\cite{Signori:2013mda,Bacchetta:2017gcc}.

The experimental values in each bin are compared against the quantity built by {\em separately} averaging the numerator and denominator of the multiplicity in Eq.~\eqref{e:mult_def} over the respective kinematics:
\begin{align}
\label{e:SIDIS_observable}
{\cal O}^{th}_\text{SIDIS}(x_{i,f},z_{i,f},|\PhT|_{i,f},Q_{i,f})
& = \fint_{Q_i}^{Q_f} dQ\, \fint_{x_i}^{x_f} dx\, \fint_{z_i}^{z_f} dz\, \fint_{|\PhT|_i}^{|\PhT|_f} d|\PhT|\ \frac{d\sigma^{\text{SIDIS}}}{dx\, dz\, d|\PhT|\, dQ} \ , \\
\nonumber
{\cal O}^{th}_\text{DIS}(x_{i,f},Q_{i,f})
& = \fint_{Q_i}^{Q_f} dQ\, \fint_{x_i}^{x_f} dx\ \frac{d\sigma^{\text{DIS}}}{dx\, dQ} \ , \\
\label{e:SIDIS_mult}
{\cal M}^{th}(x_{i,f},z_{i,f},|\PhT|_{i,f},Q_{i,f})
& = {\cal O}^{th}_\text{SIDIS}(x_{i,f},z_{i,f},|\PhT|_{i,f},Q_{i,f}) \bigg/ {\cal O}^{th}_\text{DIS}(x_{i,f},Q_{i,f}) \ .
\end{align}

The $\hermes$ collaboration provides multiplicities in bins of $|\PhT|$, whereas the $\compass$ collaboration in bins of $\PhT^2$ (see also Tab.~\ref{t:dataSIDIS}).
In both cases, the observable can be calculated as in Eq.~\eqref{e:SIDIS_observable}, but in the $\compass$ case the average is on $\PhT^2$.
Moreover, both collaborations introduce a cut on the invariant mass of the hadronic final states $W^2 = (P+q)^2$ (see Tab.~\ref{t:dataSIDIS}), which makes the upper integration limit $x_f$ a $Q$-dependent quantity.

\subsection{Transverse Momentum Distributions (TMDs)}
\label{ss:TMDs}

As a consequence of the renormalization of ultraviolet and rapidity divergences~\cite{Collins:2011zzd,Echevarria:2011epo,Grewal:2020hoc}, TMD PDFs and FFs acquire a dependence on the renormalization scale $\mu$ and on the rapidity scale $\zeta$.
The evolution of TMDs from some initial values of the scales $\mu_i$, $\zeta_i$, to some final values $\mu_f, \zeta_f$, is given by
\begin{equation}
\label{e:evolved_TMDs}
\hat{f}_1^a(x,\bT^2;\mu_f,\zeta_f) =
\hat{f}_1^a(x,\bT^2;\mu_i,\zeta_i) \,
\exp\bigg\{ \int_{\mu_i}^{\mu_f} \frac{d\mu}{\mu}\, \gamma\big(\mu,\zeta_f\big) \bigg\}\, \bigg(\frac{\zeta_f}{\zeta_i}\bigg)^{K(|\bT|, \, \mu_i)/2} \, .
\end{equation}
The anomalous dimension $\gamma$ for the renormalization-group evolution in $\mu$ reads:
\begin{equation}
\label{e:gamma_mu}
\gamma\big(\mu,\zeta\big) = \gamma_F\big(\alpha_s(\mu)\big) - \gamma_K\big(\alpha_s(\mu)\big)\, \ln\,  \frac{\sqrt{\zeta}}{\mu} \, ,
\end{equation}
where $\gamma_K$ is the cusp anomalous dimension and $\gamma_F\big(\alpha_s(\mu)\big) = \gamma\big(\mu,\mu^2\big)$ is the boundary condition~\cite{Bacchetta:2019sam}.
The Collins--Soper kernel $K$, instead, is the anomalous dimension for the evolution in $\zeta$~\cite{Collins:2011zzd}. The same structure holds for the TMD FF.
In order to avoid the insurgence of large logarithms, the scales
$\mu_i$ and $\zeta_i$ are conveniently fixed as $\mu_i =
\sqrt{\zeta_i} = \mu_b = 2e^{-\gamma_E}/|\bT|$. Since the coupling
$\alpha_s$ is computed at this scale (see Eq.~\eqref{e:gamma_mu})
the evolution of the TMD is
perturbatively meaningful only at low values of $|\bT|$ such that
the scale $\mu_b$ is sufficiently larger than the Landau pole
$\Lambda_{\rm QCD}$.
This condition can be implemented by replacing the scale $\mu_b$ with $\mu_{b_*} = 2 e^{-\gamma_E} / {b_*}$, where~\cite{Bacchetta:2017gcc}
\begin{equation}
\label{e:bTstar}
b_*(|\bT|,b_{\text{min}},b_{\text{max}}) = b_{\text{max}}\, \bigg( \frac{1 - e^{ -|\bT|^4 / b_{\text{max}}^4 }}{1 - e^{ -|\bT|^4 / b_{\text{min}}^4 }} \bigg)^{1/4} \, ,
\end{equation}
with
\begin{align}
b_{\text{max}} &= 2 e^{-\gamma_E}  \text{  GeV}^{-1} \approx 1.123 \text{  GeV}^{-1}\, ,
&
b_{\text{min}} &= 2 e^{-\gamma_E}/\mu_f \ .
\label{e:bminmax}
\end{align}
As suggested by the CSS formalism~\cite{Collins:2011zzd}, $b_*$
saturates to $b_{\text{max}}$ at large $|\bT|$ guaranteeing that
$\mu_{b_*}$ never enters the nonperturbative regime. However, this has
also the effect of introducing power corrections scaling like
$(\Lambda_{\rm QCD}/|\qT|)^k$~\cite{Catani:1996yz}, with $k>0$, that in the region
$|\qT|\simeq \Lambda_{\rm QCD}$ need to be accounted for by a
nonperturbative function. At small $|\bT|$, $b_*$ saturates
to $b_{\text{min}}$. Since $\mu_f$ is of the order of the boson virtuality $Q$,
this introduces subleading power corrections scaling like
$(|\qT|/Q)^n$, with $n>0$. Such a procedure has the advantage of facilitating a possible
matching of the TMD formula, valid for $|\qT|\ll Q$, onto the
fixed-order calculation valid for $|\qT|\simeq Q$~\cite{Bozzi:2003jy,Bozzi:2005wk,Bizon:2018foh}.
Accordingly, in the limit $|\bT| \to 0$ the Sudakov exponent vanishes.

Performing at the input scales the Operator Product Expansion (OPE) of the
TMD PDFs (TMD FFs) around $|\bT|=0$ one gets:
\begin{equation}
\label{e:TMD_matching}
\hat{f}_1^a(x,b_*;\mu_{b_*},\mu_{b_*}^2) = \sum_b\, \int_x^1 \frac{dx'}{x'}\, C^{a b}(x',b_*;\mu_{b_*},\mu_{b_*}^2)\, f_1^b\bigg( \frac{x}{x'};\mu_{b_*} \bigg)
\equiv [C \otimes f_1](x,b_*;\mu_{b_*},\mu_{b_*}^2) \, ,
\end{equation}
where the sum runs over quarks, antiquarks, and the gluon.
The matching coefficients $C$ are calculated as a perturbative expansion in powers of $\alpha_s$.

In view of the power corrections introduced by the $b_*$ prescription,
both the Collins--Soper kernel $K$~\cite{Grewal:2020hoc} and the OPE
in Eq.~(\ref{e:TMD_matching}) need to be modified to account for nonperturbative effects.
For the Collins--Soper kernel $K$, this results in a nonperturbative correction term, $g_K(\bT^2)$, for which we choose a specific functional form:
\begin{align}
\label{e:K_and_gK}
& K(|\bT|,\mu_{b_*}) = K(b_*,\mu_{b_*}) + g_K(|\bT|) \, ,
&
& g_K(\bT^2) = - g_2^2\, \frac{\bT^2}{2} \, .
\end{align}
This correction gives rise in the
evolution to a nonperturbative factor that goes like $(\zeta_f/Q_0^2)^{g_{K}/2}$ where $Q_0$ is an
arbitrary scale at which this correction is parameterised; we
set $Q_0=1$~GeV. In order not to affect the perturbative
calculation at small $|\bT|$, the term $g_K$ needs to vanish in the limit $|\bT| \to 0$.
The nonperturbative corrections to the OPE can also be parameterised by a
multiplicative function that generally depends on $x$ or $z$ and $\bT$. The net result of the inclusion of the nonperturbative corrections into
the evolved TMD PDF reads:
\begin{equation}
\label{e:solution_evolved_TMDs}
\hat{f}_1^a(x,\bT^2;\mu_f,\zeta_f) =
[C\, \otimes\, f_1](x,b_*;\mu_{b_*},\mu_{b_*}^2) \,
\exp\bigg\{ \int_{\mu_{b_*}}^{\mu_f} \frac{d\mu}{\mu}\, \gamma\big(\mu,\zeta_f\big) \bigg\}\, \bigg(\frac{\zeta_f}{\mu_{b_*}^2}\bigg)^{K(b_*, \, \mu_{b_*})/2} \, f_{1\, NP}(x, \bT^2; \zeta_f, Q_0)\, ,
\end{equation}
and the same holds for the TMD FF where one introduces
$D_{1\, NP}(z, \bT^2; \zeta, Q_0)$.  Note that the number of active
flavors $n_f$ in the perturbative quantities $\gamma$, $C$, and the
hard function ${\cal H}$ of Eqs.~\eqref{e:FUU1_def}
and~\eqref{e:FUUT_def}, is separately determined by the scales $\mu$,
$\mu_{b_*}$ and $Q$, respectively. To be more precise, given a set of
quark thresholds $\{m_1,m_2,m_3,\dots\}$, the $n_f$ associated to each
of the three scales above is computed by requiring that the scale lies
between $m_{n_f}$ and $m_{n_f+1}$. Analogously, since the collinear
distributions involved in the matching formula in
Eq.~(\ref{e:TMD_matching}) are computed at the scale $\mu_{b_*}$, the
value of $n_f$ for PDFs and FFs is chosen accordingly, \textit{i.e.}
it is the same used for the matching functions $C$.

The $f_{1\, NP}$ and $D_{1\, NP}$ factors (which we assume to be flavor-independent)
incorporate both the correction to the evolution associated to the $g_K$ function and the correction to the
respective OPE. For the TMD PDF we define
\begin{equation}
\label{e:f1NP}
f_{1\, NP}(x, \bT^2; \zeta, Q_0) =
\frac{
g_1(x)\, e^{ - g_1(x) \frac{\bT^2}{4}} +
\lambda^2\, g_{1B}^2(x)\, \bigg[ 1 - g_{1B}(x) \frac{\bT^2}{4} \bigg]\, e^{ - g_{1B}(x) \frac{\bT^2}{4}} +
\lambda_2^2\, g_{1C}(x)\, e^{ - g_{1C}(x) \frac{\bT^2}{4}}
}{
g_1(x) +  \lambda^2\, g_{1B}^2(x) + \lambda_2^2\, g_{1C}(x)
} \,
\bigg[ \frac{\zeta}{Q_0^2} \bigg]^{g_K(\bT^2)/2}\, ,
\end{equation}
and for the TMD FF the form is
\begin{align}
\label{e:D1NP}
D_{1\, NP}(z, \bT^2; \zeta, Q_0) =
\frac{
g_3(z)\, e^{ - g_3(z) \frac{\bT^2}{4z^2}} +
\frac{\lambda_F}{z^2}\, g_{3B}^2(z)\, \bigg[ 1 -
g_{3B}(z) \frac{\bT^2}{4z^2} \bigg]\, e^{ - g_{3B}(z) \frac{\bT^2}{4z^2}}
}{
g_3(z) +  \frac{\lambda_F}{z^2}\, g_{3B}^2(z)
} \,
\bigg[ \frac{\zeta}{Q_0^2} \bigg]^{g_K(\bT^2)/2}\, .
\end{align}
The nonperturbative factors $f_{1\, NP}$, $D_{1\, NP} \to 1$ for $\bT \to 0$.
The $g_i$ functions describe the dependence of the widths of the distributions
on $x$ and $z$:
\begin{align}
\label{e:gi_func_PDF}
& g_{\{1,1B,1C\}}(x) = N_{\{1,1B,1C\}} \frac{x^{\sigma_{\{1,2,3\}}}(1-x)^{\alpha^2_{\{1,2,3\}}}}{\hat{x}^{\sigma_{\{1,2,3\}}}(1-\hat{x})^{\alpha^2_{\{1,2,3\}}}} \, ,
\\
\label{e:gi_func_FF}
& g_{\{3,3B\}}(z) = N_{\{3,3B\}} \frac{(z^{\beta_{\{1,2\}}}+\delta^2_{\{1,2\}})(1-z)^{\gamma^2_{\{1,2\}}}}{(\hat{z}^{\beta_{\{1,2\}}}+\delta^2_{\{1,2\}})(1-\hat{z})^{\gamma^2_{\{1,2\}}}} \, ,
\end{align}
where $\hat{x} = 0.1$, $\hat{z} = 0.5$.

In total, the default configuration for the fit involves 21 free parameters:
one associated to the nonperturbative part of the Collins--Soper kernel (Eq.~\eqref{e:K_and_gK}),
11 related to the nonperturbative part of the TMD PDF (Eqs.~\eqref{e:f1NP},~\eqref{e:gi_func_PDF}),
and 9 for the nonperturbative part of the TMD FF (Eqs.~\eqref{e:D1NP},~\eqref{e:gi_func_FF}).  \\

The functional forms in Eqs.~\eqref{e:f1NP}-\eqref{e:gi_func_FF} are largely arbitrary.
However, an important
feature is that they are the Fourier transforms of the sum of a
Gaussian, a weighted Gaussian (multiplied by $\kperp^2$) and, in the case of
the TMD PDFs, a third Gaussian. They are therefore positive definite for all
values of $\kperp^2$.\footnote{Note, however, that the evolved TMD PDF,
Eq.~\eqref{e:solution_evolved_TMDs}, can become negative at large values of
transverse momentum. The same holds true for TMD FFs.}
The parameters $\lambda$ and $\lambda_2$ in Eq.~\eqref{e:f1NP}
are squared in order to avoid
negative contributions (there is no need to square the parameter $\lambda_F$ in Eq.~\eqref{e:D1NP}
because the fit always selects positive values for this parameter).
The widths of the Gaussians, expressed by Eqs.~\eqref{e:gi_func_PDF},~\eqref{e:gi_func_FF}, are $x$ (or $z$) dependent and vanish as $x$ (or $z$) approaches 1.
Our choice of the functional form is also inspired by model calculations of
TMD PDFs (see, e.g.,
\cite{Bacchetta:2008af,Pasquini:2008ax,Avakian:2010br,Burkardt:2015qoa,Gutsche:2016gcd,Maji:2017bcz,Alessandro:2021cbg,Signal:2021aum})
and TMD FFs (see, e.g, \cite{Bacchetta:2007wc,Matevosyan:2011vj}). Many of
these models predict the existence of terms that behave similarly to Gaussians
and weighted Gaussians. The details of the functional dependence predicted by
the models are related to the correlation between the spin of the quarks and
their transverse momentum. In the case of fragmentaton functions, a different
role can be played by different producton channels (e.g., direct production
vs. production through the decay of hadronic resonances).

Finally, for the logarithmic ordering we use the same convention
adopted in Ref.~\cite{Bacchetta:2019sam}. In particular, the orders of
truncation of the perturbative ingredients relevant to the present
analysis are summarized in Tab.~\ref{t:logcountings}. At the time of this analysis,
the full N$^3$LL accuracy could not be achieved because NNLO collinear FFs were not
available. Very recently, two analyses of collinear FFs were presented
in Refs.~\cite{Borsa:2022vvp, Khalek:2022vgy} making an extraction of
TMDs at full N$^3$LL possible. We leave this study to a future
pubblication.

\begin{table}[h!]
\begin{center}
\begin{tabular}{|c|c|c|c|c|c|}
 \hline
 Accuracy                  &  $H$ and $C$    &  $K$   and  $\gamma_F$
  &  $\gamma_K$  & PDF and $\alpha_s$ evolution & FF evolution   \\
\hline
\hline
NLL          & 0     & 1         &    2   & LO & LO\\
 \hline
N$^2$LL       & 1     & 2         &   3    & NLO &  NLO\\
 \hline
N$^3$LL$^{-}$ & 2      & 3    &  4   &  NNLO & NLO \\
 \hline
N$^3$LL & 2      & 3    &  4   &  NNLO & NNLO \\
 \hline
\end{tabular}
\caption{Truncation orders in the expansions of the perturbative
  ingredients of TMDs relevant to the logarithmic counting considered in this
  paper (see text). The last column refers to the order used for the
  evolution of the collinear FFs.}
\label{t:logcountings}
\end{center}
\end{table}
\subsection{Normalization factors for SIDIS}
\label{ss:norm_SIDIS}

In Ref.~\cite{Bacchetta:2017gcc} it was demonstrated that TMD factorization at
NLL accuracy is able to successfully reproduce the normalization and shape
of $\hermes$ SIDIS multiplicities and the shape of the available $\compass$
multiplicities. More recently, the $\compass$ collaboration published a reanalysis of their
data~\cite{COMPASS:2017mvk}. The NLL TMD predictions correctly reproduce
normalization and shape of the new data~\cite{Piacenza:2020sst}. However, when
increasing the accuracy to N$^2$LL or higher, the TMD formula severely
underestimates the
measurements~\cite{OsvaldoGonzalez-Hernandez:2019iqj,Piacenza:2020sst} by nearly
constant factors in each bin. Note that tensions between the TMD cross
sections and the associated measurements exist also at large transverse
momentum in SIDIS~\cite{Gonzalez-Hernandez:2018ipj},
DY~\cite{Bacchetta:2019tcu}, and electron-positron annihilation into
two hadrons~\cite{Moffat:2019pci}.

In the present study, as will be shown in Sec.~\ref{s:results}, we confirm that we obtain an
excellent description of both normalization and shape of the SIDIS
multiplicities at NLL and that the N$^2$LL results are much
smaller. At N$^3$LL the results increase slightly, but they are still far
from the NLL ones and, therefore, from data. At average kinematics of
the $\compass$ measurements, we obtain the following ratios of multiplicities:
\begin{align}
\frac{M_{\rm NLL}}{M_{\rm N^2LL}} &\gtrsim 2,
&
\frac{M_{\rm NLL}}{M_{\rm N^3LL}} &\gtrsim 1.5.
\end{align}

The reason for the difference between the logarithmic orders is almost entirely due to the hard factor in
Eq.~\eqref{e:FUUT_def}~\cite{Collins:2017oxh}. If we look at the explicit expression for
the hard factor\footnote{There are different
definitions for the hard factor in the literature, which are compensated by
different definitions of the matching coefficients $C$ in
Eq.\eqref{e:TMD_matching}. Here we follow the definition of Ref.~\cite{Collins:2017oxh}.} with the standard choice $\mu= Q$
\begin{equation}
{\cal H}^{\rm SIDIS} (Q,Q)=
1 + \frac{\alpha_s(Q)}{4\pi} C_F \bigg(-16+\frac{\pi^2}{3}\bigg),
\end{equation}
we can immediately see that just by introducing ${\cal O}(\alpha_s)$
corrections, at $Q= 2$ GeV with $\alpha_s\approx 0.3$, we reduce
the structure function to about 60\% of its original value.
This change is not compensated by a similar reduction in the
denominator of Eq.~\eqref{e:mult_def}: the differences between the LO and NLO
expressions of the inclusive DIS cross section are typically below 5\% and the
NLO results are actually larger than the LO ones.

If the NLL expression is much larger than the N$^2$LL and N$^3$LL ones, we may suspect that
it should overshoot the data by a factor 1.5 at least.
However, several works before the present one have shown a good
agreement with data using a parton-model
approach~\cite{Signori:2013mda,Anselmino:2013lza} or
at NLL~\cite{Echevarria:2014xaa,Sun:2014dqm}.
Moreover, the integral of the structure function over
$\qT$ is equal to the value of its Fourier transform at $\bT=0$.
Using any $b^*$ prescription with
$b_{\rm min}=2 e^{-\gamma_E}/Q $, the integral of the NLL expression
by construction corresponds to the LO expression of the collinear
SIDIS structure function, independent of the TMD nonperturbative parameters:\footnote{Note that in the absence of a $b_{\rm min}$ prescription, the integral would vanish.}
\begin{equation}
\begin{split}
\int d^2 \qT \,F_{UU,T}\big( x,z,\modqT,Q \big)
&=
x\, \sum_a e_a^2\, {\cal H}^{\rm SIDIS}(Q,Q)  \bigg(
\hat{f}_1^a(x,\bT^2;Q,Q^2)\,
\hat{D}_1^{a \to h}(z,\bT^2;Q,Q^2)\bigg)\bigg|_{\modbT=0}
\\
&\stackrel{\rm NLL}{=}
x\, \sum_a e_a^2\,
{f}_1^a(x;Q)\,
{D}_1^{a \to h}(z;Q).
\end{split}
\end{equation}

The LO collinear SIDIS predictions are known to describe the data reasonably
well and, if anything, they seem to be lower than the
data~\cite{deFlorian:2007aj,HERMES:2012uyd}. Therefore, the integral of our
NLL expression is in good agreement with data, which also indicates the
absence of a large normalization error.

If the NLL predictions describe the data well and are a factor 2 or 1.5 above
the N$^2$LL and N$^3$LL predictions, we propose to modify the
normalization of the latter to recover a good agreement with data.
An extended discussion of this issue can be
found in Ref.~\cite{Piacenza:2020sst}.
We observe that the integral of the TMD formula, valid at low $\qT$, should
reproduce only part of the full collinear cross section.
The only exception is the order ${\cal O}(\alpha_s^0)$ case, as we have seen
above, since at that order there is no contribution from gluon radiation at
high transverse momentum, beyond the TMD region.

However, at N$^2$LL or higher, in the kinematics of fixed-target SIDIS
experiments, the integral of the TMD region (\textit{i.e.}, the integral of the
so-called $W$ term in the language of Ref.~\cite{Collins:2011zzd}) is much smaller than the
corresponding collinear cross section. The missing contribution to the
integral should be recovered by
the terms in the fixed-order calculation that are not included in the TMD
resummed expression (the so-called $Y$ term).
Ideally, the $Y$ term should be
negligible in the low-$\qT$ region. This is not the case in the
experimental regions under consideration: the $Y$ term is finite but
relatively large, even at $\qT=0$.

If we consider the contribution to the integral of the $W$ term (\textit{i.e.}, the integral of
Eq.~\eqref{e:SIDIS_xsec}), with our $b_{\rm min}$ prescription,
at order $\alpha_s$ we obtain, schematically
\begin{equation}\label{eq:WintDetail}
\int d^2 \qT\,  W \biggr\vert_{{\cal O}(\alpha_s)} =
\sigma_0 \frac{\alpha_s}{4 \pi} \sum_{q} e_q^2 \Bigl[
D_1^{q \to h} \otimes C_{\rm TMD}^{qq}\otimes f_1^{q} +
D_1^{q \to h} \otimes C_{\rm TMD}^{qg}\otimes f_1^{g} +
D_1^{g \to h} \otimes C_{\rm TMD}^{gq}\otimes f_1^{q}
\Bigr](x,z,Q) ,
\end{equation}
where
\begin{equation}
\sigma_0 = \frac{4\pi^2\, \alpha^2\, z^2\, |\qT|}{x\, Q^3} Y_+.
\end{equation}
The double convolution over both $x$ and $z$ is defined as
\begin{equation}
\Bigl[ D_1^{a \to h} \otimes C^{ab}\otimes f_1^{b} \Bigr](x,z,Q) =
\frac{1}{z^2}\int_x^1 \frac{dx'}{x'} \int_z^1\frac{dz'}{z'}
D_1^{a \to h}(z';Q) C^{ab}\Bigl(\frac{x}{x'},\frac{z}{z'} \Bigr) f_1^{b}(x';Q)
\end{equation}
and the $C_{\rm TMD}$ coefficients can be found in
App.~\ref{a:Ccoeff}.

The integral in Eq.~\eqref{eq:WintDetail} should be compared to the collinear expression at the same order (see,
\textit{e.g.}, Ref.~\cite{deFlorian:1997zj})
\begin{equation}
\label{eq:SIDISintNLO}
\begin{split}
\frac{d \sigma^{\text{SIDIS}}}{d x d Q d z} \biggr\vert_{{\cal
O}(\alpha_s)}&=
\sigma_0 \frac{\alpha_s}{4 \pi} \sum_{q} e_q^2 \biggl\{\Bigl[
D_1^{q \to h} \otimes C_{1}^{qq}\otimes f_1^{q} +
D_1^{q \to h} \otimes C_{1}^{qg}\otimes f_1^{g} +
D_1^{g \to h} \otimes C_{1}^{gq}\otimes f_1^{q}
\Bigr](x,z,Q) \\
&\quad +  \frac{1-y}{1+(1-y)^2} \Bigl[
D_1^{q \to h} \otimes C_{L}^{qq}\otimes f_1^{q} +
D_1^{q \to h} \otimes C_{L}^{qg}\otimes f_1^{g} +
D_1^{g \to h} \otimes C_{L}^{gq}\otimes f_1^{q}
\Bigr](x,z,Q)  \biggr\}. \\
\end{split}
\end{equation}
The QCD coefficients $C_{1}^{a b}$
and $C_{L}^{a b}$ are calculated in perturbation theory.
The former can be written as
\begin{align}
\label{e:Ccoeff}
C_{1}^{a b} (x,z;Q,\mu) &= C_{\rm nomix}^{a
b}(x,z,Q,\mu)+ C_{\rm mix}^{a b}(x,z).
\end{align}
The coefficient $C_{\rm nomix}$ is the sum of
all those terms that contain either a
$\delta(1-x)$  or a
$\delta(1-z)$ (or both). Some of these terms are present in
$C_{\rm TMD}$, but not all. This definition holds at all orders in $\alpha_s$.
The $C_L$ matching coefficients, instead, only contain ``mixed'' contributions.
For convenience, we reproduce all coefficients at order $\alpha_s$ in App.~\ref{a:Ccoeff}.

In order to increase the size of the TMD component, we consider the
contribution of all the ``nonmixed'' terms $C_{\rm nomix}$. The reason behind
this choice is that it might be possible to include such terms into a
redefinition of the individual TMDs. Hence, we define
\begin{equation}
\label{e:nomix}
\frac{d \sigma^{\rm nomix}}{d x d Q d z} \biggr\vert_{{\cal
O}(\alpha_s)} =
\sigma_0 \frac{\alpha_s}{4 \pi} \sum_{q} e_q^2 \Bigl[
D_1^{q \to h} \otimes C_{\rm nomix}^{qq}\otimes f_1^{q} +
D_1^{q \to h} \otimes C_{\rm nomix}^{qg}\otimes f_1^{g} +
D_1^{g \to h} \otimes C_{\rm nomix}^{gq}\otimes f_1^{q}
\Bigr](x,z,Q),
\end{equation}
and similarly for higher orders,
and we introduce the following normalization factor:
\begin{equation}
\label{e:norm_SIDIS_def}
\omega(x,z,Q) = \frac{d\sigma^{\rm nomix}}{dx\, dz\, dQ}  \bigg/  \int d^2 \qT\, W  \, .
\end{equation}
We stress that these normalization factors depend only on the collinear PDFs
and FFs, are independent of the parametrization of the nonperturbative
part of the TMDs, and can be computed before performing a fit of
the latter.

At NLL, $\omega(x,z,Q) = 1$.
Beyond NLL, the prefactor becomes larger than one and guarantees that the
integral of the TMD part of the cross section reproduces most of the collinear
cross section, as suggested by the data. On the contrary, without the
enhancement due to the normalization factor, the integral of the TMD part of
the cross section would be too small, requiring a large role of the
high-transverse-momentum tail, which is not observed in the data.
The impact of the normalization factor defined in Eq.~\eqref{e:norm_SIDIS_def}
will be addressed in detail in Sec.~\ref{s:results}.

As a consequence of our procedure,
the theoretical expression for the SIDIS cross section in Eq.~\eqref{e:SIDIS_xsec}
becomes
\begin{equation}
\label{e:sidis_xsec_expr_w_norm}
\frac{d\sigma_\omega^{\text{SIDIS}}}{dx\, dz\, d|\qT|\, dQ} = \omega(x,z,Q)\, \frac{d\sigma^{\text{SIDIS}}}{dx\, dz\, d|\qT|\, dQ} \, .
\end{equation}

\section{Data selection}
\label{s:data}

\begin{figure}[h]
\centering
\includegraphics[width=0.7\textwidth]{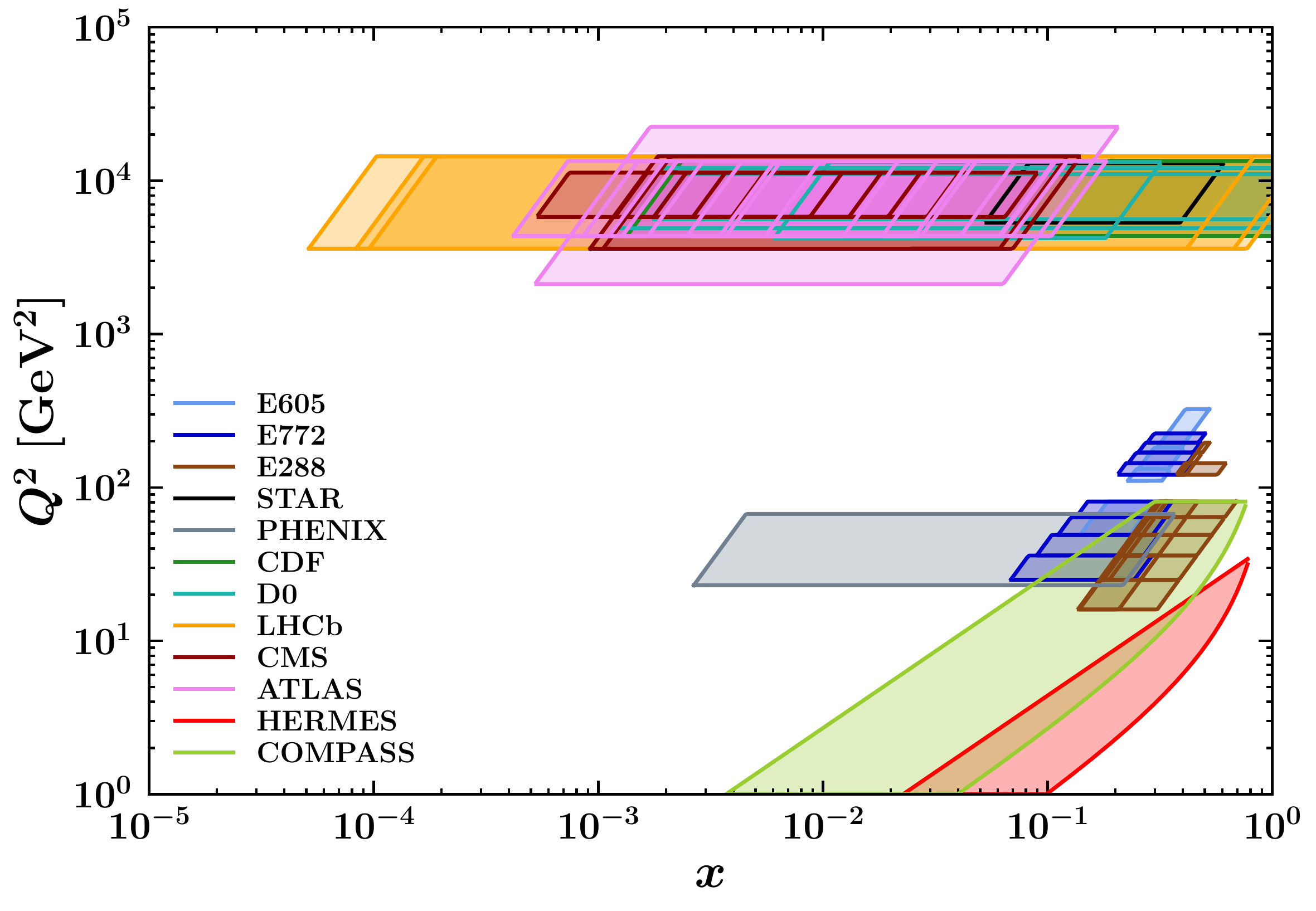}
\caption{The $x$ vs. $Q^2$ coverage spanned by the experimental data considered in this analysis (see also Tab.~\ref{t:dataDY} and Tab.~\ref{t:dataSIDIS}).}
\label{f:xQ2coverage}
\end{figure}

In this Section we describe the experimental data included in our global analysis.
We consider a large number of datasets related to DY lepton pair
production  and SIDIS, for the observables
discussed in Sec.~\ref{ss:DY_Z} and Sec.~\ref{ss:SIDIS}. The coverage in the
$x$-$Q^2$ plane spanned by these datasets is illustrated in
Fig.~\ref{f:xQ2coverage}.

The majority of datasets analyzed in the present work was already included
in the global analysis of SIDIS and DY data in
Ref.~\cite{Bacchetta:2017gcc} and in the fit of DY data discussed in
Ref.~\cite{Bacchetta:2019sam}.
For more details, we refer the reader to those references.
The new datasets included in the present analysis are:
\begin{itemize}
  \item DY di-muon production from the collision of a proton beam with an energy of 800 GeV on a $^2H$ fixed target from E772 ($\sqrt{s} = 38.8$ GeV)~\cite{E772:1994cpf};
  \item DY di-muon production from the \textsc{PHENIX} Collaboration~\cite{PHENIX:2018dwt};
  \item DY data at 13 TeV from the \textsc{CMS} Collaboration~\cite{CMS:2019raw} and the \textsc{ATLAS} Collaboration~\cite{ATLAS:2019zci}.
\end{itemize}

\subsection{Drell-Yan}
\label{ss:data_DY}

Our analysis is based on TMD factorization, which
is applicable only in the region $|\qT| \ll Q$.
Therefore, in agreement with the choices of Refs.~\cite{Bacchetta:2019sam,Scimemi:2019cmh} we impose the following cut
\begin{equation}
\label{e:DYcut}
|\qT| < 0.2\, Q \, .
\end{equation}

\begin{table}[t]
\footnotesize
\begin{center}
\renewcommand{\tabcolsep}{0.4pc}
\renewcommand{\arraystretch}{1.2}
\begin{tabular}{|c|c|c|c|c|c|c|c|}
  \hline
  Experiment & $N_{\rm dat}$ & Observable  &  $\sqrt{s}$ [GeV]& $Q$ [GeV] &  $y$ or $x_F$ & Lepton cuts & Ref. \\
  \hline
  \hline
  E605 & 50 & $E d^3\sigma/d^3 \bm{q}$ & 38.8 & 7 - 18  & $x_F=0.1$ & - & \cite{Moreno:1990sf} \\
  \hline
  E772 & 53 & $E d^3\sigma/d^3 \bm{q}$ & 38.8 & 5 - 15  & $0.1 < x_F < 0.3$ & - & \cite{E772:1994cpf} \\
  \hline
  E288 200 GeV & 30 & $E d^3\sigma/d^3 \bm{q}$ &  19.4  & 4 - 9  & $y=0.40$ & - & \cite{Ito:1980ev} \\
  \hline
  E288 300 GeV & 39 & $E d^3\sigma/d^3 \bm{q}$ &  23.8  & 4 - 12  & $y=0.21$ & - & \cite{Ito:1980ev} \\
  \hline
  E288 400 GeV & 61 & $E d^3\sigma/d^3 \bm{q}$ &  27.4  & 5 - 14  & $y=0.03$ & - & \cite{Ito:1980ev} \\
  \hline
  STAR 510 & 7 & $d\sigma/d |\qT|$ & 510  & 73 - 114  & $|y|<1$ & \makecell{$p_{T\ell} > 25$~GeV\\ $|\eta_\ell|<1$} & - \\
  \hline
  PHENIX200 & 2 & $d\sigma/d |\qT|$ & 200 & 4.8 - 8.2  & $1.2 < y < 2.2$ & - & \cite{PHENIX:2018dwt} \\
  \hline
  CDF Run I & 25 & $d\sigma/d |\qT|$ & 1800 & 66 - 116  & Inclusive & - & \cite{Affolder:1999jh} \\
  \hline
  CDF Run II & 26 & $d\sigma/d |\qT|$ & 1960 & 66 - 116  &  Inclusive & - & \cite{Aaltonen:2012fi} \\
  \hline
  D0 Run I & 12 & $d\sigma/d |\qT|$ & 1800 & 75 - 105  &  Inclusive & - & \cite{Abbott:1999wk} \\
  \hline
  D0 Run II & 5 & $(1/\sigma)d\sigma/d |\qT|$ & 1960 & 70 - 110  & Inclusive & - & \cite{Abazov:2007ac} \\
  \hline
  D0 Run II $(\mu)$ & 3 & $(1/\sigma)d\sigma/d |\qT|$ & 1960 & 65 - 115  & $|y|<1.7$ & \makecell{$p_{T\ell} > 15$~GeV\\$|\eta_\ell|<1.7$} & \cite{Abazov:2010kn} \\
  \hline
  LHCb 7 TeV & 7 & $d\sigma/d |\qT|$ & 7000 & 60 - 120  & $2<y<4.5$ & \makecell{$p_{T\ell} > 20$ GeV\\$2<\eta_\ell<4.5$} & \cite{Aaij:2015gna} \\
  \hline
  LHCb 8 TeV & 7 & $d\sigma/d |\qT|$ & 8000 & 60 - 120  & $2<y<4.5$ & \makecell{$p_{T\ell} > 20$ GeV\\$2<\eta_\ell<4.5$} & \cite{Aaij:2015zlq} \\
  \hline
  LHCb 13 TeV & 7 &  $d\sigma/d |\qT|$ & 13000 & 60 - 120  & $2<y<4.5$ & \makecell{$p_{T\ell} > 20$ GeV\\$2<\eta_\ell<4.5$} & \cite{Aaij:2016mgv} \\
  \hline
  CMS 7 TeV & 4 & $(1/\sigma)d\sigma/d |\qT|$  & 7000 & 60 - 120  & $|y|<2.1$ & \makecell{$p_{T\ell} > 20$ GeV\\$|\eta_\ell|<2.1$} & \cite{Chatrchyan:2011wt} \\
  \hline
  CMS 8 TeV & 4 & $(1/\sigma)d\sigma/d |\qT|$ & 8000 & 60 - 120  & $|y|<2.1$ & \makecell{$p_{T\ell} > 15$ GeV\\$|\eta_\ell|<2.1$} & \cite{Khachatryan:2016nbe} \\
  \hline
  CMS 13 TeV & 70 & $d\sigma/d |\qT|$ & 13000 & 76 - 106 & \makecell{$|y|<0.4$ \\ $0.4<|y|<0.8$ \\ $0.8<|y|<1.2$\\$1.2<|y|<1.6$\\$1.6<|y|<2.4$} & \makecell{$p_{T\ell} > 25$~GeV\\$|\eta_\ell|<2.4$} & \cite{CMS:2019raw} \\
  \hline
  ATLAS 7 TeV & \makecell{6\\6\\6}& $(1/\sigma)d\sigma/d |\qT|$ & 7000 & 66 - 116 & \makecell{$|y|<1$ \\ $1<|y|<2$ \\ $2<|y|<2.4$}  & \makecell{$p_{T\ell} > 20$~GeV\\$|\eta_\ell|<2.4$} & \cite{Aad:2014xaa} \\
  \hline
  \makecell{ATLAS 8 TeV \\ on-peak} & \makecell{6\\6\\6\\6\\6\\6} & $(1/\sigma)d\sigma/d |\qT|$ & 8000 & 66 - 116  & \makecell{$|y|<0.4$ \\ $0.4<|y|<0.8$ \\ $0.8<|y|<1.2$\\$1.2<|y|<1.6$\\$1.6<|y|<2$\\$2<|y|<2.4$} & \makecell{$p_{T\ell} > 20$~GeV\\$|\eta_\ell|<2.4$} & \cite{Aad:2015auj} \\
  \hline
  \makecell{ATLAS 8 TeV \\ off-peak} & \makecell{4 \\ 8} & $(1/\sigma)d\sigma/d |\qT|$ & 8000 & \makecell{46 - 66 \\ 116 - 150} & $|y|<2.4$ & \makecell{$p_{T\ell} > 20$ GeV\\$|\eta_\ell|<2.4$} & \cite{Aad:2015auj} \\
  \hline
  ATLAS 13 TeV & 6 & $(1/\sigma)d\sigma/d |\qT|$ & 13000 & 66 - 116 & $|y|<2.5$ & \makecell{$p_{T\ell} > 27$ GeV\\$|\eta_\ell|<2.5$} & \cite{ATLAS:2019zci} \\
  \hline
  \hline
  Total & 484 & & & & & & \\
  \hline
\end{tabular}
\caption{
  Breakdown of the DY datasets considered in this analysis.
  For each dataset, the table includes information on: the number of data
  points ($N_{\rm dat}$) that survive the nominal cut on $|\qT|$ (see Eq.~\eqref{e:DYcut}), the
  observable delivered, the center-of-mass energy $\sqrt{s}$, the
  range(s) in invariant mass $Q$, the angular variable (either $y$ or
  $x_F$), possible cuts on the single final-state leptons, and the
  published reference (when available). The total number of DY data points amounts to 484.
  }
\label{t:dataDY}
\end{center}
\end{table}

Table~\ref{t:dataDY} summarizes all the DY datasets included in our analysis.
For some DY datasets the experimental observable is given within a fiducial
region. This means that kinematic cuts on transverse momentum
$p_{T\ell}$ and pseudo--rapidity $\eta_\ell$ of the single final-state leptons are
enforced (values reported in the next--to--last column of
Tab.~\ref{t:dataDY}). For more details we refer the reader to
Ref.~\cite{Bacchetta:2019sam}.
The second column of Tab.~\ref{t:dataDY} reports, for each experiment, the
number of data points
($N_{\text{dat}}$) that
survive the kinematic cuts.
The total number of DY data points considered in this work is 484.
Note that for \textsc{E605} and \textsc{E288} at 400 GeV we have excluded the bin in $Q$ containing the $\Upsilon$ resonance ($Q\simeq 9.5$~GeV).

As can be seen in Tab.~\ref{t:dataDY}, the cross sections are released in different forms: some of them are normalized to the total (fiducial) cross
section while others are not.
When necessary, the required total cross section $\sigma$ is computed using the code
$\dynnlo$~\cite{Catani:2007vq,Catani:2009sm} with the \textsc{MMHT14} collinear PDF
set, consistently with the perturbative order of the differential cross
section (see also Tab.~\ref{t:logcountings}). More precisely, the total cross
section is computed at NLO for NNLL accuracy, and NNLO for N$^3$LL$^{-}$
accuracy.
The values of the total cross sections at different orders can be found in
Table 3 of Ref.~\cite{Bacchetta:2019sam}. For the \textsc{ATLAS} dataset at 13 TeV, the
value of the fiducial cross section is 694.3 pb at NLO and 707.3 pb at
NNLO.

\subsection{SIDIS}
\label{ss:data_SIDIS}

The identification of the TMD region in SIDIS is not a trivial task
and may
be subject to revision as new data appears and the theoretical description is
improved, as discussed in dedicated
studies~\cite{Boglione:2016bph,Boglione:2019nwk,Boglione:2022gpv}.

First of all, a cut in the virtuality $Q$ of the exchanged photon is necessary to
respect the condition $Q \gg \Lambda_{\text{QCD}}$ needed for
perturbation theory to be applicable. In this way also mass
corrections and higher twist corrections can be neglected.
In this work, we require that $Q > 1.4$~GeV. Studies of SIDIS in collinear kinematics employ similar
cuts~\cite{deFlorian:2017lwf,Khalek:2022vgy}.

In order to restrict ourselves to the SIDIS current fragmentation region and
interpret the observables in terms of parton distribution and fragmentation functions,
we apply a cut in the
kinematic variable $z$ by requiring $0.2 < z < 0.7$.  The lower
limit is the same used in the study of collinear fragmentation
functions~\cite{deFlorian:2017lwf,Khalek:2022vgy}.  We used a slightly more
restrictive upper limit, to avoid contributions from exclusive channels and to
focus on a region where the collinear fragmentation functions have small
relative uncertainties.

For what concerns the cut on transverse momentum, our baseline choice is
\begin{equation}
\label{e:SIDIScut}
|\PhT| < \text{min}\big[\, \text{min}[ c_1\, Q, c_2\, zQ] + c_3 \, \text{GeV} ,\, zQ \big] \, ,
\end{equation}
with fixed parameters $c_1 = 0.2$, $c_2 = 0.5$ and $c_3 = 0.3$. This choice is more restrictive than a similar one made in Ref.~\cite{Bacchetta:2017gcc}, but less restrictive than the one made in
 Ref.~\cite{Scimemi:2019cmh}. It allows for many data points with $|\PhT| \ll Q$ but also with $0.2\, Q < |\qT| < Q$. In
 Sec.~\ref{s:results}, we will discuss variations of
 the baseline SIDIS cut in Eq.~\eqref{e:SIDIScut} that give
 phenomenological support to our choice.

As for the datasets included in the present analysis,
the main difference with
Ref.~\cite{Bacchetta:2017gcc} is that we include the new
release of $\compass$ data~\cite{COMPASS:2017mvk}.
In this dataset, the vector--boson contributions have been subtracted. For the
$\hermes$ dataset we consistently select the vector--meson--subtracted dataset (\verb!.vmsub! set).
Moreover, we select the \verb!zxpt!-\verb!3D!-binning for $\hermes$
multiplicities, since it presents a finer binning in $|\PhT|$.
The breakdown of the entire SIDIS dataset included in the present analysis is reported in Tab.~\ref{t:dataSIDIS}.

\begin{table}[t]
\footnotesize
\begin{center}
\renewcommand{\tabcolsep}{0.4pc}
\renewcommand{\arraystretch}{1.2}
\begin{tabular}{|c|c|c|c|c|c|c|c|c|}
  \hline
  Experiment & $N_{\rm dat}$ & Observable & Channels & $Q$ [GeV] & $x$ & $z$ & Phase space cuts & Ref. \\
  \hline
  \hline
  HERMES & 344 & $M(x,z,|\PhT|,Q)$ & \makecell{$p \rightarrow \pi^+$ \\ $p \rightarrow \pi^-$ \\$p \rightarrow K^+$ \\ $p \rightarrow K^-$ \\ $d \rightarrow \pi^+$ \\ $d \rightarrow \pi^-$ \\$d \rightarrow K^+$ \\ $d \rightarrow K^-$ \\} & 1 - $\sqrt{15}$ & \makecell{ \\ $0.023<x<0.6$ \\ (6 bins) \\ \\} & \makecell{$0.1<z<1.1$ \\ (8 bins)} & \makecell{$W^2 > 10$ GeV$^2$\\$ 0.1<y<0.85$} & \cite{HERMES:2012uyd} \\
  \hline
  COMPASS & 1203 & $M(x,z,\PhT^2,Q)$ & \makecell{$d \rightarrow h^+$ \\ $d \rightarrow h^-$ \\} & \makecell{ 1 - 9 \\ (5 bins) \\} & \makecell{ \\ $0.003<x<0.4$ \\ (8 bins) \\ \\} & \makecell{ $0.2<z<0.8$ \\ (4 bins) \\ } & \makecell{$W^2 > 25$ GeV$^2$\\$ 0.1<y<0.9$} & \cite{COMPASS:2017mvk} \\
  \hline
  \hline
  Total & 1547 & & & & & & & \\
  \hline
\end{tabular}
\caption{
  Breakdown of the SIDIS datasets included in this analysis. For each
  dataset, the table includes information on: the number of data
  points ($N_{\rm dat}$) surviving the nominal cut on $|\PhT|$, the
  observable delivered, the SIDIS channel, the range(s) in photon invariant mass $Q$, the ranges in the kinematic variables $x$ and $z$, possible cuts on the single final-state lepton, and the
  public reference (when available).
  The total number of SIDIS data points amounts to 1547.
  }
\label{t:dataSIDIS}
\end{center}
\end{table}

The second column of Tab.~\ref{t:dataSIDIS} shows the number of data points
($N_{\text{dat}}$)  that respect the kinematic  cuts for each dataset, with a
total number of 1547 data points.

In conclusion, the total number of DY and SIDIS data points surviving our kinematic cuts is 2031.

\subsection{Error treatment}
\label{ss:errors}

The considered experimental datasets are released with a set of systematic and statistical uncertainties. As already pointed out in Ref.~\cite{Bacchetta:2019sam}, a proper treatment of the experimental uncertainties is extremely important in order to obtain a reliable extraction of TMDs.
Thus, we choose to treat systematic uncertainties as fully correlated only if, in the corresponding publication, it is explicitly specified that they are correlated.
The statistical uncertainties, instead, are always considered as uncorrelated.

At variance with Ref.~\cite{Bacchetta:2019sam}, in this analysis we do
not make use of the iterative $t_0$-prescription~\cite{Ball:2009qv} for the
treatment of correlated normalization uncertainties. This
prescription is usually introduced to avoid the underestimation of the
predictions caused by the so--called D'Agostini bias~\cite{DAGOSTINI1994306}.
After performing the fit with and without the
$t_0$-prescription, we found that our analysis is not affected by the
D'Agostini bias and therefore we saw no reason to introduce the $t_0$-prescription in the computation of the $\chi^2$.

On top of experimental systematic uncertainties, there can be several sources of systematic theoretical errors.
The first one comes from the choice of the underlying collinear parton distribution and fragmentation functions.
In our case we choose the
\textsc{MMHT2014}~\cite{Harland-Lang:2014zoa} collinear PDFs and the
\textsc{DSS} collinear FFs.
Since the $\hermes$ collaboration provides multiplicities for pions and kaons separately (see Tab.~\ref{t:dataSIDIS}), we use \textsc{DSS14}~\cite{deFlorian:2014xna} for $\pi^\pm$ and \textsc{DSS17}~\cite{deFlorian:2017lwf} for $K^\pm$.
Given the nature of the PDF and FF set used in this analysis, their
uncertainties are computed using the Hessian method~\cite{Pumplin:2001ct,deFlorian:2009vb,deFlorian:2014xna}.
We observed that PDF and FF uncertainties are significantly correlated
across bins. In order to account for this correlation, we decomposed
the corresponding uncertainties into a fully correlated part that
amounts to 80\% of the total, while we treated the remaining 60\% as
uncorrelated.\footnote{Notice that, using this decomposition, the sum in quadrature of correlated and
uncorrelated parts reproduce the original uncertainties.}

Moreover, for the observables measured by $\compass$ one needs a
collinear set of FFs for unidentified charged hadrons. Since the
dedicated \textsc{DSS07} FF set for unidentified charged
hadrons~\cite{deFlorian:2007ekg} does not provide an estimate of the
uncertainties, we computed the $\compass$ multiplicities by using the
sum of the \textsc{DSS14} and \textsc{DSS17} sets for pions and
kaons.\footnote{We thus assumed that the yield due to other hadronic
  species, such as protons, $\Lambda$, etc., is negligible as compared
  to the sum of kaons and pions. As argued in
  Ref.~\cite{Bertone:2018ecm}, the contribution to the total yield due
  to hadrons heavier that pions and kaons is indeed marginal.} The
associated uncertainty is calculated by propagating to the
multiplicity the Hessian errors associated to each of the two hadronic
component. As pointed out in Ref.~\cite{Scimemi:2019cmh}, the choice of
specific sets for the collinear distributions may have a sizeable impact
on the final result. In our analysis, we did not consider alternative
collinear sets and postpone this study to a future
publication. Likewise, we leave for future work the study of other
sources of theoretical uncertainties such as higher-twist corrections,
TMD flavor dependence and the choice of the perturbative scales.

\section{Results}
\label{s:results}

In this section, we present the results obtained for the extraction of
unpolarized quark TMDs from a global analysis including both DY and SIDIS data
(see Sec.~\ref{s:data}). This work represents an important upgrade with
respect to Ref.~\cite{Bacchetta:2019sam}. In fact, in the SIDIS process a
single hadron is measured in the final state, allowing us to extract
information about fragmentation functions. Therefore, the final result of this
work is the simultaneous extraction of both TMD PDFs and FFs at N$^3$LL$^-$.
At the moment, this is the most precise extraction of TMDs that has been achieved on more than two thousand data points.
In Sec.~\ref{ss:fitqual} we present the quality of the fit, in
Sec.~\ref{ss:tmds} we discuss the extracted TMD distributions, and in
Sec.~\ref{ss:fitvar} we investigate the effect of variations on the baseline
fit configuration.

\subsection{Fit quality}
\label{ss:fitqual}

In this section, we discuss the quality of the baseline fit performed at
N$^3$LL$^-$ imposing the cuts on $|\qT|/Q$ as discussed in
Sec.~\ref{s:data}. The error analysis is performed with the so-called
bootstrap method, which consists in fitting an ensemble of 250 Monte Carlo
replicas of the experimental data (see Ref.~\cite{Bacchetta:2017gcc} for more
details).

The most complete statistical information about the TMDs is given by the full
ensemble of 250 replicas.
For some purposes, however, it is useful to define a single, representative
result instead of the full replica ensemble. In order to estimate the quality of our fit, the most appropriate indicator is
the $\chi^2$ value of the best fit to the central (not fluctuated)
experimental data ($\chi^2_0$). We refer to
this fit as the ``central replica."

It is possible to analyze also the average of the $\chi^2$ over all replicas
($\langle \chi^2 \rangle$) as well as the $\chi^2$ of the
mean replica ($\chi_m^2$),
constructed as the average of all replicas~\cite{Bacchetta:2019sam}. These
values should be very close to each other.

Tab.~\ref{t:chitable} reports the breakdown of the $\chi_0^2$ values normalized
to the number of data points ($N_{\text{dat}}$) for DY and SIDIS datasets. As
already discussed in Ref.~\cite{Bacchetta:2019sam},
in the presence of bin-by-bin correlated uncertainties the total $\chi^2$ can be expressed as the sum of two contributions
\begin{eqnarray}
\label{e:chi2terms}
\chi^2 = \sum_i^N \left( \frac{\text{exp}_i - \overline{\text{th}}_i}{\sigma_i} \right)^2 + \chi^2_{\lambda} = \chi^2_D + \chi^2_{\lambda} \; ,
\end{eqnarray}
where $\chi^2_D$ is given by the standard formula for $N$ experimental
data points exp$_i$ and statistical and uncorrelated uncertainties
$\sigma_i^2 = \sigma_{i, \rm stat}^2 + \sigma_{i, \rm uncor}^2$, but involving theoretical predictions shifted by the correlated uncertainties,
\begin{equation}
\label{e:th_shifts}
\overline{\text{th}}_i = \text{th}_i + \sum_{\alpha=1}^k
\lambda_\alpha \, \sigma_{i, \rm corr}^{(\alpha)} \; ,
\end{equation}
where $\sigma_{i, \rm corr}^{(\alpha)}$ is the $\alpha$-th (100\%) correlated uncertainty associated with the $i$-th experimental data
point and $\lambda_\alpha$ is the so-called nuisance parameter. In Eq.~\eqref{e:chi2terms}, the $\chi^2_{\lambda}$ is a penalty term due to the presence of correlated uncertainties and is completely determined by the nuisance parameters,
\begin{equation}
\label{e:chilambda}
\chi^2_\lambda =  \sum_{\alpha=1}^k \lambda_\alpha^2 \; .
\end{equation}
The optimal value of the nuisance parameters is obtained by minimizing the total $\chi^2$ in Eq.~\eqref{e:chi2terms} with respect to them.
Because the shifted predictions in Eq.~\eqref{e:th_shifts} provide
a better visual assessment of the fit quality, we will consistently
display them for all observables used in the fit.

From Tab.~\ref{t:chitable}, the global $\chi_0^2$ value is 1.06,
indicating that the description of the whole dataset is very good.\footnote{We
also obtain $\langle \chi^2 \rangle = 1.08 \pm 0.01$ and
$\chi_m^2 = 1.07$.} This means
that the fit is able to simultaneously describe experimental data coming from
two different processes over a wide kinematic range. As can be seen in
Tabs.~\ref{t:dataDY} and \ref{t:dataSIDIS}, the low-energy dataset comes from
fixed-target DY experiments and SIDIS observables, while the high-energy
dataset comes from collider experiments at the LHC and Tevatron at energies
higher by more than two orders of magnitude.

It is important to notice that the correlated penalty term $\chi^2_{\lambda}$
gives a significant contribution to the total $\chi^2_0$ value. This means that
the shifts induced by correlated uncertainties are often
large.
The $\chi^2_{\lambda}/N_{\text{dat}} = 0.29$ obtained in this analysis is
larger than the one in Ref.~\cite{Bacchetta:2019sam}, mainly because
of the different treatment of the theoretical uncertainties  related to collinear PDFs and FFs,
which are considered here as 80\% correlated.

\begin{table}[t]
\footnotesize
\begin{center}
\renewcommand{\tabcolsep}{0.4pc}
\renewcommand{\arraystretch}{1.2}
\begin{tabular}{|l|c|c|c|c|}
  \hline
  \multicolumn{1}{|c|}{ } & \multicolumn{4}{|c|}{N$^3$LL$^-$} \\
  \hline
  Data set & $N_{\rm dat}$ & $\chi^2_D$  &  $\chi^2_{\lambda}$ & $\chi^2_0$ \\
  \hline
  \hline
  CDF Run I & 25 & 0.45 & 0.09 & 0.54 \\
  \hline
  CDF Run II & 26 & 0.995 & 0.004 & 1.0 \\
  \hline
  D0 Run I & 12 & 0.67 & 0.01 & 0.68 \\
  \hline
  D0 Run II & 5 & 0.89 & 0.21 & 1.10 \\
  \hline
  D0 Run II $(\mu)$ & 3 & 3.96 & 0.28 & 4.2 \\
  \hline
  {\it Tevatron total} & 71 & 0.87 & 0.06 & 0.93 \\
  \hline
  LHCb 7 TeV & 7 & 1.24 & 0.49 & 1.73 \\
  \hline
  LHCb 8 TeV & 7 & 0.78 & 0.36 & 1.14 \\
  \hline
  LHCb 13 TeV & 7 & 1.42 & 0.06 & 1.48 \\
  \hline
  {\it LHCb total} & 21 & 1.15 & 0.3 & 1.45 \\
  \hline
  ATLAS 7 TeV & 18 & 6.43 & 0.92 & 7.35 \\
  \hline
  ATLAS 8 TeV & 48 & 3.7 & 0.32 & 4.02 \\
  \hline
  ATLAS 13 TeV & 6 & 5.9 & 0.5 & 6.4 \\
  \hline
  {\it ATLAS total} & 72 & 4.56 & 0.48 & 5.05 \\
  \hline
  CMS 7 TeV & 4 & 2.21 & 0.10 & 2.31 \\
  \hline
  CMS 8 TeV & 4 & 1.938 & 0.001 & 1.94 \\
  \hline
  CMS 13 TeV & 70 & 0.36 & 0.02 & 0.37 \\
  \hline
  {\it  CMS total} & 78 & 0.53 & 0.02 & 0.55 \\
  \hline
  PHENIX 200 & 2 & 2.21 & 0.88 & 3.08 \\
  \hline
  STAR 510 & 7 & 1.05 & 0.10 & 1.15 \\
  \hline
  \hline
  DY collider total & 251 & 1.86 & 0.2 & 2.06 \\
  \hline
  \hline
  E288 200 GeV & 30 & 0.35 & 0.19 & 0.54 \\
  \hline
  E288 300 GeV & 39 & 0.33 & 0.09 & 0.42 \\
  \hline
  E288 400 GeV & 61 & 0.5 & 0.11 & 0.61 \\
  \hline
  E772 & 53 & 1.52 & 1.03 & 2.56 \\
  \hline
  E605 & 50 & 1.26 & 0.44 & 1.7 \\
  \hline
  \hline
  DY fixed-target total & 233 & 0.85 & 0.4 & 1.24 \\
  \hline
  \hline
  HERMES ($p \rightarrow \pi^+$) & 45 & 0.86 & 0.42 & 1.28 \\
  \hline
  HERMES ($p \rightarrow \pi^-$) & 45 & 0.61 & 0.31 & 0.92 \\
  \hline
  HERMES ($p \rightarrow K^+$) & 45 & 0.49 & 0.04 & 0.53 \\
  \hline
  HERMES ($p \rightarrow K^-$) & 37 & 0.18 & 0.13 & 0.31 \\
  \hline
  HERMES ($d \rightarrow \pi^+$) & 41 & 0.68 & 0.45 & 1.13 \\
  \hline
  HERMES ($d \rightarrow \pi^-$) & 45 & 0.63 & 0.35 & 0.97 \\
  \hline
  HERMES ($d \rightarrow K^+$) & 45 & 0.2 & 0.02 & 0.22 \\
  \hline
  HERMES ($d \rightarrow K^-$) & 41 & 0.14 & 0.08 & 0.22 \\
  \hline
  {\it HERMES total} & 344 & 0.48 & 0.23 & 0.71 \\
  \hline
  COMPASS ($d \rightarrow h^+$) & 602 & 0.55 & 0.31 & 0.86 \\
  \hline
  COMPASS ($d \rightarrow h^-$) & 601 & 0.68 & 0.3 & 0.98 \\
  \hline
  {\it COMPASS total} & 1203 & 0.62 & 0.3 & 0.92 \\
  \hline
  \hline
  SIDIS total & 1547 & 0.59 & 0.28 & 0.87 \\
  \hline
  \hline
  {\bf Total} & {\bf 2031} & {\bf 0.77} & {\bf 0.29} & {\bf 1.06} \\
  \hline
\end{tabular}
\caption{
  Breakdown of the values of $\chi^2$ normalized to the number of data
  points $N_{\text{dat}}$ that survive the kinematic cuts for all
  datasets considered in our baseline fit.
  The $\chi^2_D$ refers to uncorrelated
  uncertainties, $\chi^2_\lambda$ is the penalty term due to correlated
  uncertainties (see Eq.~(\ref{e:chi2terms})), $\chi^2_0$ is the sum of $\chi^2_D$ and $\chi^2_\lambda$. All
  $\chi^2$ values refer to the central replica (see text).
  }
\label{t:chitable}
\end{center}
\end{table}

\subsubsection{SIDIS}
\label{sss:qualitySIDIS}

For $\hermes$ and $\compass$ multiplicities, which represent about 75\% of the total number of data points considered in
this work, the description obtained by our
global analysis is very good. From Tab.~\ref{t:chitable}, the values of
$\chi^2_0/N_{\text{dat}}$ are almost always smaller than 1.

In the case of $\hermes$ multiplicities, we note that the largest contribution to
the $\chi^2_0$ comes from the $\pi^+$ channel, for both proton and deuteron
targets. This result is consistent with the findings in both
Refs.~\cite{Signori:2013mda, Bacchetta:2017gcc} and~\cite{Scimemi:2019cmh}.
Since kaon
multiplicities are affected by larger statistical errors, and the collinear
FFs for kaons display large uncertainties, the corresponding $\chi^2_0$ value is
lower.

\begin{figure}[h]
\centering
\includegraphics[width=0.8\textwidth]{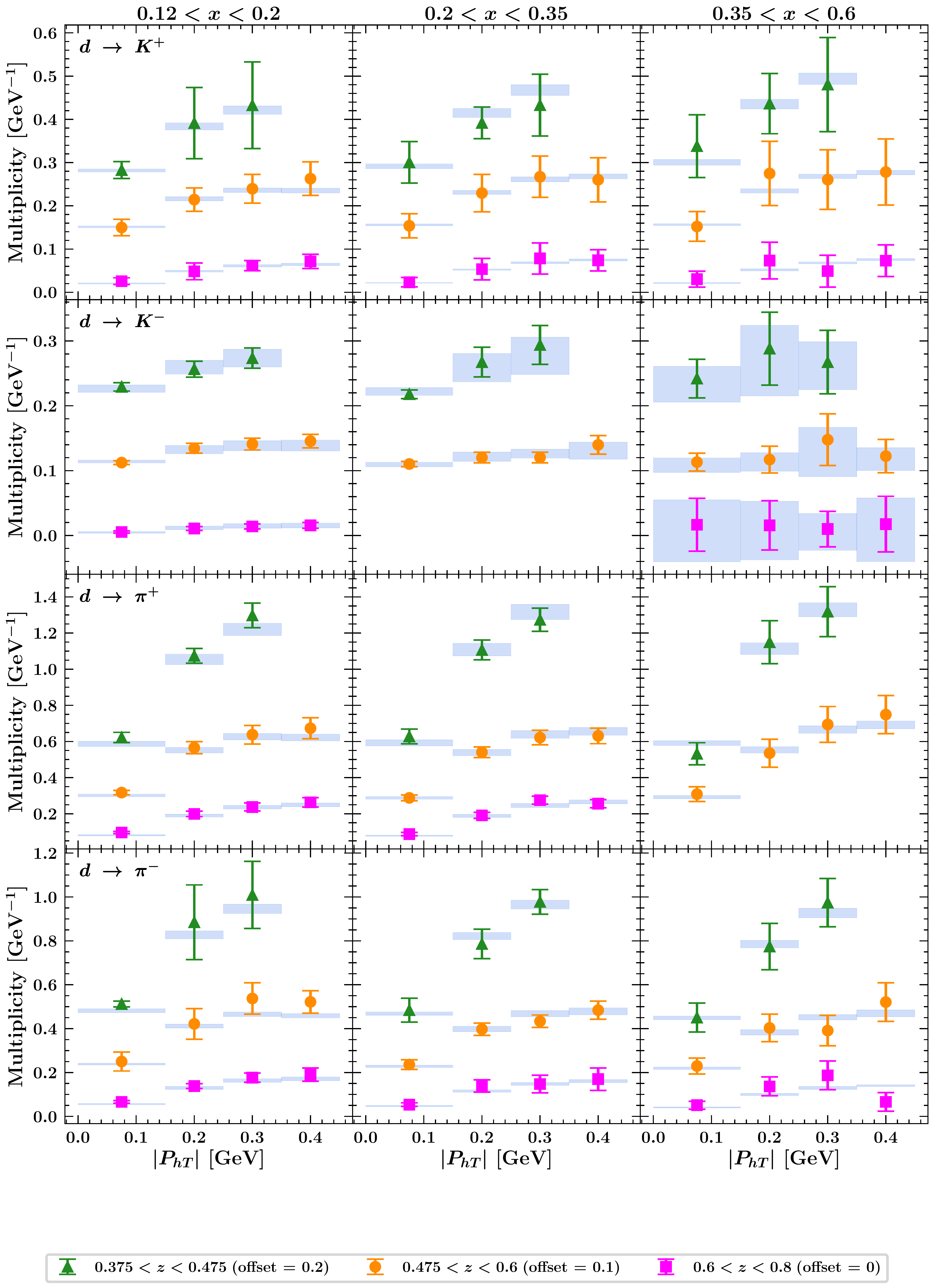}
\caption{Comparison between data and theoretical predictions for the $\hermes$ multiplicities for the production of charged pions and kaons off a deuteron target for different $x$ and $z$ bins as a function of the transverse momentum $|\PhT|$ of the final-state hadron. For better visualization, each $z$ bin is shifted by the indicated offset.}
\label{f:HER_plot_D}
\end{figure}

The comparison between theoretical results for the SIDIS multiplicities of
Eq.~\eqref{e:SIDIS_mult} and $\hermes$ data for the production of charged
pions and kaons off a deuteron target is shown in
Fig.~\ref{f:HER_plot_D}. Each column corresponds to a specific $x$ bin. Each
row corresponds to a specific final-state channel. The results are displayed
as functions of the transverse momentum $|\PhT|$ of the measured final-state
hadron. Points with different markers and colors correspond to different
representative $z$
bins, and are offset for a better visualization as indicated in the plot legend. The
light blue rectangles are the theoretical results and correspond to
the 68\% Confidence-Level (CL) band (namely excluding the
largest and smallest 16\% of the replicas).

We note that for $K^-$ production the central column, corresponding to the $0.2 < x < 0.35$ bin, does not include the magenta points for the highest $0.6 < z < 0.8$ bin because of the kinematic cut in $z$. Similarly, in all panels there are only three green points (for the lowest $z$ bin) because of the $z$-dependent cut in Eq.~\eqref{e:SIDIScut}, which leads to the exclusion of a larger number of $|\qT|$ bins at lower values of $z$. The theoretical results display larger uncertainties for $K^-$ production (second row) because of the combined effect of larger experimental errors and larger uncertainties in the kaon collinear FFs.

\begin{figure}[h]
\centering
\includegraphics[width=0.8\textwidth]{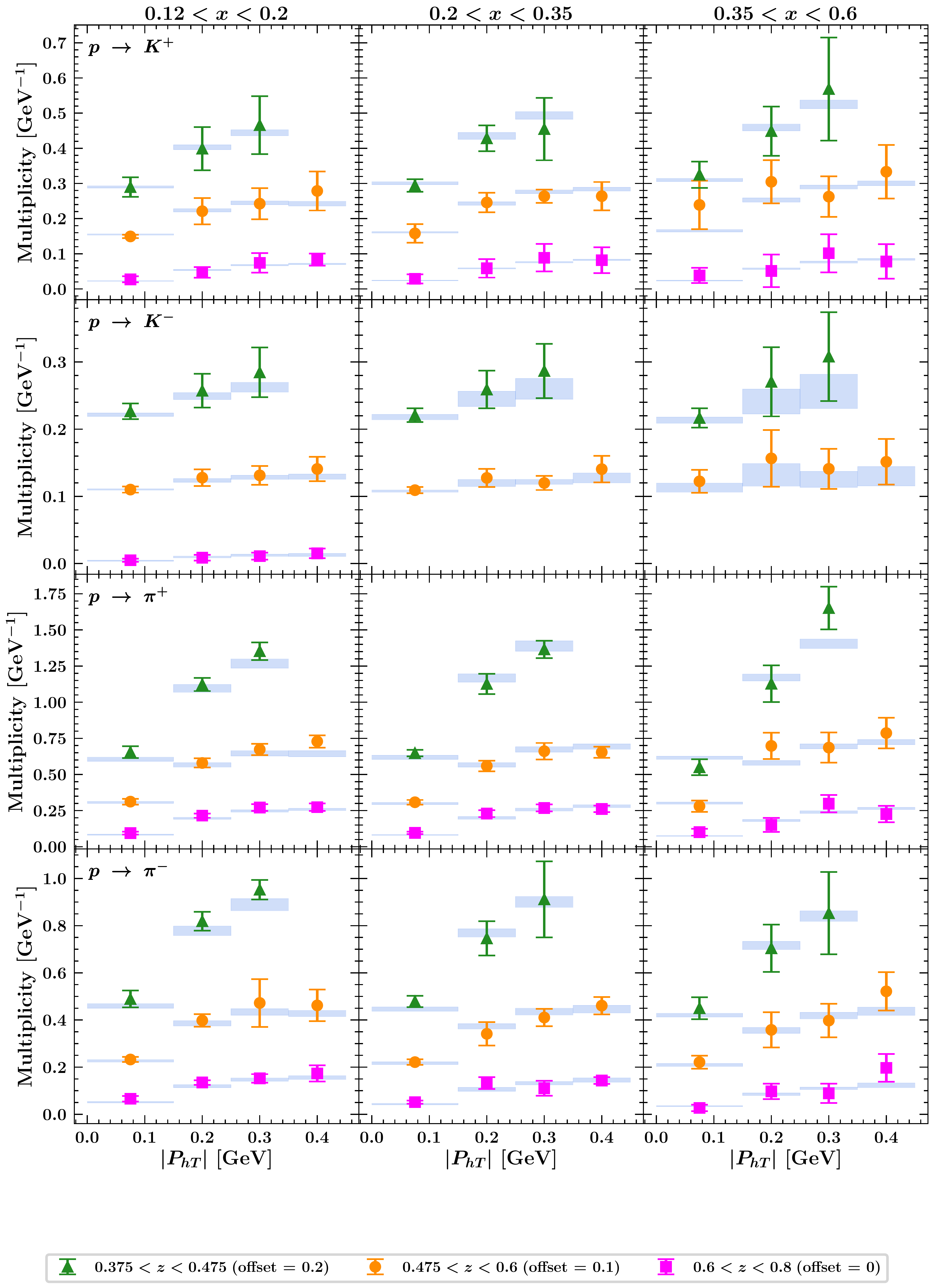}
\caption{Same conventions and notation as in previous figure but for charged pions and kaons off proton target.}
\label{f:HER_plot_P}
\end{figure}

Fig.~\ref{f:HER_plot_P} refers to the same $\hermes$ multiplicities with same conventions and notation as in Fig.~\ref{f:HER_plot_D} but off a proton target. We remark that for $K^-$ production the kinematic cuts have a more drastic effect because the magenta points for the $|\PhT|$ distributions at the largest $z$ bin are excluded for the two largest $x$ bins considered (central and rightmost panels of second row from top).

\begin{figure}[h]
\centering
\includegraphics[width=1.0\textwidth]{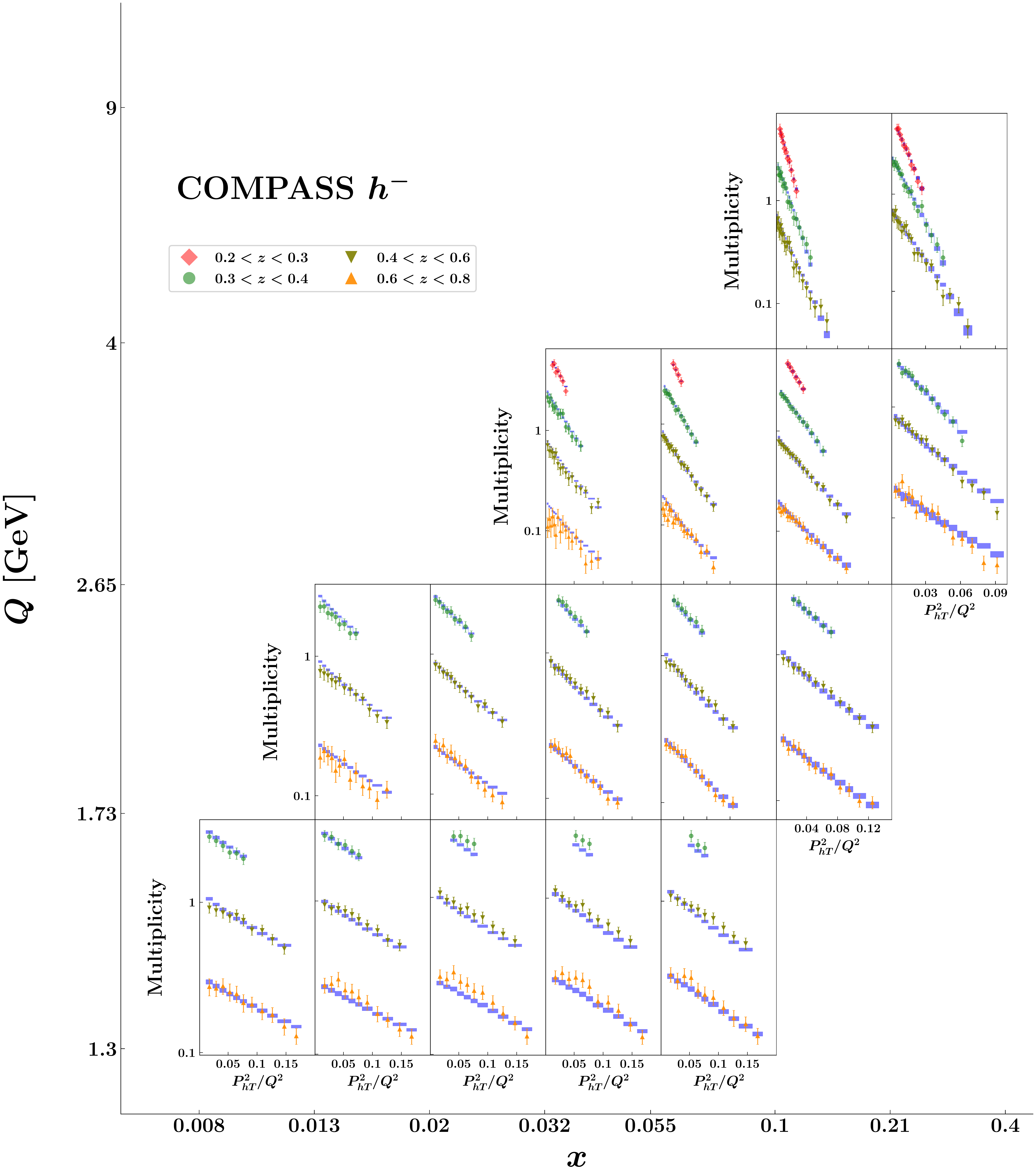}
\caption{Comparison between data and theoretical predictions for the $\compass$ multiplicities for the production of negatively charged hadrons off a deuteron target. For each $Q, x$ bin,  the multiplicities are displayed as functions of $\PhT^2/Q^2$ for different $z$ bins surviving kinematic cuts, as indicated in the legend.}
\label{f:COMP_plot_M}
\end{figure}

In Fig.~\ref{f:COMP_plot_M}, we show the result of our fit for the
$\compass$ SIDIS multiplicities for the production of unidentified
negatively charged hadrons off a deuteron target. For each $Q$ and $x$
bin, each panel displays the multiplicity on a logarithmic scale as a
function of $\PhT^2/Q^2$. Again, points with different markers and
colors correspond to different representative $z$ bins, as indicated
in the plot legend. As before, the light-blue rectangles correspond to the 68\% CL
theoretical results. The results for unidentified negatively charged hadrons $h^-$ are obtained by simply adding the results for negatively charged pions and kaons, $h^- \sim \pi^- + K^-$.

We note that the agreement is good for almost all bins, which is reflected in small $\chi^2_0$ values in Tab.~\ref{t:chitable}. The situation worsens for the lowest $Q$ bin ($1.3 < Q < 1.73$ GeV), particularly for $x \gtrsim 0.02$.
We also remark that for some $Q$ and $x$ bins the theoretical uncertainties for the largest $z$ bin compatible with our kinematic cuts ($0.6 < z < 0.8$) are significantly larger than for other $z$ bins because of much larger uncertainties in the collinear FFs.
Finally, looking at the table of panels from top to bottom, we realize that the lowest $z$-bin distributions (red  diamonds) are present only for the largest $Q$ bins, and vice-versa, because of the kinematic cut in Eq.~\eqref{e:SIDIScut}

\begin{figure}[h]
\centering
\includegraphics[width=1.0\textwidth]{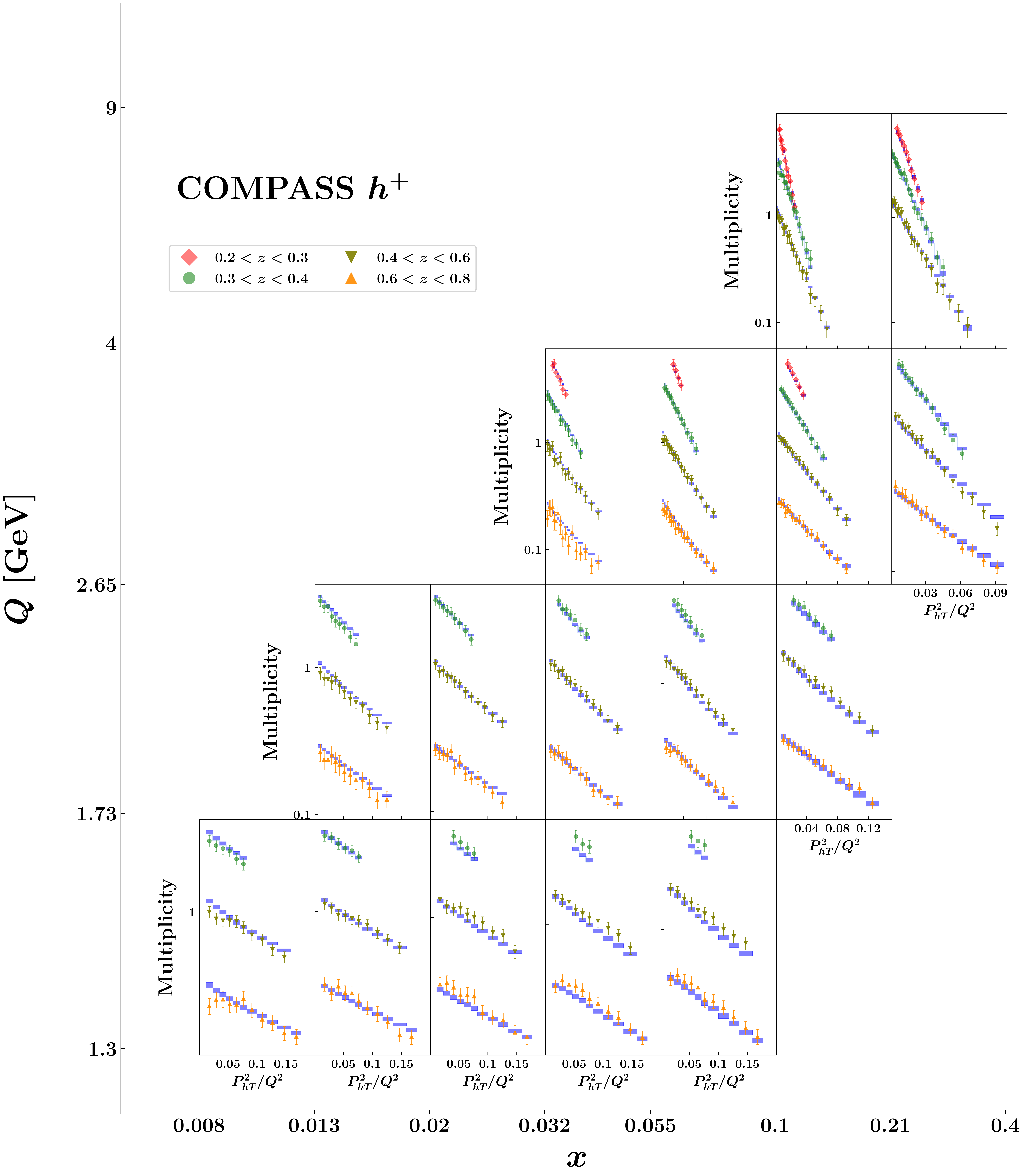}
\caption{Same conventions and notation as in previous figure but for unidentified positively charged hadrons off deuteron target.}
\label{f:COMP_plot_P}
\end{figure}

Fig.~\ref{f:COMP_plot_P} refers to the same $\compass$ multiplicities
with same conventions and notation as in Fig.~\ref{f:COMP_plot_M} but
for unidentified positively charged hadrons $h^+$. Again, the
light-blue rectangles correspond to the 68\% CL theoretical results, and are obtained by adding the results for positively charged pions and kaons, $h^+ \sim \pi^+ + K^+$. Comments similar to Fig.~\ref{f:COMP_plot_M} can be made about the agreement between data and theory.

\subsubsection{Drell-Yan}
\label{sss:qualityDY}

DY data represents approximately 25\% of the full set of analyzed
data.
From Tab.~\ref{t:chitable} it is evident that most of low-energy DY data from
fixed-target experiments (\textsc{E605, E288, E772}), but also from
\textsc{PHENIX} and \textsc{STAR}, can be fitted with low $\chi^2$ values,
much lower than high-energy DY data from collider experiments like,
\textit{e.g.}, those at
the LHC. As already pointed out in Ref.~\cite{Bacchetta:2019sam}, this most
likely originates from the fact that low-energy DY data are affected by larger
errors and collinear PDFs at these kinematics have larger
uncertainties.

From Tab.~\ref{t:chitable}, we also note that the quality of our fit for the
\textsc{ATLAS} datasets is poor. In particular, the description worsens for
the first two low-rapidity bins of both \textsc{ATLAS} 7 TeV and
\textsc{ATLAS} 8 TeV datasets, the worst case being at $|y| < 1$ for
\textsc{ATLAS} 7 TeV. Several effects might be responsible for this
result. Since the experimental observable is a normalized cross section,
systematic errors cancel in the ratio producing measurements with very small
error bars. Fitting these data is very difficult, also because small
theoretical effects can give significant contributions to the
$\chi^2$. Moreover, different implementations of phase-space cuts on the final-state leptons could lead to modifications
in both the shape and the normalization of the theoretical observable (see,
\textit{e.g.}, Ref.~\cite{Chen:2022cgv,Buonocore:2021tke,Camarda:2021jsw}). We
leave this issue for future studies. At variance with
Ref.~\cite{Scimemi:2019cmh}, we obtain our results without excluding any extra
data points on top of the ones exceeding the maximum value of $|\qT|/Q$ in
Eq.~\eqref{e:DYcut}.

It is interesting to comment the results of the fit for those datasets that were not included in the previous analysis of Ref.~\cite{Bacchetta:2019sam} (see Sec.~\ref{s:data}). For \textsc{E772}, we are able to obtain a good description only for data points above the peak of the $\Upsilon$ resonance. For $Q < 9$ GeV, the quality of the fit worsens. At variance with Ref.~\cite{Scimemi:2019cmh}, we keep the $Q < 9$ GeV bins because there is no evident motivation to exclude them.

The new \textsc{CMS} dataset at $\sqrt{s} = 13$ TeV is important
because it extends the kinematic coverage considered in
Ref.~\cite{Bacchetta:2019sam}. This dataset is very nicely described,
even better than the ones at the smaller center-of-mass energies
$\sqrt{s}=7, 8$ TeV. This is probably due to the fact that the
\textsc{CMS} 13 TeV dataset is densely binned in
rapidity.

\begin{figure}[h]
\centering
\includegraphics[width=1.0\textwidth]{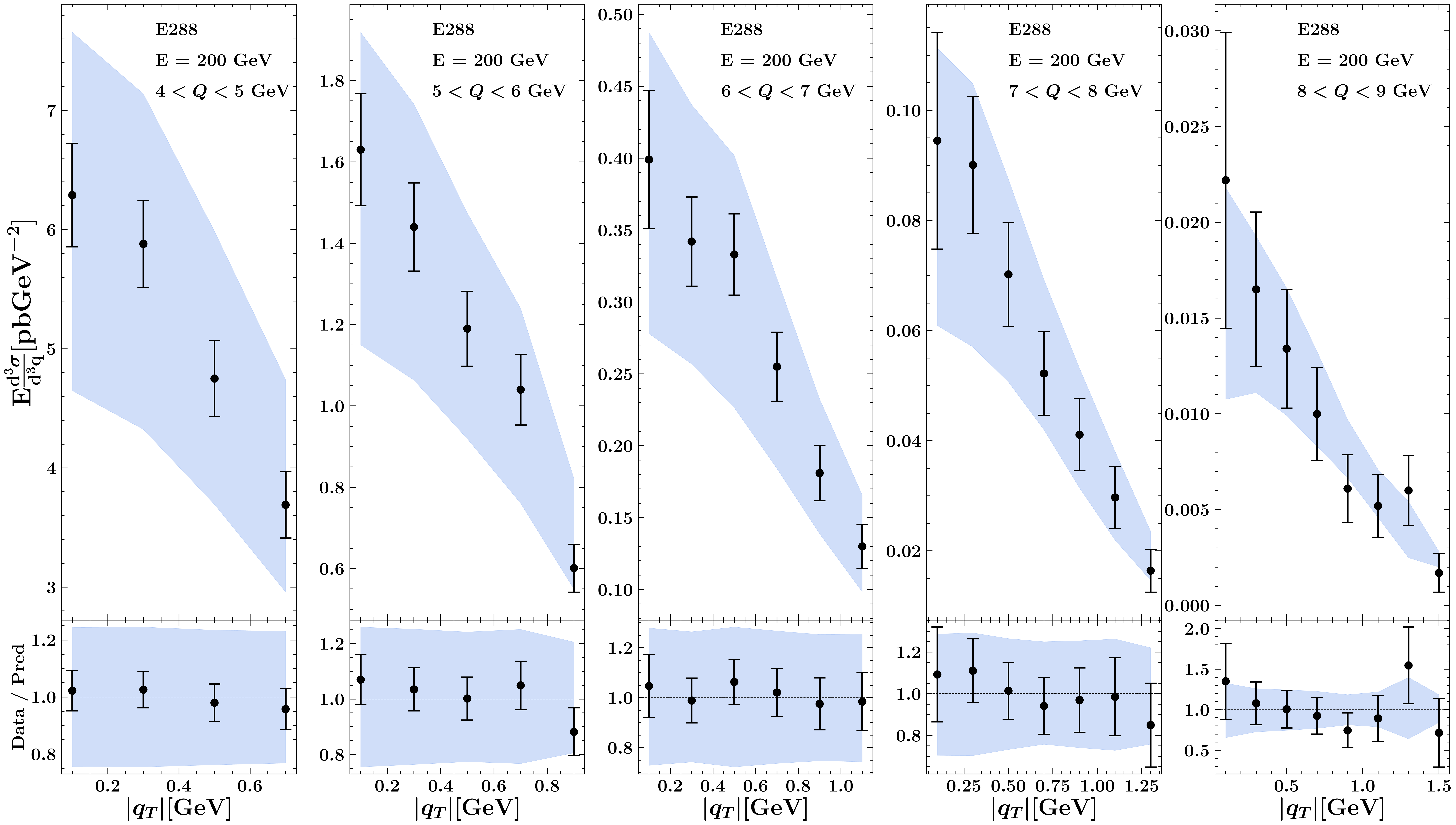}
\caption{Upper panel: comparison between data and theoretical predictions for the DY cross section differential in $|\qT|$ for the \textsc{E288} dataset at $E_{beam} = 200$ GeV for different $Q$ bins; uncertainty bands correspond to the 68\% CL. Lower panel: ratio between experimental data and theoretical cross section.} 
\label{f:E288_plot}
\end{figure}

\begin{figure}[h]
\centering
\includegraphics[width=1.0\textwidth]{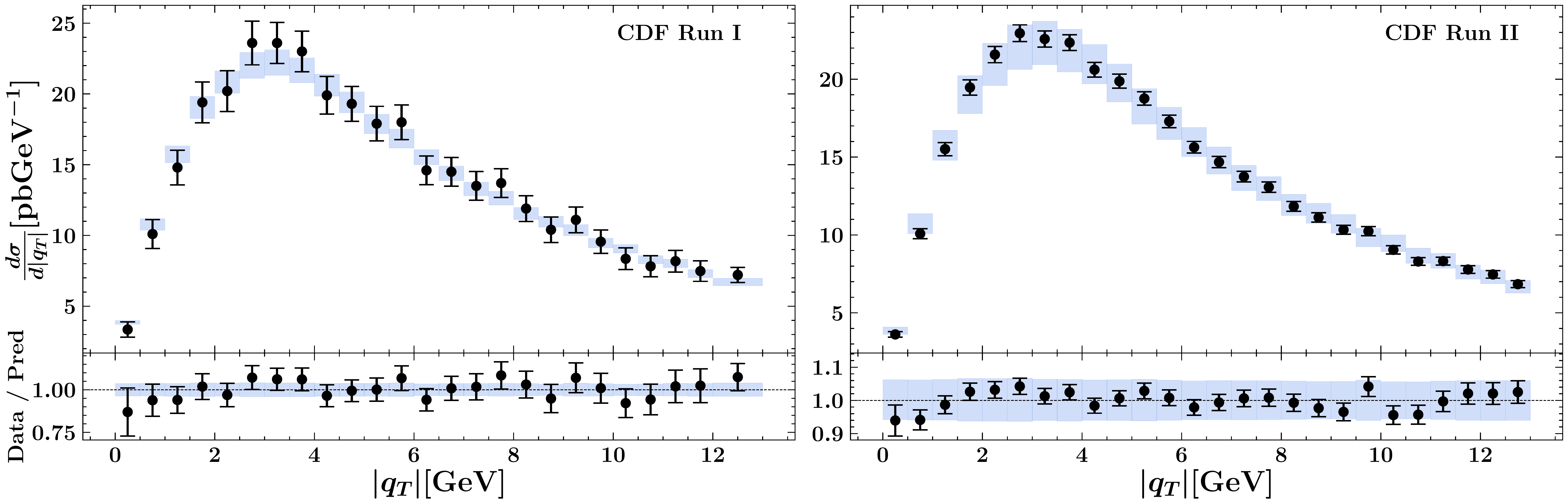}
\caption{Upper panels: comparison between experimental data and theoretical
  predictions for the cross section differential in $|\qT|$ for $Z$ bosons produced in
  $p\bar{p}$ collisions at the Tevatron from \textsc{CDF} Run I (left panel) and run II (right panel); uncertainty bands correspond to the 68\% CL. Lower panel: ratio between
  experimental data and theoretical results.}
\label{f:CDF_plot}
\end{figure}

\begin{figure}[h]
\centering
\includegraphics[width=1.0\textwidth]{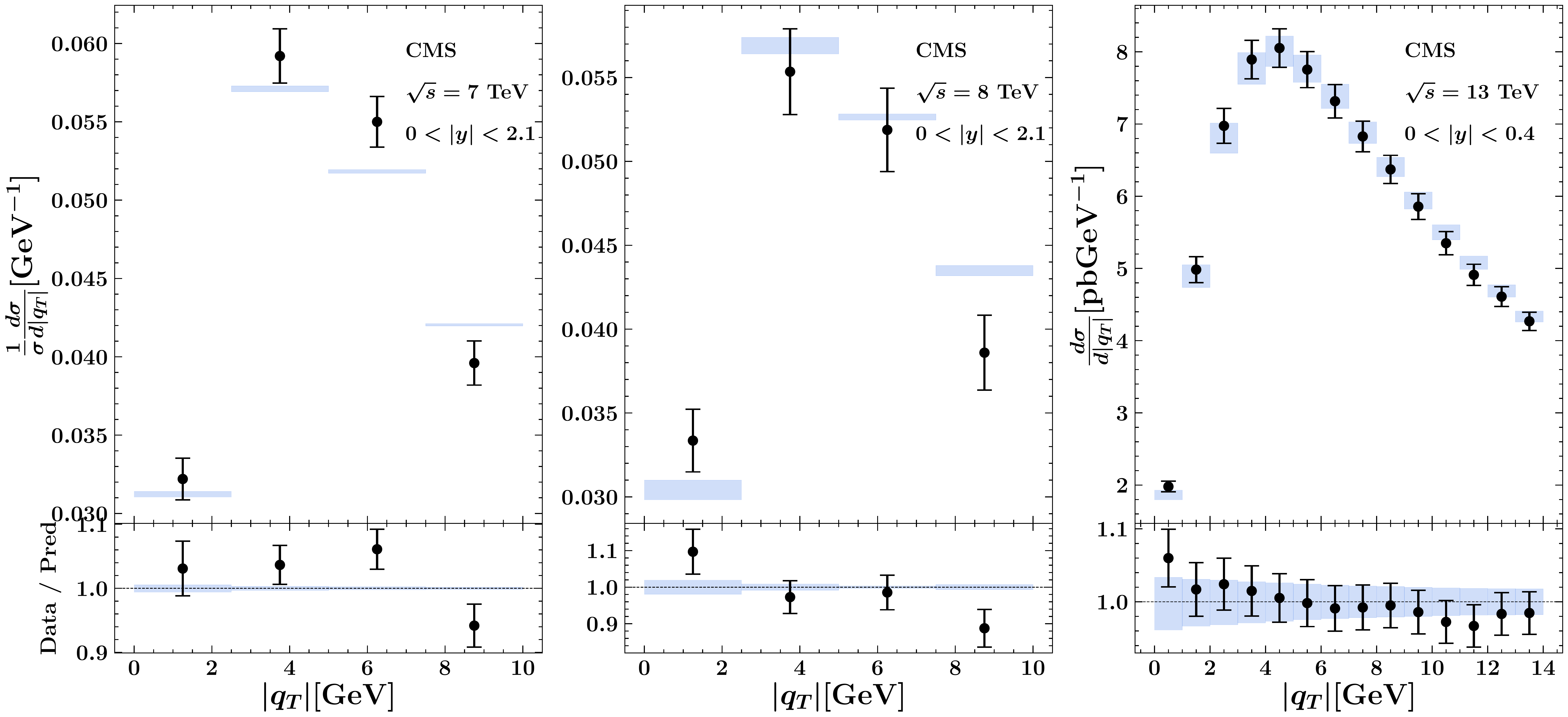}
\caption{Same as in previous figure but for $Z$ boson production in $pp$ collisions measured by the CMS Collaboration. From left to right: increasing $\sqrt{s} =$ 7,  8, 13 TeV, respectively. For $\sqrt{s} =$ 7, 8 TeV, the results are normalized to the fiducial cross section.}
\label{f:CMS_plot}
\end{figure}

\begin{figure}[h]
\centering
\includegraphics[width=1.0\textwidth]{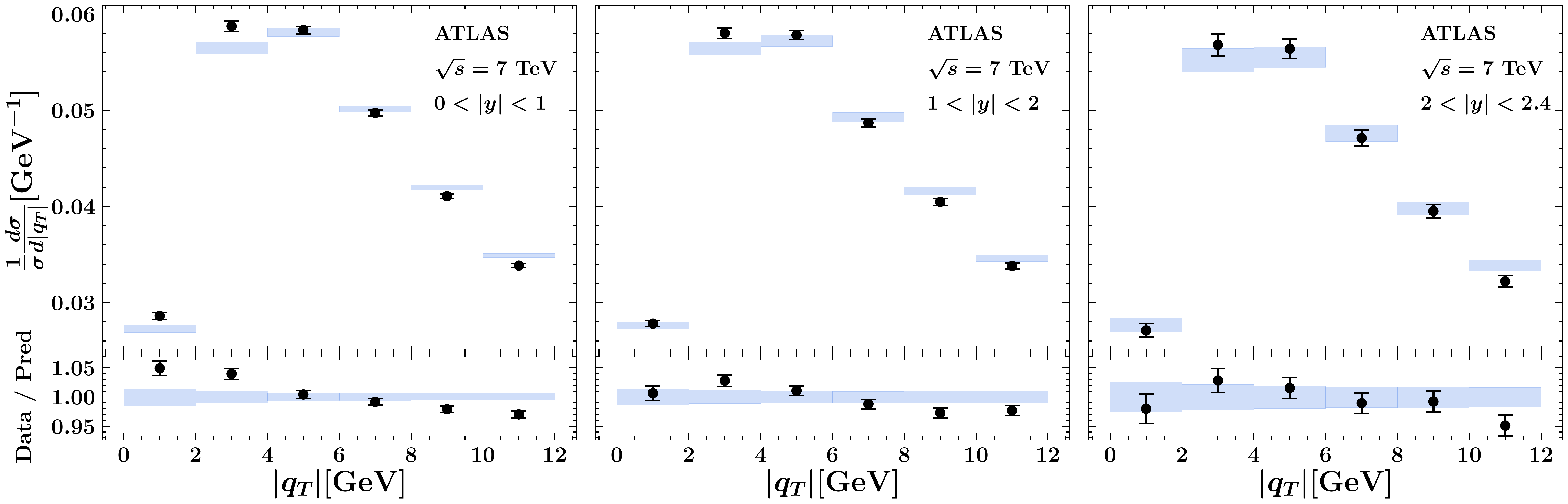}
\caption{Same as in the left and central panels of previous figure, but
  for \textsc{ATLAS} kinematics at $\sqrt{s} = 7$ TeV. From left to right, results
  at increasing rapidity.}
\label{f:ATLAS_plot}
\end{figure}

In order to visualize the quality of our fit, in
Figs.~\ref{f:E288_plot}-\ref{f:ATLAS_plot} we present the comparison
between experimental data and theoretical results for a representative
selection of the DY dataset. In the upper panels of each plot, we
display the cross section differential in $|\qT|$, while in the lower
panels we show the ratio of data to theory. As for the SIDIS case, the
light-blue bands are the 68\% CL theoretical results.

Fig.~\ref{f:E288_plot} displays the DY cross section for
the \textsc{E288} dataset at beam energy $E_{beam} = 200$ GeV for
different bins in $Q$. For DY fixed-target observables, we calculate the cross
section at mean values of $|\qT|$ (not integrating upon the bin limits). Hence,
we display the 68\% CL uncertainty as a band
rather than a series of rectangles. We remark that the uncertainty band is
larger for lower $Q$ bins; this trend is induced by larger correlated
uncertainties for smaller invariant masses of the lepton pair.

In Fig.~\ref{f:CDF_plot}, we compare the theoretical results for the
cross section for DY in $p\bar{p}$ collisions at the Tevatron.  Black
data points in the left panel correspond to the results of Run I of
the \textsc{CDF} experiment, while in the right panel the results for
Run II are reported. The lower panels show the corresponding ratio of
experimental data to theoretical results. The latter are displayed as
light-blue rectangles, each one corresponding to the integral of the
cross section within the corresponding bin limits. The size of the
rectangle is given by the 68\% CL. The quality of the fit for
\textsc{CDF} data is comparable to the one in Ref.~\cite{Bacchetta:2019sam}.

In Fig.~\ref{f:CMS_plot}, we compare the theoretical results for the
DY cross section in $pp$
collisions at the LHC. From left to right, the black data points
refer to the measurements by the \textsc{CMS} Collaboration at increasing
$\sqrt{s} =$ 7, 8, 13 TeV. For $\sqrt{s} =$ 7, 8 TeV, the
results are normalized to the fiducial cross section. As in previous figures,
the lower panels display the ratio of experimental data to
the 68\% CL theoretical results.
 For $\sqrt{s} =$ 7, 8 TeV, the quality of the fit is comparable to that in Ref.~\cite{Bacchetta:2019sam}. For the new dataset at $\sqrt{s} = 13$ TeV, the agreement in the displayed rapidity bin is excellent, but remains very good also for higher rapidities.

In Fig.~\ref{f:ATLAS_plot}, we compare the theoretical results for the DY
cross section in $pp$ collisions normalized to the fiducial cross
section for the \textsc{ATLAS} data at $\sqrt{s} = 7$
TeV. From left to right, we consider three representative bins at
increasing rapidity $|y|$. The leftmost one corresponds to the worst described bin in our global fit, with
$\chi^2 / N_{\text{data}} = 13.5$.  As in previous figures, the lower part of
each panel displays the ratio of experimental data to the 68\% CL theoretical
results.
The quality of the fit increases at more forward rapidities (from left to
right). The same trend is observed at $\sqrt{s} = 8$ TeV, but not for
\textsc{CMS} at 13 TeV.

\subsection{TMD distributions}
\label{ss:tmds}

We now discuss the TMD distributions extracted from our baseline fit with
N$^3$LL$^-$ accuracy. Tab.~\ref{t:fitparams} displays the list of our 21
fitting parameters with their mean value and standard deviation.
The majority of the parameters is well constrained. The only parameter that is
compatible with zero is $\gamma_2$.

\begin{table}[h]
\footnotesize
\begin{center}
\renewcommand{\tabcolsep}{0.4pc}
\renewcommand{\arraystretch}{1.2}
\begin{tabular}{|c|c|}
  \hline
  \textbf{Parameter} & \textbf{Average over replicas} \\
  \hline
  $g_2 \ [\text{GeV}]$ & 0.248 $\pm$ 0.008 \\
  \hline
  $N_1 \ [\text{GeV}^2]$ & 0.316 $\pm$ 0.025 \\
  \hline
  $\alpha_1$ & 1.29 $\pm$ 0.19 \\
  \hline
  $\sigma_1$ & 0.68 $\pm$ 0.13 \\
  \hline
  $\lambda \ [\text{GeV}^{-1}]$ & 1.82 $\pm$ 0.29 \\
  \hline
  $N_3 \ [\text{GeV}^2]$ & 0.0055 $\pm$ 0.0006 \\
  \hline
  $\beta_1$ & 10.23 $\pm$ 0.29 \\
  \hline
  $\delta_1$ & 0.0094 $\pm$ 0.0012 \\
  \hline
  $\gamma_1$ & 1.406 $\pm$ 0.084 \\
  \hline
  $\lambda_F \ [\text{GeV}^{-2}]$ & 0.078 $\pm$ 0.011 \\
  \hline
  $N_{3B} \ [\text{GeV}^2]$ & 0.2167 $\pm$ 0.0055 \\
  \hline
  $N_{1B} \ [\text{GeV}^2]$ & 0.134 $\pm$ 0.017 \\
  \hline
  $N_{1C} \ [\text{GeV}^2]$ & 0.0130 $\pm$ 0.0069 \\
  \hline
  $\lambda_2 \ [\text{GeV}^{-1}]$ & 0.0215 $\pm$ 0.0058 \\
  \hline
  $\alpha_2$ & 4.27 $\pm$ 0.31 \\
  \hline
  $\alpha_3$ & 4.27 $\pm$ 0.13 \\
  \hline
  $\sigma_2$ & 0.455 $\pm$ 0.050 \\
  \hline
  $\sigma_3$ & 12.71 $\pm$ 0.21 \\
  \hline
  $\beta_2$ & 4.17 $\pm$ 0.13 \\
  \hline
  $\delta_2$ & 0.167 $\pm$ 0.006 \\
  \hline
  $\gamma_2$ & 0.0007 $\pm$ 0.0110 \\
  \hline
\end{tabular}
\caption{
  Average and standard deviation over the Monte Carlo replicas of the free parameters fitted to the data.
  }
\label{t:fitparams}
\end{center}
\end{table}

The $\lambda$ parameter measures the relative weight of the first Gaussian and the weighted-Gaussian in the nonperturbative part of the TMD PDF in Eq.~\eqref{e:f1NP}. The value of this parameter is close to 2, indicating that the contribution of the weighted-Gaussian component is important. In Eq.~\eqref{e:f1NP}, the parameter $\lambda_2$ measures the relative weight of the first Gaussian and the third Gaussian; this parameter is small but not compatible with zero, which means that also this component of the TMD PDF is important to reach a good description of experimental data.

Our parametrization
of the nonperturbative part of TMD FFs in Eq.~\eqref{e:D1NP} contains just the
combination of a Gaussian and a weighted Gaussian: this is sufficient to describe the data in an accurate way.

The $\lambda_F$ parameter measures the relative weight of the two components; its value is close to 0.1, indicating that the contribution of the weighted Gaussian is small. Nevertheless, it has non-trivial consequences on the tail of the TMD FF, as we will show below.

The $g_2$ parameter is a key
ingredient to the extraction of the Collins--Soper kernel, discussed in
Sec.~\ref{sss:cskernel}. The same parameter was used in the analysis
of Ref.~\cite{Bacchetta:2017gcc}. It is interesting to observe that
the value obtained in the present global fit is smaller by almost a factor
of 4 with respect to Ref.~\cite{Bacchetta:2017gcc}. This may be due to
the higher theoretical accuracy of the present analysis and to the role of
the very precise high-energy Drell--Yan measurements, which also determine the very
small
standard deviation of $g_2$.

\begin{figure}[h]
\centering
\includegraphics[width=0.6\textwidth]{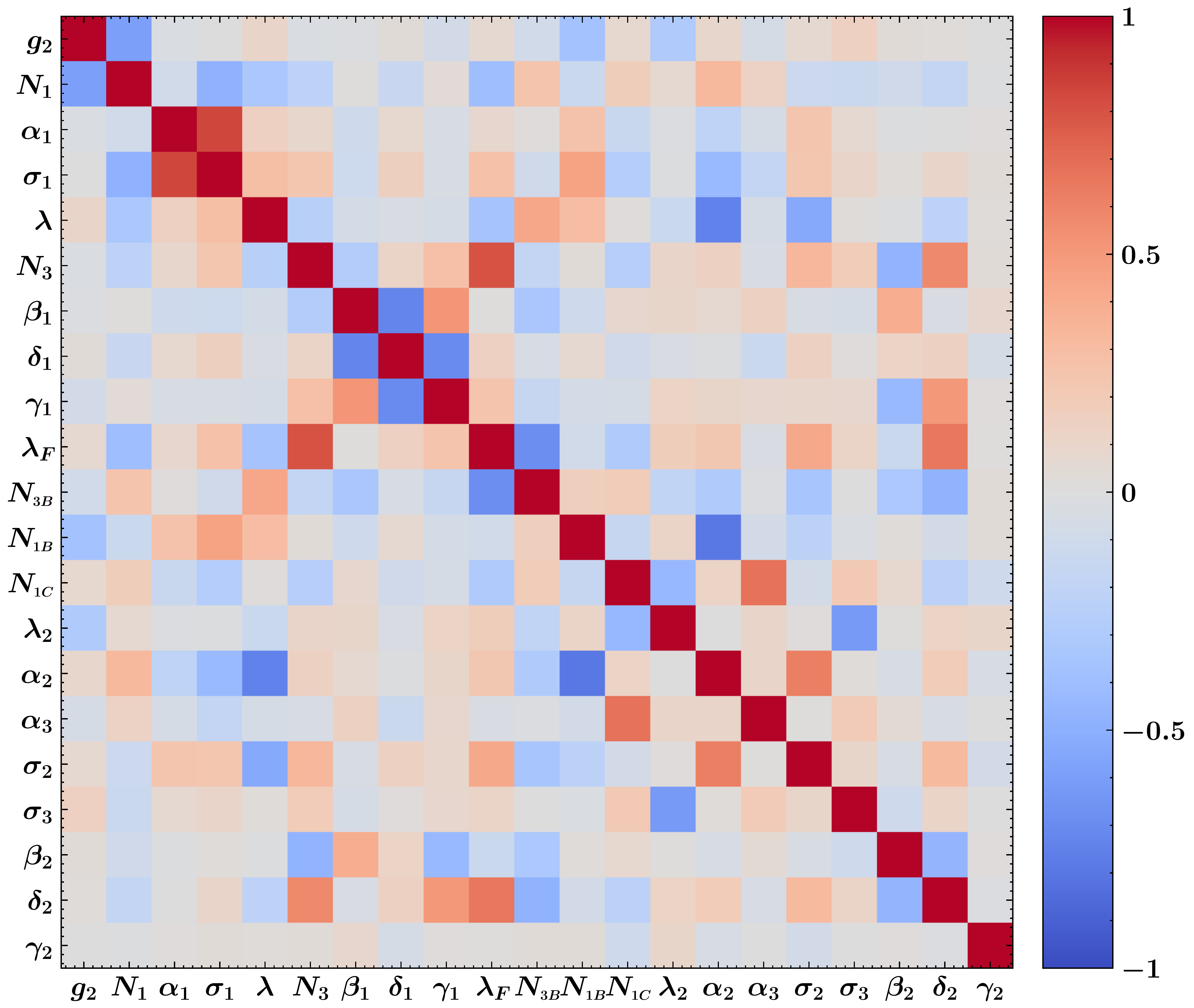}
\caption{Graphical representation of the correlation matrix for the fitted parameters.}
\label{f:params}
\end{figure}

In Fig.~\ref{f:params}, we show a graphical representation of the correlations
among the 21 fitting parameters. Using the color code indicated in the legend,
it is easy to realize that the nondiagonal elements are very small except for
some (anti--)correlation among the $\beta_1$, $\delta_1$ and $\gamma_1$
parameters that control the $z$--dependent width of the Gaussians in the TMD
FF (see Eqs.~\eqref{e:D1NP},~\eqref{e:gi_func_FF}). The overall absence of
large correlations suggests that the model parametrization of the non
perturbative parts of TMDs is appropriate.

\begin{figure}[h]
\centering
\includegraphics[width=1.0\textwidth]{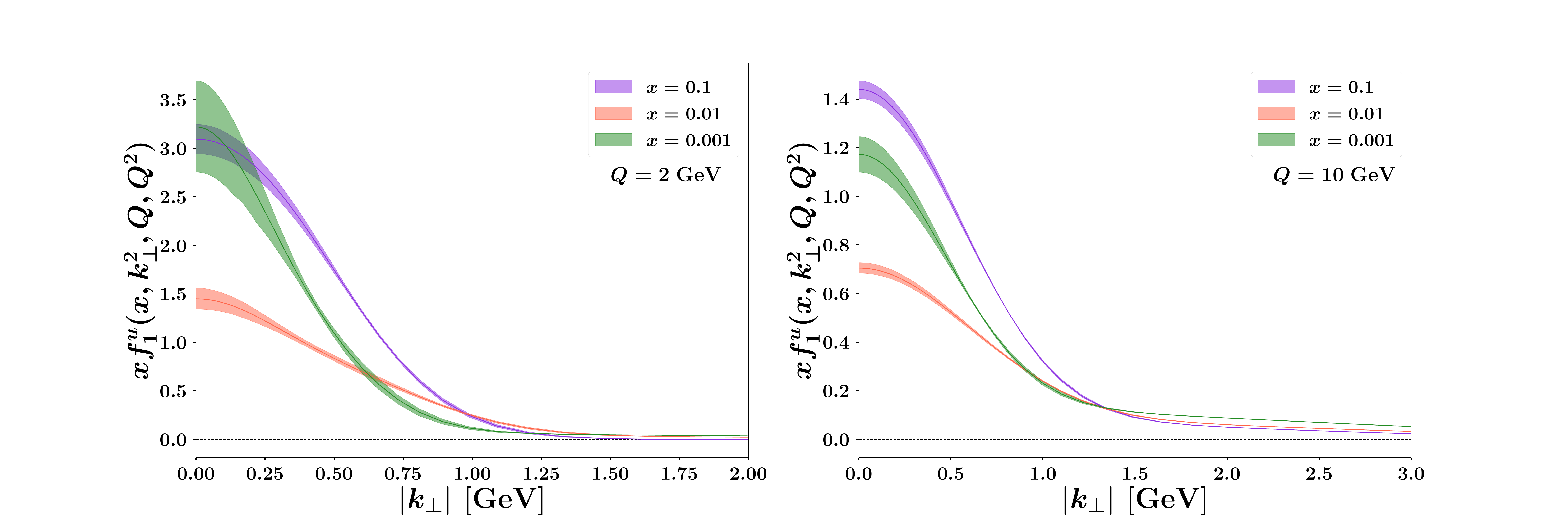}
\caption{The TMD PDF of the up quark in a proton at $\mu = \sqrt{\zeta} = Q = 2$ GeV (left panel) and 10 GeV (right panel) as a function of the partonic transverse momentum $|\kperp|$ for $x$ = 0.001, 0.01 and 0.1. The uncertainty bands represent the 68\% CL.}
\label{f:tmdpdf}
\end{figure}

In Fig.~\ref{f:tmdpdf}, we show the unpolarized TMD PDF for the up quark in the
proton at $\mu = \sqrt{\zeta} = Q = 2$ GeV (left panel) and 10 GeV (right
panel) as a function of the quark transverse momentum $|\kperp|$ for three
different values of $x$, namely $x$ = 0.001, 0.01, and 0.1. The bands
correspond to the 68$\%$ CL.

The TMD seems to be wider at intermediate $x = 0.01$, but has also a high tail
at $x= 0.001$. As already mentioned, a significant role is played by the
weighted Gaussian and by the second Gaussian in Eq.~\eqref{e:f1NP}. This may be
a sign of the presence of contributions from different quark flavors and/or
from different spin configurations (see Sec.~\ref{ss:TMDs}).

It is worth noticing that in both left and right panels the TMD PDF at $x =
0.001$ shows the largest error band, particularly at low $|\kperp|$. This is
due to the lack of experimental points in that kinematic
region (see Fig.~\ref{f:xQ2coverage}).
Future data from the Electron-Ion Collider (EIC) are expected to play an important role in getting a
better description of the TMD PDFs at low $x$~\cite{AbdulKhalek:2021gbh,AbdulKhalek:2022erw}.

\begin{figure}[h]
\centering
\includegraphics[width=1.0\textwidth]{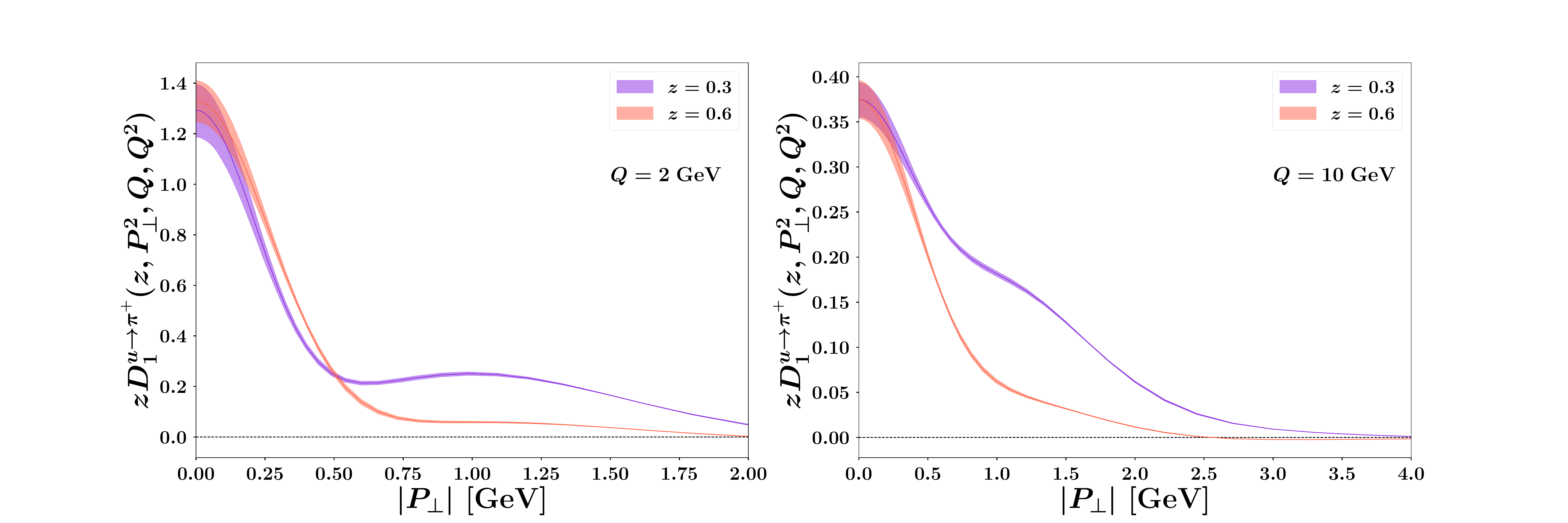}
\caption{The TMD FF for an up quark fragmenting into a $\pi^+$ at $\mu = \sqrt{\zeta} = Q = 2$ GeV (left panel) and 10 GeV (right panel) as a function of the hadron transverse momentum $|\Pperp|$ for $z$ = 0.3 and 0.6. The uncertainty bands represent the 68$\%$ CL.}
\label{f:tmdff}
\end{figure}

In Fig.~\ref{f:tmdff}, we show the TMD FF for the up quark fragmenting
into a $\pi^+$ at $\mu = \sqrt{\zeta} = Q = 2$ GeV (left panel) and 10
GeV (right panel) as a function of the pion transverse momentum
$|\Pperp|$ (with respect to the fragmenting quark axis) for two
different values of $z= $ 0.3 and 0.6. As in the previous figure, the
uncertainty bands correspond to the 68$\%$ CL. In both left and right panels, an additional
structure clearly emerges at intermediate $P_{\perp}$, especially at $z= 0.3$, which is induced by the weighted Gaussian in Eq.~\eqref{e:D1NP}.
Further investigations on this topic are needed, and data from electron-positron
annihilations would be valuable to
better explore these features.

We stress that the error bands displayed in
Figs.~\ref{f:tmdpdf}-\ref{f:tmdff} reflect the uncertainty on the
fitted parameters (see Eqs.~\eqref{e:f1NP}-\eqref{e:D1NP}) that are
determined by taking into account the uncertainty on the collinear PDFs
and FFs as discussed in Sec.~\ref{ss:errors}. However, since the fits
are performed using the central set of the collinear distributions, all TMD replicas
have the same integral in $\kperp$ (\textit{i.e.}, their values at
$\bT=0$ are the same). As a consequence, the plots in
Figs.~\ref{f:tmdpdf}-\ref{f:tmdff} only partially account for the error of the
collinear distributions.


\subsubsection{Collins--Soper kernel}
\label{sss:cskernel}

It is interesting to study the Collins--Soper
kernel~\cite{Collins:1981uk,Collins:2011zzd}
that drives the evolution of TMDs in terms of the
rapidity scale $\zeta$. Recent discussions of this crucial component of the TMD
formalism have been presented in
Refs.~\cite{Vladimirov:2020umg,Martinez:2022gsz} and estimates based on
lattice QCD have been proposed in Refs.~\cite{Schlemmer:2021aij,
Shanahan:2021tst, LPC:2022ibr}.


The Collins--Soper kernel, as written in
Eq.~\eqref{e:K_and_gK}, is composed of two parts. The first part can be
calculated perturbatively at N$^k$LL accuracy, and is computed at $b_\ast$:

\begin{equation}
\label{eq:PertCS}
K(b_*(|\mathbf{b}_T|),\mu) =
\sum_{n=0}^{k-1} \left(\frac{\alpha_s(\mu_{b_*})}{4 \pi}\right)^{n+1} K^{(n,0)}- \sum_{n=0}^{k}\gamma_K^{(n)} \int_{\mu_{b_*}}^{\mu}\frac{d\mu'}{\mu'} \left(\frac{\alpha_s(\mu')}{4 \pi}\right)^{n+1} \, ,
\end{equation}
where $K^{(n,0)}$ and $\gamma_K^{(n)}$ are coefficients of the perturbative
expansion (see, e.g., Ref.~\cite{Collins:2017oxh}). Note that the integral on the r.h.s. is directly computed
by means of numerical integration, thus providing a fully resummed result.
The second part, denoted as $g_K$, 
cannot be computed in perturbation theory 
and is one of the results of our fit. Only the full Collins--Soper kernel
can be compared with other works.

\begin{figure}[h]
\centering
\includegraphics[width=0.7\textwidth]{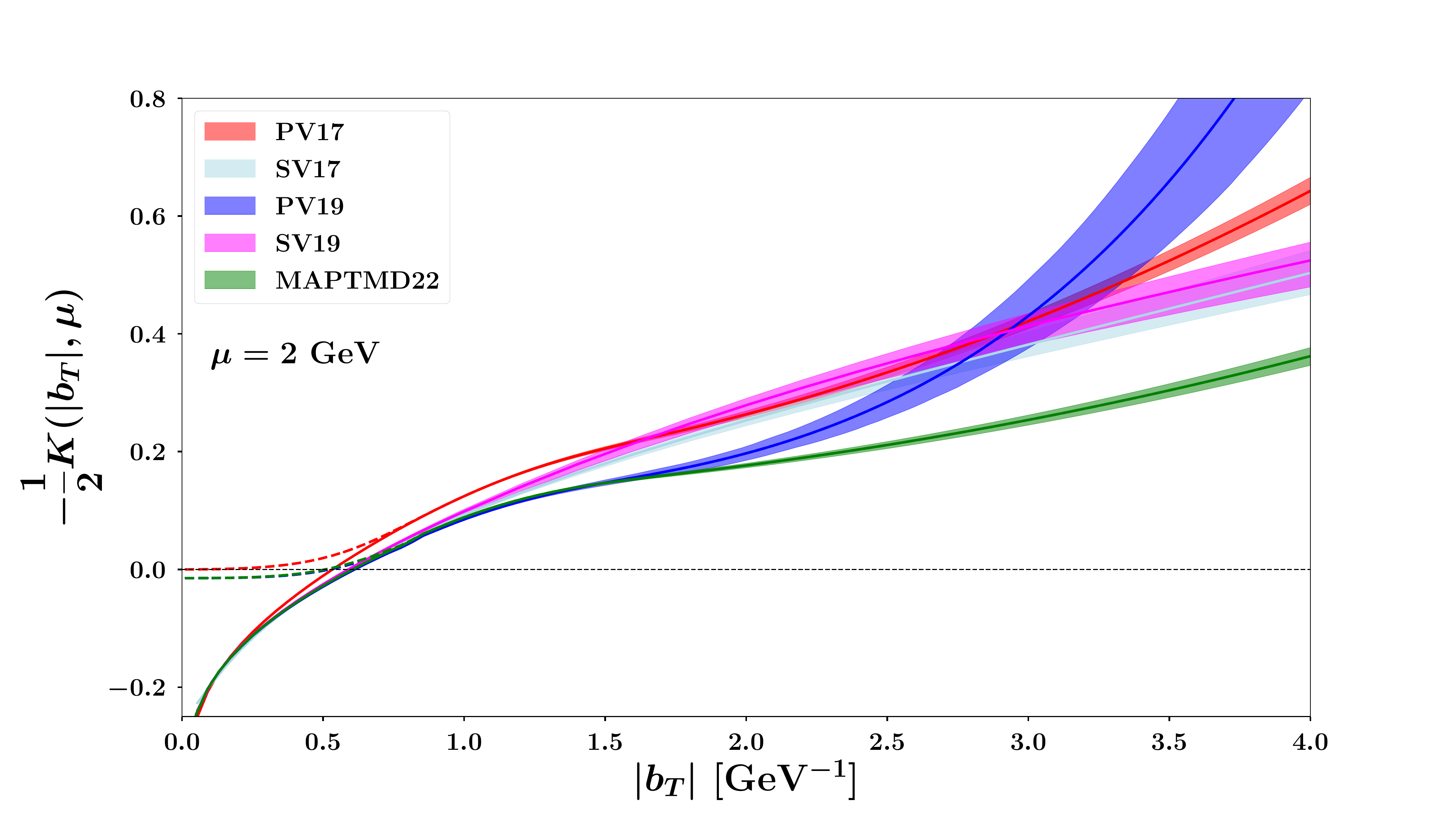}
\caption{The Collins--Soper kernel as a function of $|\bT|$ at a
scale $\mu=2$ GeV from the present analysis (MAPTMD22), compared with
the PV17~\cite{Bacchetta:2017gcc}, SV17~\cite{Scimemi:2017etj}, PV19~\cite{Bacchetta:2019sam}, and SV19~\cite{Scimemi:2019cmh} analyses. For
the MAPTMD22, PV17, and PV19 curves, the uncertainty bands represent the 68\% CL. The corresponding dashed lines show
the effect of including the $b_{\text{min}}$-prescription (see text).}
\label{f:cskernel}
\end{figure}

In Fig.~\ref{f:cskernel}, we show the Collins--Soper kernel as a function of
$|\bT|$ by conventionally keeping the scale $\mu$ fixed at 2 GeV, for our
present analysis (MAPTMD22, green band) and for
four other analyses in the
literature~\cite{Bacchetta:2017gcc,Scimemi:2017etj,Bacchetta:2019sam,Scimemi:2019cmh}.
The solid lines at low $|\bm{b}_T|$ follow the perturbative result. For MAPTMD22, PV19~\cite{Bacchetta:2019sam} and PV17~\cite{Bacchetta:2017gcc}, they
correspond to setting $b_{\mathrm{min}} = 0$ for sake of comparison with the other
SV19~\cite{Scimemi:2019cmh}, SV17~\cite{Scimemi:2017etj} results.
The slight differences between the curves are due to the different logarithmic
accuracies of the perturbative calculations: the PV17 analysis was performed at NLL,
the SV17 analysis at N$^2$LL, the PV19, SV19 and MAPTMD22 at N$^3$LL.
The size of the bands around the solid lines corresponds to one standard deviation
of the parameter $g_2$ around its best-fit value.
The $b_{\ast}$ prescription modifies the curves starting from
$|\bT| \approx 1$ GeV$^{-1}$. The behavior at high $|\bT|$ is driven by $g_K$ and is
different for the various analyses.

The dashed curves show the effect of using our prescription $b_{\mathrm{min}}
= 2e^{-\gamma_E} / \mu \approx 1.123/\mu$ in MAPTMD22, PV19 and PV17. This implies that at low $|\bT|$
the Collins--Soper kernel saturates to a finite value, as indicated by the dashed
lines.  As the scale increases, this
modification occurs at lower and lower values of $|\bT|$ and becomes less
relevant.

\subsubsection{Average squared transverse momenta}
\label{sss:avtm}

The average squared transverse momenta $\langle \kperp^2 \rangle(x,Q)$,
$\langle \Pperp^2 \rangle(z,Q)$ are calculated with the Bessel weighting
technique suggested in Refs.~\cite{Boer:2011xd,Boer:2014bya}.

In the case of the TMD PDF for a quark $q$ in the proton at $\mu = \sqrt{\zeta} = Q$, one has~\cite{Boer:2011xd,Boer:2014bya}:
\begin{align}
\label{e:avkp2}
\langle \kperp^2 \rangle^{q}(x,Q) & =
\frac{\int d^2 \kperp\, \kperp^2\, f_1^{q}(x,\kperp^2,Q,Q^2)}{\int d^2 \kperp\, f_1^{q}(x,\kperp^2,Q,Q^2)} =
\frac{2M^2\, \hat{f}_1^{q\, (1)}(x,|\bT|,Q,Q^2)}{\hat{f}_1^{q}(x,|\bT|,Q,Q^2)}\bigg|_{\modbT=0} \, ,
\end{align}
where the Fourier transform $\hat{f}_1^q$ of the TMD PDF has been defined in Eq.~\eqref{eq:FTdef} and the first Bessel moment of the TMD PDF $\hat{f}_1^{q\, (1)}$ is defined as~\cite{Boer:2011xd}:
\begin{equation}
\label{e:BesMom_f1}
\hat{f}_1^{q\, (1)}(x,|\bT|,Q,Q^2) = \frac{2\pi}{M^2}\,
\int_0^{+\infty} d|\kperp|\, \frac{\kperp^2}{|\bT|}\, J_1\big( |\kperp| |\bT|
\big)\, f_1^q(x,\kperp^2,Q,Q^2) =
-\frac{2}{M^2} \frac{\partial}{\partial \bT^2}
   \hat{f}_1^{q\,}(x,|\bT|,Q,Q^2)  \, .
\end{equation}

In order to obtain meaningful values for the average squared transverse momenta, \emph{i.e.}, finite, positive, and not dominated by the perturbative tails of the TMDs,
we shift the value of $|\bT|$ from 0 to a value well inside the nonperturbative region~\cite{Boer:2014bya}.
In this way the Bessel functions $J_{0,1}$ tame the power-law behavior of the TMD at large transverse momentum. 
The choice of the specific value for $\modbT$ is of course arbitrary
and has a significant effect on the associated numerical values. We
choose $\modbT=1.5\, b_{\text{max}}$ that guarantees that the average squared transverse momenta are positive across the $x$, $Q$ values considered in the fit.
Accordingly, Eq.~(\ref{e:BesMom_f1}) becomes:
\begin{equation}
\label{e:avkp2_reg}
\langle \kperp^2 \rangle^{q}_r (x,Q) =
\frac{2M^2\, \hat{f}_1^{q\, (1)}(x,|\bT|,Q,Q^2)}{\hat{f}_1^{q}(x,|\bT|,Q,Q^2)}\bigg|_{\modbT=1.5\, b_{\text{max}}} \, ,
\end{equation}
where the subscript $r$ stands for \emph{regularized}. We have checked that
the results are consistent when choosing either the integral or differential
expressions in Eq.~\eqref{e:BesMom_f1}.

The same arguments apply to the \emph{regularized} average squared transverse momentum produced during the hadronization of the quark $q$ into the final state hadron $h$~\cite{Boer:2011xd,Boer:2014bya,Bacchetta:2019qkv}:
\begin{align}
\label{e:avPp2_reg}
\langle \Pperp^2 \rangle^{q \to h}_r (z,Q) & =
\frac{2\, z^2\, M_h^2\, \hat{D}_1^{q \to h\, (1)}(z,|\bT|,Q,Q^2)}{\hat{D}_1^{q \to h}(z,|\bT|,Q,Q^2)}\bigg|_{\modbT=1.5\, b_{\text{max}}} \, ,
\end{align}
where the Fourier transform $\hat{D}_1^{q \to h}$ of the TMD FF is defined in Eq.~\eqref{eq:FTdefFF} and the first Bessel moment of the TMD FF $\hat{D}_1^{q \to h\, (1)}$ is defined as~\cite{Bacchetta:2019qkv}:
\begin{equation}
  \begin{split}
\label{e:BesMom_D1}
\hat{D}_1^{q \to h\, (1)}(z,|\bT|,Q,Q^2) &= \frac{2\pi}{M_h^2}\,
\int_0^{+\infty} \frac{d|\Pperp|}{z}\, \frac{|\Pperp|}{z}\,
\frac{|\Pperp|}{z|\bT|}\, J_1\big( |\bT| |\Pperp|/z \big)\, D_1^{q \to
  h}(z,\Pperp^2,Q,Q^2)
\\
&=
-\frac{2}{M_h^2} \frac{\partial}{\partial \bT^2}
\hat{D}_1^{q\to h}(z,|\bT|,Q,Q^2) \, .
\end{split}
\end{equation}

\begin{figure}[h]
\centering
\includegraphics[width=0.85\textwidth]{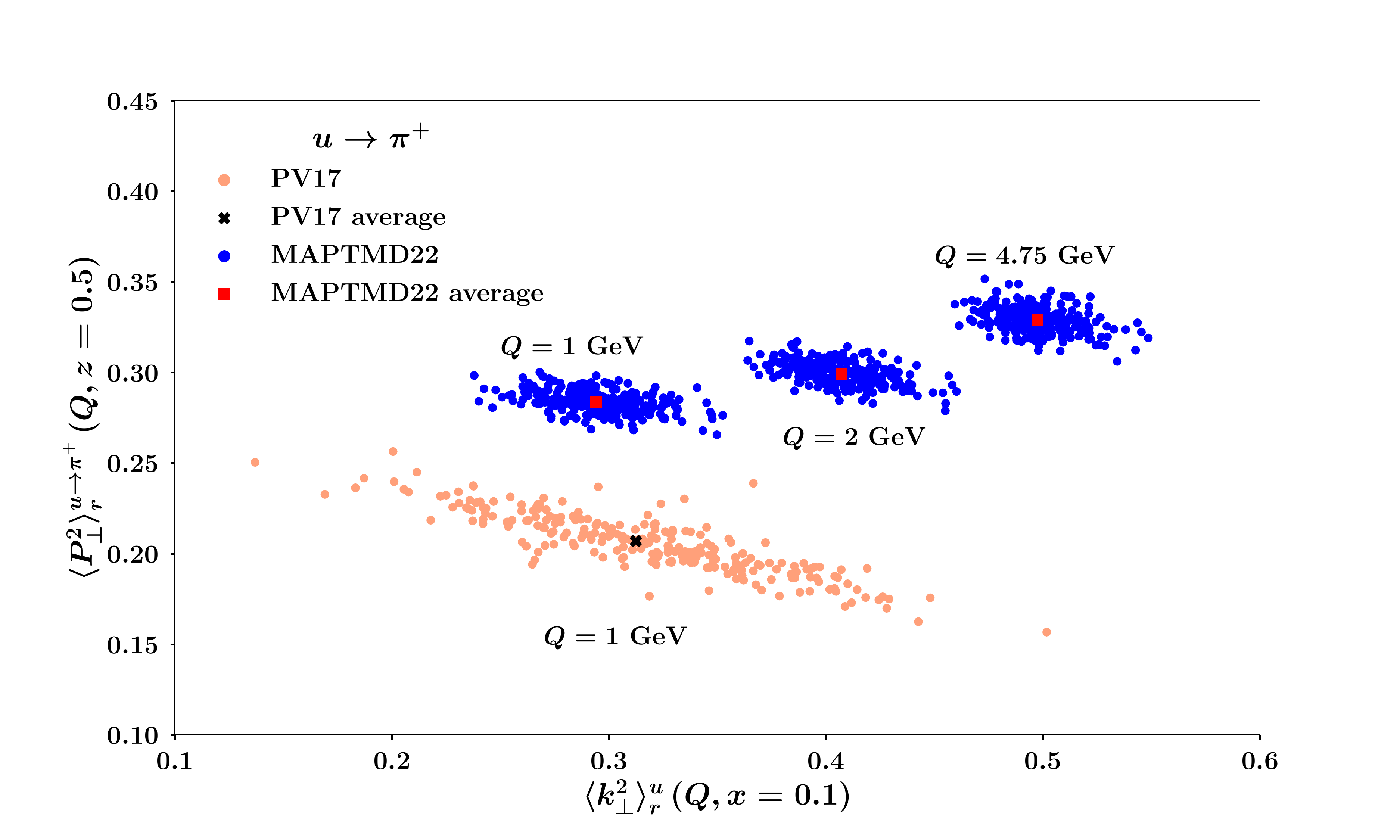}
\caption{Scatter plot of average squared transverse momenta for the TMD PDF of the up quark at $x=0.1$ and for the TMD FF of the $u\to \pi^+$ fragmentation at $z=0.5$. Orange circles for the PV17 analysis~\cite{Bacchetta:2017gcc} at NLL at $Q = 1$ GeV; the black cross represents the average. Blue circles for this analysis (MAPTMD22) at N$^3$LL$^-$ and at $Q =$ 1, 2, 4.75 GeV; the red squares represent the average values for each considered $Q$ value.}
\label{f:scatter_avtm}
\end{figure}

In Fig.~\ref{f:scatter_avtm}, we show the scatter plot of
$\langle \kperp^2 \rangle^u_r$ (up quark contribution) at $x=0.1$ vs.
$\langle \Pperp^2 \rangle_r^{u \to \pi^+}$ (``favored'' fragmentation) at $z=0.5$.
The blue circles (denoted  by MAPTMD22) correspond to the single replicas while the red
square is the average over all replicas for the N$^3$LL$^-$ analysis.
Three different values of $Q =$ 1, 2, 4.75 GeV are included to show the evolution of the average transverse momenta with the scale.
The orange circles indicate the PV17 replicas~\cite{Bacchetta:2017gcc}
at NLL and $Q = 1$ GeV with the black cross being the average.
No regularization is needed for the values extracted in the PV17 analysis, since the involved TMDs at $Q=1$ reduce entirely to their nonperturbative components.
By comparing MAPTMD22 at $Q = 1$ GeV to PV17, we observe that the
former produces much less anti-correlation between $\langle \kperp^2
\rangle$ and $\langle \Pperp^2 \rangle$ than the latter,
probably because of the inclusion of very precise DY data.

\subsection{Variations on the fit configurations}
\label{ss:fitvar}

In this subsection we discuss the results obtained by modifying some
of the baseline settings. In Sects.~\ref{sss:fitNNLL}
and~\ref{sss:fitNLL} we present fits at NNLL and NLL accuracy,
respectively, comparing them to the baseline N$^3$LL$^-$. Finally, in
Sec.~\ref{sss:qTcut} we study the impact of adopting different cuts
in $|\qT|$ on the SIDIS dataset.

\begin{table}[h]
\footnotesize
\begin{center}
\renewcommand{\tabcolsep}{0.4pc}
\renewcommand{\arraystretch}{1.2}
\begin{tabular}{|l|c|c|c|c|c|c|}
  \hline
  \multicolumn{1}{|c|}{ } & \multicolumn{2}{|c|}{N$^3$LL$^-$} & \multicolumn{2}{|c|}{NNLL} & \multicolumn{2}{|c|}{NLL} \\
  \hline
  Data set & $N_{\rm dat}$ & $ \langle \chi^2 \rangle \pm \delta \langle \chi^2 \rangle $ & $N_{\rm dat}$ & $ \langle \chi^2 \rangle \pm \delta \langle \chi^2 \rangle $ & $N_{\rm dat}$ & $ \langle \chi^2 \rangle \pm \delta \langle \chi^2 \rangle $ \\
  \hline
  \hline
  ATLAS & 72 & 5.01 $\pm$ 0.26 & / & / & / & / \\
  \hline
  PHENIX 200 & 2 & 3.26 $\pm$ 0.31 & 2 & 0.81 $\pm$ 0.11 & / & / \\
  \hline
  STAR 510 & 7 & 1.16 $\pm$ 0.04 & 7 & 0.99 $\pm$ 0.03 & / & / \\
  \hline
  Other sets & 170 & 0.83 $\pm$ 0.01 & 170 & 2.37 $\pm$ 0.11 & / & / \\
  \hline
  \hline
  DY collider & 251 & 2.06 $\pm$ 0.07 & 179 & 2.3 $\pm$ 0.1 & / & / \\
  \hline
  \hline
  E772 & 53 & 2.48 $\pm$ 0.12 & 53 & 2.05 $\pm$ 0.22 & / & / \\
  \hline
  Other sets & 180 & 0.87 $\pm$ 0.04 & 180 & 0.71 $\pm$ 0.04 & 180 & 0.81 $\pm$ 0.04 \\
  \hline
  \hline
  DY fixed-target & 233 & 1.24 $\pm$ 0.04 & 233 & 1.01 $\pm$ 0.05 & 180 & 0.81 $\pm$ 0.04 \\
  \hline
  \hline
  HERMES & 344 & 0.71 $\pm$ 0.04 & 344 & 1.1 $\pm$ 0.06 & 344 & 0.51 $\pm$ 0.02 \\
  \hline
  COMPASS & 1203 & 0.95 $\pm$ 0.02 & 1203 & 0.6 $\pm$ 0.06 & 1203 & 0.41 $\pm$ 0.01 \\
  \hline
  \hline
  SIDIS & 1547 & 0.89 $\pm$ 0.02 & 1547 & 0.71 $\pm$ 0.05 & 1547 & 0.43 $\pm$ 0.01 \\
  \hline
  \hline
  Total & 2031 & 1.08 $\pm$ 0.01 & 1959 & 0.89 $\pm$ 0.01 & 1727 & 0.47 $\pm$ 0.01 \\
  \hline
\end{tabular}
\caption{
  Comparison of $\chi^2$ values normalised to the number of data points
  $N_{\text{dat}}$ for fits at different perturbative accuracies. The $\langle \chi^2 \rangle$ and  $\delta \langle \chi^2 \rangle$ are the average and standard deviation of the $\chi^2$ values of all replicas.
  }
\label{t:chivariations}
\end{center}
\end{table}

\subsubsection{Global fit at NNLL}
\label{sss:fitNNLL}

The baseline fit presented in the previous section is performed at N$^3$LL$^-$
(see Tab.~\ref{t:logcountings}).
As already emphasized in Ref.~\cite{Bacchetta:2019sam}, the inclusion
of perturbative corrections up to N$^3$LL is crucial to achieve an
optimal description of some of the most recent experimental
measurements, such as those by the LHC. However, it might be useful to
extract unpolarized TMDs at lower perturbative orders. One of the
reasons is that such sets can be used in global analyses of polarized
TMDs where it is not possible to reach the same level of accuracy.

This is the case of the Sivers TMDs where the computation of the polarized cross section
for the Sivers effect presently cannot go beyond the NNLL level~\cite{Bacchetta:2020gko,Echevarria:2020hpy,Bury:2021sue}, hence demanding unpolarised TMDs at the same level of accuracy.

To this aim, we perform a new global fit at NNLL. However, when
lowering the perturbative accuracy, it is possible to obtain
acceptably good fits only by excluding those datasets whose precision
requires the highest theoretical accuracy. Specifically, we found that
only by removing the \textsc{ATLAS} dataset we were able to achieve an
acceptable global description at NNLL accuracy. As a matter of fact,
in Tab.~\ref{t:chivariations} the value of $\chi^2$ in this configuration, namely for fixed-target DY and SIDIS, is lower than at N$^3$LL$^-$ where \textsc{ATLAS} data is included.

Because of the difference in the perturbative accuracy as well as in the
dataset, we do not expect to get compatible values for the best fit
parameters between the NNLL and N$^3$LL$^-$ fits. For instance, we
obtain $\lambda = 12 \pm 10$ GeV$^{-1}$ and $\lambda_F = 340 \pm 280$
GeV$^{-2}$ at NNLL, to be compared to $\lambda = 1.8 \pm 0.3$
GeV$^{-1}$ and $\lambda_F = 0.08 \pm 0.01$ GeV$^{-2}$ at N$^3$LL$^-$.

The $\lambda$ and $\lambda_F$ parameters control the relative weight
of the weighted Gaussian in the non perturbative
part of the TMD PDF and FF, respectively, and control the size of the
DY and SIDIS spectrum at middle to large values of $|\qT|$.  The large
values obtained in the NNLL imply that the weighted Gaussian dominates
for both TMD PDF and FF parametrizations. This behavior may be
partially induced by the lack of perturbative corrections of the NNLL
fit with respect to the N$^3$LL$^-$ one, which are compensated by
nonperturbative effects.

\subsubsection{Global fit at NLL}
\label{sss:fitNLL}

We performed also a global analysis at NLL accuracy. Similarly to the
NNLL fit discussed above, it might be useful to have also a global fit at NLL accuracy for contexts where it is
not possible to reach the highest accuracy.  However, in order to
obtain acceptable $\chi^2$ values at this order, we had to exclude all
collider DY data and the E772 fixed-target DY dataset. Thus, we
reduced the dataset to the SIDIS data and the remaining fixed-target
DY data, \textit{i.e.}, \textsc{E605} and \textsc{E288}.

We point out that in our approach the integral of the $W$-term at
NLL is equal by construction to the SIDIS $\qT$-integrated collinear
cross section. As a consequence, the value of the prefactor $\omega$
in Eq.~\eqref{e:sidis_xsec_expr_w_norm} is automatically equal to 1.
Moreover, for this fit we consistently used \textsc{MMHT2014} at LO
for the collinear PDFs and we used \textsc{DSS} at NLO for
collinear FFs. As shown in Tab.~\ref{t:logcountings}, at NLL both
collinear PDFs and FFs should be evaluated at LO, but we choose FFs at
NLO because no recent extractions at LO are currently available.

As can be seen in Tab.~\ref{t:chivariations}, we obtain low $\chi^2$
values for all included datasets. It is interesting to compare this
result with that of the PV17 analysis~\cite{Bacchetta:2017gcc}. In
that work, the normalization of $\compass$ data was fixed by the first
bin in $\PhT^2$. Here we demonstrate that we can obtain an excellent
description of the most recent $\compass$ data without any adjustment
of the normalization.

\subsubsection{Cut in $|q_T|/Q$}
\label{sss:qTcut}

A crucial ingredient of any phenomenological analysis of TMDs is the
introduction of the kinematic cut $|\qT|/Q \ll 1$ on the dataset to ensure
that TMD factorization is valid. Our default choices are discussed in
Sec.~\ref{s:data}. It is interesting to study how the global quality
of our fit changes upon variations of this cut in order to get
quantitative information on the range of validity of TMD
factorization.

To this purpose, we changed the parameters $c_1$, $c_2$, and $c_3$ in
Eq.~\eqref{e:SIDIScut} devised for the SIDIS data, while keeping the
constraint $|\qT|/Q< 0.2$ for the DY data. We refer the reader to
Ref.~\cite{Bacchetta:2019sam} for an analogous study of the effect of
the $|\qT|/Q$ cut on the DY data.

We consider five different configurations for the cut on SIDIS data:
\begin{itemize}

\item[(a)] A first and most conservative cut is performed by fixing the z-independent
  upper value $|\qT|/Q < 0.4$ which can be obtained by setting $c_1=c_2=0.4$ and $c_3=0$ in Eq.~\eqref{e:SIDIScut};

           \item[(b)] A second cut by setting $c_1 = 0.15$,
             $c_2 = 0.4$ and $c_3 = 0.2$ in Eq.~\eqref{e:SIDIScut};

           \item[(c)] The cut of our baseline fit with $c_1 = 0.2$,
             $c_2 = 0.5$ and $c_3 = 0.3$;

           \item[(d)] A fourth cut with $c_1 = 0.2$, $c_2 = 0.6$ and
             $c_3 = 0.4$ but without imposing $|\qT| < Q$
             (\textit{i.e.} removing the outermost ``min'' in
             Eq.~\eqref{e:SIDIScut});

           \item[(e)] A fifth cut inspired by the PV17
             analysis~\cite{Bacchetta:2017gcc}, namely the same as in the previous case but with $c_1 = 0.2$,
             $c_2 = 0.7$ and $c_3 = 0.5$; this is the least conservative choice.
\end{itemize}

\begin{figure}[h]
\centering
\includegraphics[width=0.6\textwidth]{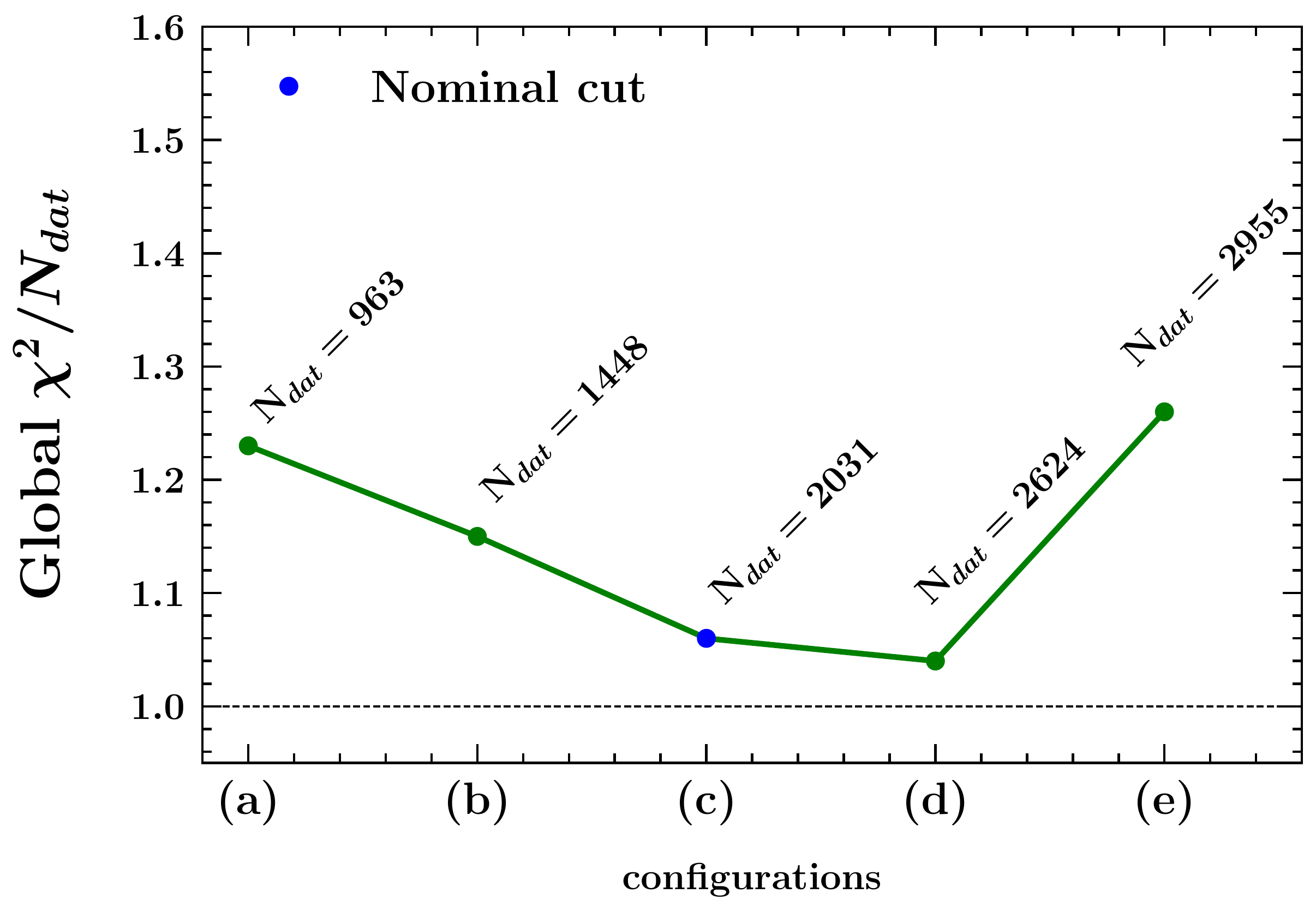}
\caption{Global $\chi^2 / N_{\text{dat}}$ for different configurations of the kinematic cut on SIDIS data sets (see text). The blue point corresponds to the reference cut used in the present baseline fit.}
\label{f:cuts}
\end{figure}

In Fig.~\ref{f:cuts}, we show how the global $\chi^2 / N_{\text{dat}}$ changes when considering the five configurations described above. By observing that more conservative choices do not
necessarily correspond to better $\chi^2$ values, we conclude that the
TMD formalism is able to describe SIDIS data that fails to fulfil
the formal requirement $|\qT|/Q \ll 1$. In this respect, we notice
that the global $\chi^2 / N_{\text{dat}}$ of cut (d) is
smaller than the baseline fit, despite including a larger amount of
data, some of which at $|\qT| \gtrsim Q$, \textit{i.e.}, well outside the region where TMD factorization is valid.

\begin{figure}[h]
\centering
\includegraphics[width=0.7\textwidth]{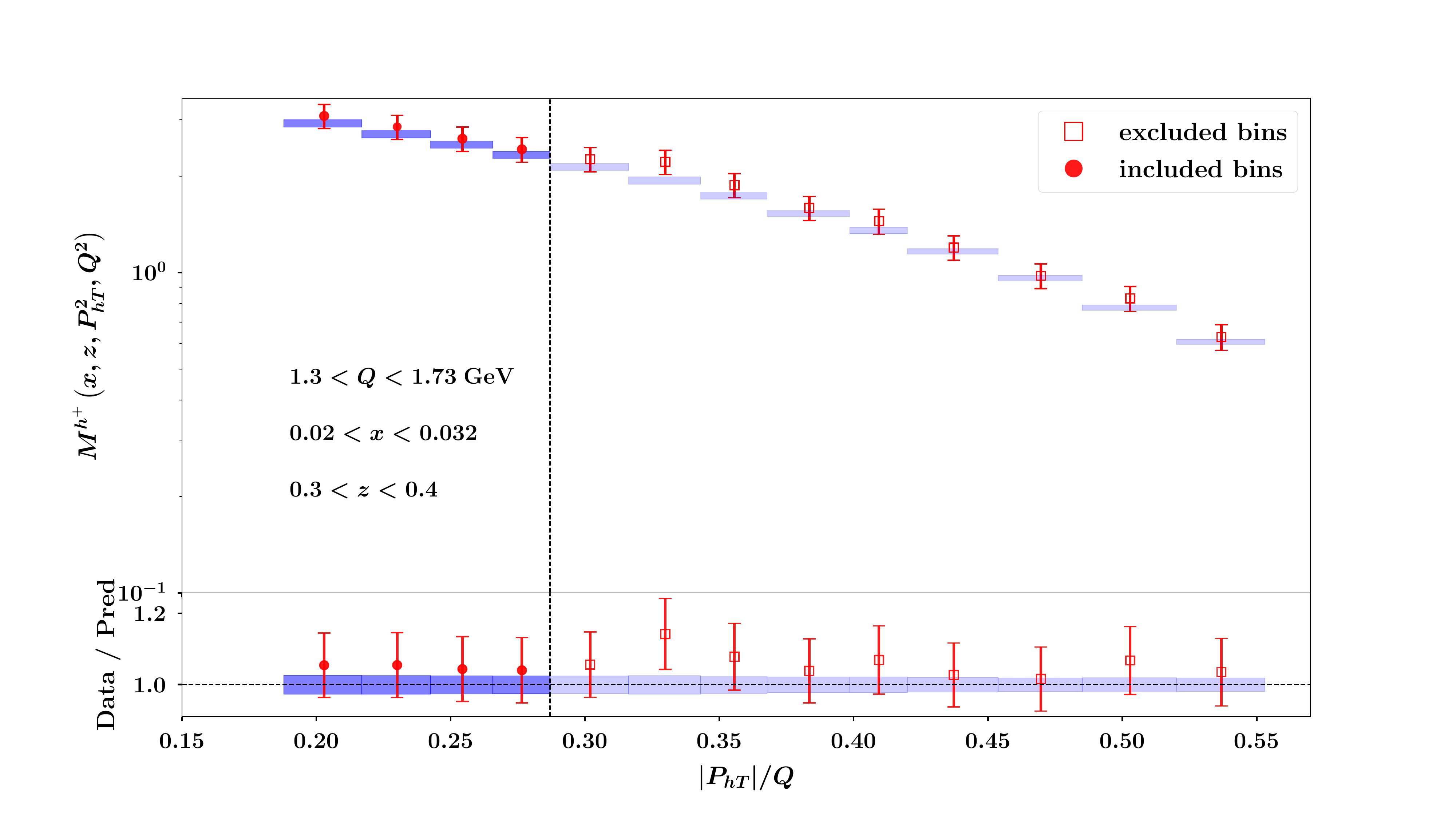}
\caption{Comparison between $\compass$ multiplicities and theoretical results for the SIDIS production of unidentified positively charged hadrons off a deuteron target at $1.3<Q<1.73$ GeV, $0.02<x<0.032$ and $0.3<z<0.4$ as a function of $|\PhT|/Q$. Upper panel: light-blue rectangles for baseline fit at 68\% CL, empty squares for data points not included in the baseline fit. Lower panel: ratio between experimental data and theoretical results.}
\label{f:COMPASS_excluded}
\end{figure}

In Fig.~\ref{f:COMPASS_excluded}, we better illustrate the situation by showing the comparison between
$\compass$ data and theoretical predictions from our baseline
fit for the SIDIS multiplicity for positively charged hadrons as a
function of $|\PhT|/Q$ in the bin $1.3<Q<1.73$ GeV, $0.02<x<0.032$,
$0.3<z<0.4$. The upper panel of the plot displays the multiplicity while the lower panel shows the ratio of experimental data to the theoretical predictions. The solid circles indicate data points
included in the fit while empty squares refer to those that do not survive the
cut. Remarkably, the agreement remains very
good up to $|\PhT|/Q \simeq 0.5$, well beyond the largest
$|\PhT|$ allowed by the cut. We also stress that this behavior is not
specific of the considered bin but is a general feature also of other bins, as well as of the $\hermes$ dataset.

In conclusion, from our analysis it emerges that the validity of the
TMD formalism in the kinematic region covered by $\compass$ and
$\hermes$
seems to extend well beyond the customary cut $|\qT|/Q \ll 1$.

This evidence justifies in a quantitative way our choice for the cut
$|\qT|/Q$ in Eq.~\eqref{e:SIDIScut} for the baseline fit, and explains
why we obtain values of $\chi^2 / N_{\text{dat}}$ close to one also
with less conservative cuts. Moreover, it suggests that the
applicability of TMD factorization in SIDIS might be defined in
terms of $|\PhT|$ rather than $|\qT|$, calling for more extensive
studies in this direction.

\section{Conclusions and outlook}
\label{s:conclusions}

In this article, we presented an extraction of unpolarized Transverse-Momentum Dependent Parton Distribution Functions and Fragmentation Functions (TMD PDFs and TMD FFs, respectively), which we refer to as MAPTMD22.  

We analyzed 2031 data points collected by several experiments: 251 data points from Drell--Yan (DY) production  measured at Tevatron, LHC and RHIC, 233 points from fixed-target DY (see Tab.~\ref{t:dataDY}) and 1547 data points from Semi-Inclusive Deep Inelastic Scattering (SIDIS) measured by the \hermes \ and \compass \ collaborations (see Tab.~\ref{t:dataSIDIS}).

Our description of the experimental observables is based on TMD factorization at a perturbative accuracy defined as N$^3$LL$^-$, in the sense that we use TMD evolution at the N$^3$LL level, hard factor and matching coefficients at order $\alpha^2_s$, collinear PDFs at NNLO, and collinear FFs at NLO (see Tab.~\ref{t:logcountings}).

We constructed the full TMDs by combining the perturbative components, regularized by means of the $b_{\ast}$ prescription defined in Eq.~\eqref{e:bTstar}, and nonperturbative parts described as a sum of Gaussians and weighted Gaussians (see Eqs.~\eqref{e:f1NP} and \eqref{e:D1NP}). We used 21 free parameters: 11 related to the TMD PDF, 9 for the TMD FF, and one for the Collins--Soper kernel driving the TMD evolution. We assumed these parameters to be the same for all quark flavors.

The TMD formalism is applicable only if observed transverse momenta are much smaller than the hard scale $Q$ of the considered process.
The details of our data selection
are explained in Sec.~\ref{s:data}. In the DY case, for the final lepton pair
transverse momentum $\qT$ we adopted a simple criterion, $|\qT| < 0.2\, Q$,
which is common to other works in the literature. In the SIDIS case, our
selection criterion is more restrictive than the one made in
Ref.~\cite{Bacchetta:2017gcc},  but less restrictive than the one made in
Ref.~\cite{Scimemi:2019cmh}.
For SIDIS data, where the detected hadron momentum is $\PhT \approx - z \qT$,
our cut includes many data points with $|\PhT| \ll Q$ but also with $0.2\, Q <
|\qT| < Q$.

We also tested the quality of our fit by exploring other criteria for the $|\qT|/Q$ cut. Surprisingly, we find that fits of quality comparable to the baseline result can be obtained with less conservative cuts such that $|\qT| \gtrsim Q$ for several data points. We believe that the definition of the range of applicability of the TMD factorization formula deserves further studies.


We found it difficult to reproduce the normalization of SIDIS data measured in fixed-target experiments at moderate to low scales. The formalism works well at NLL order, but severely underestimates the data at the N$^2$LL order and above. The reason can be traced back to the fact that the integral upon transverse momentum of the TMD cross section coincides with the known collinear result only at NLL, while it misses other contributions at higher orders. A rigorous solution of the problem would imply the calculation of all these contributions (including the so-called $Y$-term in the language of Ref.~\cite{Collins:2011zzd}), which are currently not under control, and are beyond the scope of this work. We decided to modify our predictions by including pre-computed normalization factors that are independent of the fit, and that are identified by comparing the integral upon transverse momenta of the TMD formula to the corresponding collinear calculation, as explained in Sec.~\ref{ss:norm_SIDIS}.

The fit is performed with the bootstrap method, leading to an ensemble of 250
replicas. We reach a very good overall agreement with data, with
$\chi^2/N_{\rm dat} = 1.06$ for the central replica (defined in Sec.~\ref{ss:fitqual}). The description of all individual datasets is satisfactory, except for a few cases, in particular for the ATLAS data (see Tab.~\ref{t:chitable} and Fig.~\ref{f:ATLAS_plot}). The discrepancy with the ATLAS data may be due to corrections not fully included in our analysis (see, {\it e.g.}, Ref.~\cite{Chen:2022cgv}) which, in spite of being small, can have a large impact in the comparison with very high precision data.

The resulting TMDs are shown in Figs.~\ref{f:tmdpdf} and~\ref{f:tmdff}. It is interesting to note that they deviate from a simple Gaussian shape (especially for the FFs) and the TMD shape changes in a nontrivial way as a function of $x$ or $z$. Tables with grids of the obtained TMDs will be made publicly available at the \textsc{NangaParbat} website\footnote{\href{https://github.com/MapCollaboration/NangaParbat}{https://github.com/MapCollaboration/NangaParbat}} and the \textsc{TMDlib} website\footnote{\href{https://tmdlib.hepforge.org}{https://tmdlib.hepforge.org}}~\cite{Abdulov:2021ivr}.

The availability of TMD analyses at increasing precision will be essential in
guiding detector design and feasibility studies for the future Electron Ion
Collider, and will also have a strong impact on precision measurements of
observables particularly sensitive to the hadron structure, such as $W$ mass
measurements at hadron colliders (see, {\it e.g.}, Refs.\cite{Bacchetta:2018lna,Bozzi:2019vnl,CDF:2022hxs}).

Our work can be extended by improving the perturbative accuracy of the
analysis (see, {\it e.g.}, Refs.~\cite{Camarda:2021ict,Chen:2022cgv}), including
theoretical uncertainties in terms
of suitable scale variations (along the lines of Ref.~\cite{Bertone:2022sso}),
improving the treatment of correlated uncertainties,
and investigating possible flavor dependence of TMDs~\cite{Signori:2013mda,Bury:2022czx}.

\begin{appendix}

\section{Explicit expression of $C$ coefficients at order $\alpha_s$}
\label{a:Ccoeff}

For completeness, in this Appendix we report the explicit expression of the
coefficients mentioned Sec.~\ref{ss:norm_SIDIS}.
The $C^{ab}_{\mathrm{TMD}}$ coefficients of Eq.~\eqref{eq:WintDetail} are given by:

\begin{align}
\begin{split}
C_{\rm TMD}^{q q}(x,z) &= 2 C_F \biggl(-8 \delta (1-x) \delta (1-z)
+ \delta (1-x)\Bigl[ 2 L_2(z) +(1-z) \Bigr]
+ \delta (1-z) (1-x) \biggr),
\end{split}
\\
C_{\rm TMD}^{g q}(x,z) &= 2 C_F \biggl( P_{gq}(z) \delta(1-x)\ln\left(z(1-z) \right) + z \delta(1-x) \biggr),
\\
C_{\rm TMD}^{q g}(x,z) &= 2 T_F \biggl( \delta(1-z) 2x(1-x) \biggr).
\end{align}



They correspond to the combination of the matching coefficients $C^{ab}$ included
in the TMD PDFs and FFs in Sec.~\ref{ss:TMDs}. The following $C^{ab}_{\mathrm{nomix}}$
coefficients are the ones included in the computation of the SIDIS normalization factors of Eq.~\eqref{e:nomix}:

\begin{align}
\begin{split}
C_{\rm nomix}^{q q}(x,z;Q,\mu) &=
C_{\rm TMD}^{q q}(x,z)+ 2 C_F \Biggl[
\delta (1-x)\Biggl(P_{qq}(z) \ln \frac{Q^2}{\mu^2} + L_1(z) -L_2(z) \Biggr)
\\ &\quad
+ \delta (1-z)\Biggl(P_{qq}(x) \ln \frac{Q^2}{\mu^2} + L_1(x) - L_2(x) \Biggr)
\Biggr],
\end{split}
\\
C_{\rm nomix}^{g q}(x,z;Q,\mu) &= C_{\rm TMD}^{g q}(x,z)+ 2 C_F
  \delta(1-x) P_{gq}(z)\ln \frac{Q^2}{\mu^2}
,
\\
C_{\rm nomix}^{q g}(x,z;Q,\mu) &= C_{\rm TMD}^{q g}(x,z)+ 2 T_F
\delta(1-z) P_{qg}(x)\ln\left(\frac{Q^2}{\mu^2} \frac{1-x}{x}\right).
\end{align}

Finally, the missing contributions to Eq.~\eqref{e:Ccoeff}, $C^{ab}_{\mathrm{mixed}}$, are given by:

\begin{align}
\begin{split}
C_{\rm mixed}^{q q}(x,z;Q,\mu) &=
2 C_F \Biggl[
2\frac{1}{(1-x)_+}\frac{1}{(1-z)_+} -\frac{1+z}{(1-x)_+} -\frac{1+x}{(1-z)_+}
+ 2 (1+xz)
\Biggr],
\end{split}
\\
C_{\rm mixed}^{g q}(x,z;Q,\mu) &= 2 C_F \Biggl[
P_{gq}(z)\frac{1}{(1-x)_+} + 2 (1+x-xz) -\frac{1+x}{z}
\Biggr],
\\
C_{\rm mixed}^{q g}(x,z;Q,\mu) &= 2 T_F
P_{qg}(x) \left[\frac{1}{(1-z)_+} +\frac{1}{z} -2 \right]
.
\end{align}

In the above expressions $\mu$ denotes the factorization scale for PDFs or FFs, while $L_{1,2}$ are abbreviations for the following logarithmic functions:
\begin{equation}
L_1(\xi) \equiv \left(1+\xi^2 \right) \left(\frac{\ln(1-\xi)}{1-\xi} \right)_+, \qquad L_2(\xi) \equiv \frac{1+\xi^2}{1-\xi} \ln\xi.
\end{equation}

\end{appendix}

\begin{acknowledgments}
We thank Mariaelena Boglione, Osvaldo Gonzalez Hernandez, Emanuele Nocera, Ignazio Scimemi, Alexey Vladimirov for stimulating discussions.
This work is supported by the European Union's Horizon 2020 programme under grant agreement No. 824093 (STRONG2020).
AS acknowledges support from the European Commission through the Marie Sk\l{}odowska-Curie Action SQuHadron (grant agreement ID: 795475).
CB is supported by the DOE contract DE-AC02-06CH11357.
\end{acknowledgments}
\bibliographystyle{JHEP}
\bibliography{MAPTMD22.bib}
\end{document}